\documentstyle[psfig,floats,aps,twocolumn]{revtex} 
 
\def\begeq{\begin{equation}} 
\def\endeq{\end{equation}} 
\def\beglett{\begin{mathletters}} 
\def\endlett{\end{mathletters}} 
\def\Btil{\tilde B} 
\def\btil{\tilde b} 
\def\dtil{\tilde D} 
\def\Dtil{\tilde D} 
\def\ctil{\tilde c} 
\def\ctilinf{\tilde c_\infty} 
\def\ktil{\tilde k} 
\def\ltil{\tilde l} 
\def\mtil{\tilde m} 
\def\Mtil{\tilde M} 
\def\qtil{\tilde q} 
\def\util{\tilde u} 
\def\Stil{\tilde S} 
\def\rhotil{\tilde\rho} 
\def\lapl{{\vec\nabla}^2} 
\def\lmin{\lambda_{\rm min}} 
\def\dtmin{\Delta T_{\rm min}} 
\def\pikap{\pi\kappa} 
\def\uav{\left<u\right>} 
\def\utilav{\left<\util\right>} 
\def\Kav{\left<K\right>} 
\def\xiav{\left<\xi\right>} 
\def\order{{\cal O}} 
\def\Re{{\rm Re}} 
\def\Im{{\rm Im}} 

\begin{document} 
 
\draft

\twocolumn[\hsize\textwidth\columnwidth\hsize\csname 
@twocolumnfalse\endcsname

\title{Eutectic Colony Formation: A Stability Analysis} 
 
\author{Mathis Plapp and Alain Karma} 
 
\address{Physics Department and Center for Interdisciplinary 
  Research on Complex Systems, \\ 
  Northeastern University, Boston MA 02115} 
 
\date{December 23, 1998}

\maketitle

\begin{abstract} 
Experiments have widely shown that a  
steady-state lamellar eutectic 
solidification front is destabilized on a scale much larger than 
the lamellar spacing by the rejection of a dilute ternary impurity
and forms two-phase cells commonly referred to as ``eutectic 
colonies''. We extend
the stability analysis of Datye and Langer {\rm [}V.~Datye and J.~S.~Langer, 
Phys. Rev. B {\bf 24}, 4155 (1981){\rm ]} for a binary eutectic to
include the effect of a ternary impurity.
We find that the expressions for the critical onset 
velocity and morphological instability wavelength 
are analogous to those for the classic Mullins-Sekerka 
instability of a monophase planar interface, 
albeit with an effective surface tension that depends on the 
geometry of the lamellar interface and, nontrivially, on 
interlamellar diffusion. A qualitatively new aspect of this
instability is the occurrence of oscillatory
modes due to the interplay between the destabilizing effect of
the ternary impurity and the dynamical feedback
of the local change in lamellar spacing on the front motion.
In a transient regime, these modes lead to the formation of
large scale oscillatory microstructures
for which there is recent experimental evidence  
in a transparent organic system. 
Moreover, it is shown that the 
eutectic front dynamics on a scale larger than the lamellar spacing
can be formulated 
as an effective monophase interface free boundary problem  
with a modified Gibbs-Thomson condition 
that is coupled to a slow evolution equation for the lamellar spacing. 
This formulation provides additional physical insights into the 
nature of the instability and a simple means to calculate 
an approximate stability spectrum. 
Finally, we investigate the influence  
of the ternary impurity on a short wavelength oscillatory instability 
that is already present at off-eutectic compositions in binary eutectics.
\end{abstract}

]

\section{Introduction} 
\nobreak 
\noindent 
The interfacial patterns  
that arise naturally during the solidification 
of eutectic alloys have attracted  
widespread interest for several 
decades from both fundamental and practical 
viewpoints. At a fundamental level,  
the main theoretical challenge lies in understanding 
the complex spatiotemporal dynamics of  
phase boundaries (solid-liquid and solid-solid) 
resulting from the competition of 
two thermodynamically stable solid phases growing 
simultaneously into a metastable liquid phase. In 
particular, one basic question is how to understand the  
nature of the morphological instability of the  
simplest spatially periodic steady-state that gives 
rise to this rich dynamics. From a practical viewpoint, 
the composite microstructure formed by lamellae or rods 
of these two solid phases growing simultaneously from the melt leads to  
interesting materials where the properties of two different solids can be  
advantageously combined. Moreover, the typical size of the microstructure 
pattern is about an order of magnitude smaller than in dendritic 
alloys, leading to superior mechanical properties. Consequently, 
understanding the solidification processing  
conditions that yield a particular eutectic 
microstructure is a goal of direct  
technological relevance. 
 
\begin{figure} 
\centerline{ 
  \psfig{file=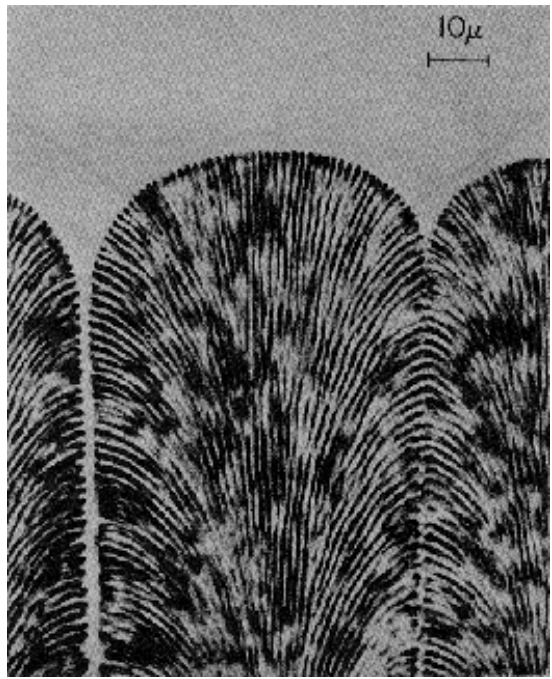,width=.4\textwidth}} 
\medskip 
\caption{Eutectic colonies in the transparent organic  
alloy CBr$_4$-C$_2$Cl$_6$ (from Ref. \protect\cite{Hunt66}),
grown by a directional solidification experiment.
The growth direction is from bottom to top.
The envelope of this two-phase structure  
(i.e. the solidification front on a scale 
much larger than the lamellar spacing) closely resembles 
the typical monophase cells observed during the directional 
solidification of a dilute binary alloy.} 
\label{fighunt} 
\end{figure} 
Since the early investigations of eutectic alloys, it has been 
remarked that besides the fine lamellae or rods, there 
may exist cellular structures, termed colonies 
\cite{Weart58,Chilton61,Kraft61,Hunt66,Gruzleski68,Rumball68,Rohatgi69,Bullock71,Rinaldi72}. 
Their size is typically 10 to 100 times the lamellar spacing. 
In Fig. \ref{fighunt} we show an experimental picture of 
colonies obtained in a directional solidification experiment 
\cite{Hunt66}. In their overall shape, the two-phase cells 
are remarkably similar to the monophase (i.e. single solid phase) 
solidification cells formed in standard 
directional solidification of a dilute binary alloy.  
This analogy is further supported 
by the experimental finding that colonies appear only when 
a ternary impurity, rejected by both solid phases, is  
present \cite{Chilton61,Bullock71}. In contrast, in binary eutectics  
the large scale solidification front stays planar for a 
range of compositions around the eutectic point. This suggests 
that the mechanism of the instability is similar to the 
classical Mullins-Sekerka instability of a monophase 
solidification front \cite{Mullins64}. Indeed, its onset is  
relatively well described quantitatively 
by the constitutional supercooling criterion 
\cite{Tiller58}, according to which the interface becomes 
unstable when the ternary impurity concentration gradient  
in the liquid ahead of the interface exceeds a critical value 
set by the ratio $G/v_p$ of the temperature gradient and 
the pulling speed of the sample. 
However, the spatiotemporal character of the  
linear modes associated with this instability 
has never been investigated.  
In particular, it has remained 
unclear how the Mullins-Sekerka analysis has to be modified 
to account for the composite structure of the interface. 
To answer this question, we present in this paper a linear 
stability analysis of a lamellar eutectic solidification 
front in the presence of a ternary impurity. 
 
It is useful to first briefly review the progress 
accomplished to date on the related problem  
of binary eutectic stability (without a ternary
impurity). The approach that we shall adopt 
here builds on earlier work in this context. 
Theoretical developments have mostly  
focused on the lamellar morphology in thin-film  
geometry, as then the problem can be treated as quasi  
two-dimensional. For a stability analysis, one must first 
obtain a steady-state solution: a shape-preserving solidification  
front propagating at constant velocity. Studies of this  
problem \cite{Tiller58,Brandt45,Zener46,Hillert57,Jackson66} 
led to the insight that there exists a family of steady-state 
solutions that can be parametrized by the lamellar spacing $\lambda$. 
The average undercooling $\Delta T$ of the solid-liquid interface  
with respect to the eutectic temperature depends on $\lambda$,  
and the curve $\Delta T$ versus $\lambda$ presents 
a minimum for a certain spacing $\lmin$. The experimentally observed 
spacings are usually close to $\lmin$ \cite{Trivedi88}. 
 
A hypothesis attributed to  
Cahn by Jackson and Hunt \cite{Jackson66} 
is that the lamellae always grow 
normal to the envelope of the solidification front.  
This hypothesis seems to work well in practice 
when the surface energy anisotropy of the solid-liquid 
and solid-solid phase boundaries is small enough  
to prevent locking of the lamellae to preferred growth directions. 
When used to analyze heuristically the long-wavelength stability of 
a eutectic front, this hypothesis 
leads to the conclusion that the lamellar structure 
is unstable for lamellar spacings below $\lmin$. 
Under Cahn's hypothesis, the lamellar spacing in a concave part of 
the solid-liquid interface decreases as the interface 
advances. Consequently, if the average spacing falls below $\lmin$,  
the local undercooling will increase in such a way that 
thinner lamellae fall further behind the front, leading finally to 
lamella termination. On the other hand, for spacings larger than $\lmin$, 
the opposite occurs: finer lamellae grow faster than 
wider ones and the concavity of the eutectic  
front is smoothed out. 
 
This argument can determine only a lower bound for $\lambda$. To 
assess stability for $\lambda>\lmin$, a more involved analysis is 
required. Several authors tried to adapt the linear stability analysis 
of Mullins and Sekerka for single-phase solidification \cite{Mullins64} 
to eutectic systems. The eutectic problem, however, is considerably 
more difficult because the basic steady-state solution is already 
periodic in space. Moreover, the presence of {\it mobile} trijunction  
points between three phases complicates enormously the  
stability calculation by ruling out  
a smooth sinusoidal perturbation.  
For this reason, early attempts to average over the properties 
of the two solid phases \cite{Cline68}, or to consider  
perturbations with immobile trijunctions 
\cite{Hurle68,Strassler74,Cline79},  
did not produce consistent results  
(see Ref. \cite{Karma96} for a more detailed discussion). 
 
The most complete analytical  
stability analysis of a eutectic interface has 
been performed by Datye and Langer (DL) \cite{Datye81}.  
Their calculation is a perturbation analysis of the  
Jackson-Hunt (JH) \cite{Jackson66} steady-state solution, using as  
basic variables the coordinates of the trijunction points both  
parallel and perpendicular to the interface. They first 
calculate an approximate solution to the diffusion equation for a 
perturbed lamellar interface. The assumption of local equilibrium 
at the solid-liquid interface and the use of Cahn's hypothesis then allow  
one to obtain an eigenvalue problem for the linear  
growth modes and to extract the 
stability spectrum of the interface. In the limit where the wavelength 
of the perturbation is large compared to the lamellar spacing  
(referred to hereafter as the ``long-wavelength limit''), 
a simplified calculation confirms JH's conclusion that lamellar 
spacings below $\lmin$ are unstable \cite{Langer80}. In addition, 
the DL analysis predicted the occurrence of an oscillatory instability 
with a wavelength twice the lamellar spacing for sufficiently  
off-eutectic compositions ($2\lambda$-O instability). 
 
The existence of this short-wavelength instability was later 
confirmed by numerical simulations of eutectic 
front dynamics using a random walk  
algorithm \cite{Karma87} and, more recently, 
a boundary integral approach \cite{Karma96,Sarkissian96}. 
The latter study pinpointed the existence 
of additional short-wavelength instabilities, one of which 
(tilt bifurcation) was previously known 
\cite{Karma87,Kassner91}, 
and made specific quantitative predictions  
for the CBr$_4$-C$_2$Cl$_6$ organic 
system that have been validated by a detailed comparison 
with experiments \cite{Faivre92,Ginibre97}.  
As an additional result, which is relevant for the present 
analysis, the boundary integral study revealed
that the stability predictions of the DL analysis are quite 
accurate for lamellar spacings close to $\lmin$, and only become 
inaccurate for larger spacings where the JH description of the 
diffusion field breaks down. 
 
In summary, according to both theory and experiment,  
a planar lamellar eutectic front in a binary alloy 
is completely stable for compositions 
sufficiently close to the eutectic composition, and for lamellar 
spacings near $\lmin$. To understand the instability leading 
to colony formation, one must therefore include a ternary impurity. To date,  
few theoretical studies of ternary systems have been available. Rinaldi,  
Sharp, and Flemings derived a generalized constitutional supercooling 
criterion for ternary systems and used it to interpret their experiments 
\cite{Rinaldi72}. McCartney, Hunt, and Jordan adapted the JH 
steady-state analysis to include impurities. They interpreted  
the formation of colonies as the result of a Mullins-Sekerka  
instability driven by the ternary impurity \cite{McCartney80}, 
but did not carry out a detailed stability analysis. 
 
In this paper, we extend Datye  
and Langer's linear stability approach, based on 
a Jackson-Hunt approximation of the diffusion field, 
to include the effect of a ternary impurity. For the 
reasons mentioned above, we expect this approach to yield 
relatively accurate predictions for spacings close to 
$\lmin$, which is typically the dynamically  
selected range of interest in experiments. 
We obtain the full linear stability spectrum of 
the steady-state lamellar eutectic  
front growing in two dimensions. 
Our final result is quite complicated, but  
can be substantially simplified for a model alloy with a symmetric  
phase diagram, solidified at its eutectic composition.  
From the study of this special case, we can identify all  
important factors that determine the stability of the front. 
In particular, we find that the interlamellar eutectic diffusion  
field gives a stabilizing contribution with a functional form 
similar to the usual capillary term. Thus, this 
contribution leads effectively to a  
``renormalization'' of the capillary length. Using 
this insight, we are able to  
reformulate the stability problem by treating the large
scale dynamics of the eutectic front 
as an `effective monophase interface', 
as suggested by Fig. \ref{fighunt}, with 
a Gibbs-Thomson condition that is coupled to an equation 
of motion for the local lamellar spacing. A similar
type of approach has been used previously to analyze the
long-wavelength modes of cellular arrays 
during directional solidification of
dilute binary alloys \cite{Karma89}. More recent
numerical work, however, has shown that the 
modes that limit the range of stable cell spacings at
low velocity are oscillatory and non-oscillatory 
instabilities with a wavelength
equal to twice the cell spacing \cite{Kopetal}
that have been observed in experiments \cite{Gor}.
In contrast, here, no short-wavelength instabilities are
present near the eutectic composition for $\lambda$ close
tp $\lmin$. Therefore, the effective interface approach
provides an accurate description of 
the interface dynamics in the limit of 
perturbation wavelengths much 
larger than the lamellar spacing. Moreover,  
it can be extended to derive a simplified  
expression for the stability spectrum by 
incorporating phenomenologically  
the effect of surface tension. This formula is 
found to predict all important  
features of the instability and to
yield reasonably good
quantitative predictions.

Our calculation both confirms the 
expectations based on the analogy between 
a two-phase eutectic front in the presence 
of a ternary impurity and a monophase front, and 
at the same time yields a
surprising result. Namely, we find that 
the expressions for the onset  
velocity and wavelength of the instability 
are analogous to those for a monophase 
front with a surface tension renormalized by 
the geometry of the lamellar front 
and interlamellar diffusion. Thus, as far as these 
quantities are concerned, the lamellar structure leads to 
quantitative differences, but no new qualitative features 
of the instability. 
The new ingredient, however, which could not have been 
expected on the basis of the analogy with a monophase front, is 
that the instability is oscillatory. The origin of this difference  
is due to the additional degrees of freedom associated with the 
underlying lamellar structure of the interface.  
According to Cahn's hypothesis, 
the change in the local lamellar spacing is determined by the  
shape of the front. The spacing, in turn, is related to the 
local interface temperature. As a consequence of the interplay  
between this effect and the instability driven by the impurities, 
long-wavelength perturbations may oscillate during growth or 
form traveling waves. There indeed seems to be recent experimental 
evidence for such large scale oscillatory behavior  
near the onset of colony formation 
in a transparent organic model alloy \cite{Faivre98}. 
 
We also investigate the  
influence of the ternary impurity on the 
short wavelength ($2\lambda$-O)  
oscillatory instability that is already 
present in a binary eutectic. The main result is that this 
instability is enhanced by the impurity boundary layer, 
which leads to a reduced composition range for stable 
lamellar growth even below constitutional supercooling. 
 
The structure of this paper is as follows. In Sec. II, we 
introduce the basic sharp-interface 
equations. We then summarize in Sec. III  
the Jackson-Hunt approach and apply it to calculate 
the steady-state solution in the presence of 
a ternary impurity. In Sec. IV,  
we review the principles of the DL approach and calculate the 
additional terms arising from the presence  
of an impurity. Section V is devoted to a detailed discussion of  
the stability spectrum at the eutectic composition 
in a model phase diagram that is symmetric about this composition. 
In Sec. VI, we reformulate 
the stability problem in terms of 
an effective interface approach and derive an approximate 
expression for the stability spectrum for an arbitrary 
phase diagram and material parameters. 
In Sec. VII, we discuss how the off-eutectic 
short-wavelength oscillatory instability is affected 
by the ternary impurity. 
Finally, we summarize our main  
results in Sec. VIII. 
 
\section{Basic equations} 
\nobreak 
\noindent 
We study the solidification of a binary eutectic alloy  
containing a small amount of ternary impurity. Let $c$ 
denote the concentration (in molecules per unit volume) 
of one of the constituents of the binary eutectic, and 
$\ctil$ the concentration of the ternary impurity. As we 
restrict our attention to small impurity concentrations, 
we shall assume that these two quantities can be treated 
as independent variables. In other words, we assume 
that the phase diagram of the binary eutectic is only 
slightly altered by the presence of the impurity.  
The two solid phases are denoted by $\alpha$ and $\beta$; 
$c_\alpha$ and $c_\beta$ are the concentrations limiting  
the eutectic plateau in the binary phase diagram,  
$\Delta c = c_\beta-c_\alpha$,  
and $c_E$ is the eutectic composition.  
 
\begin{figure} 
\centerline{ 
  \psfig{file=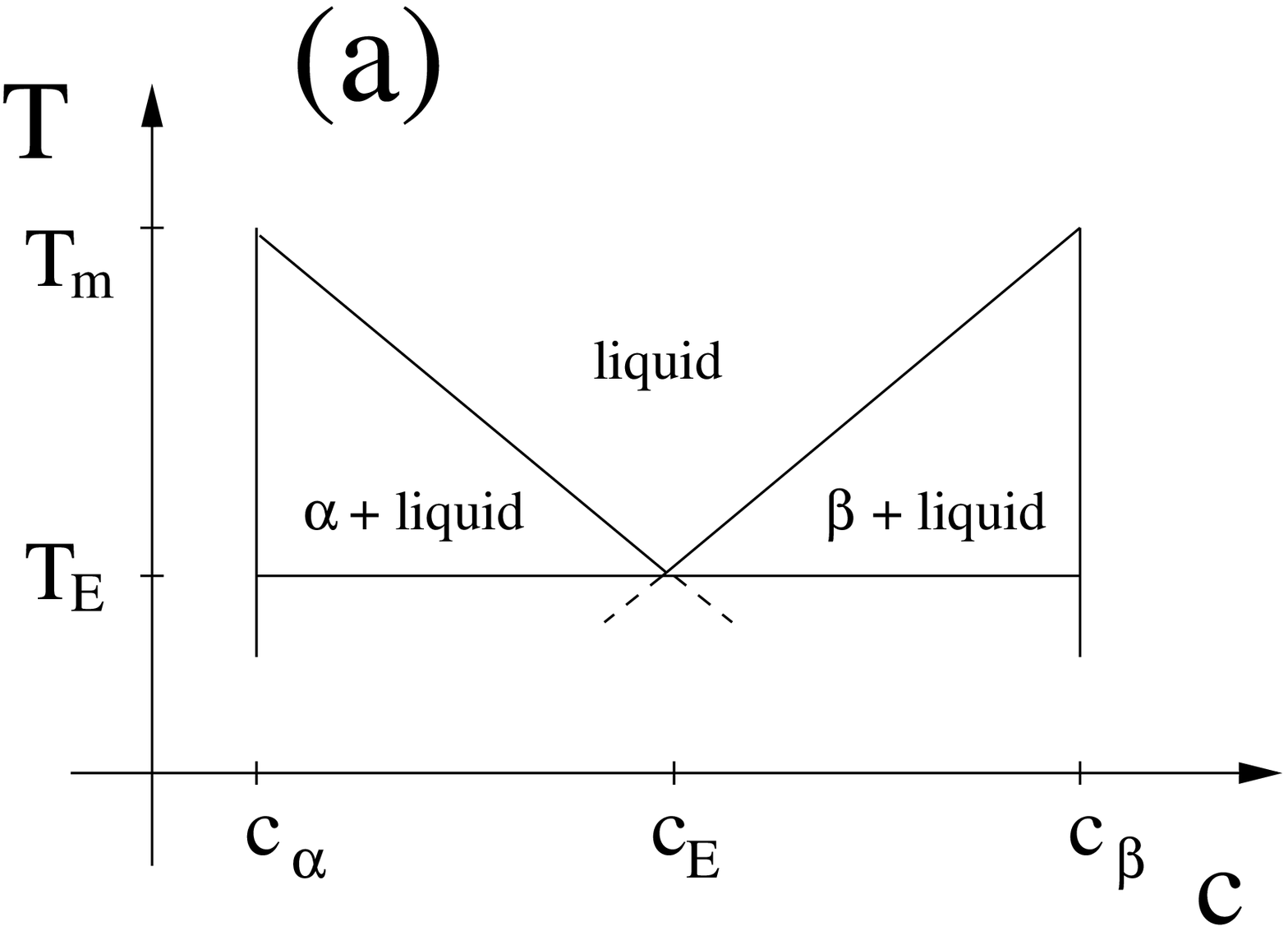,width=.25\textwidth}\hfill 
  \psfig{file=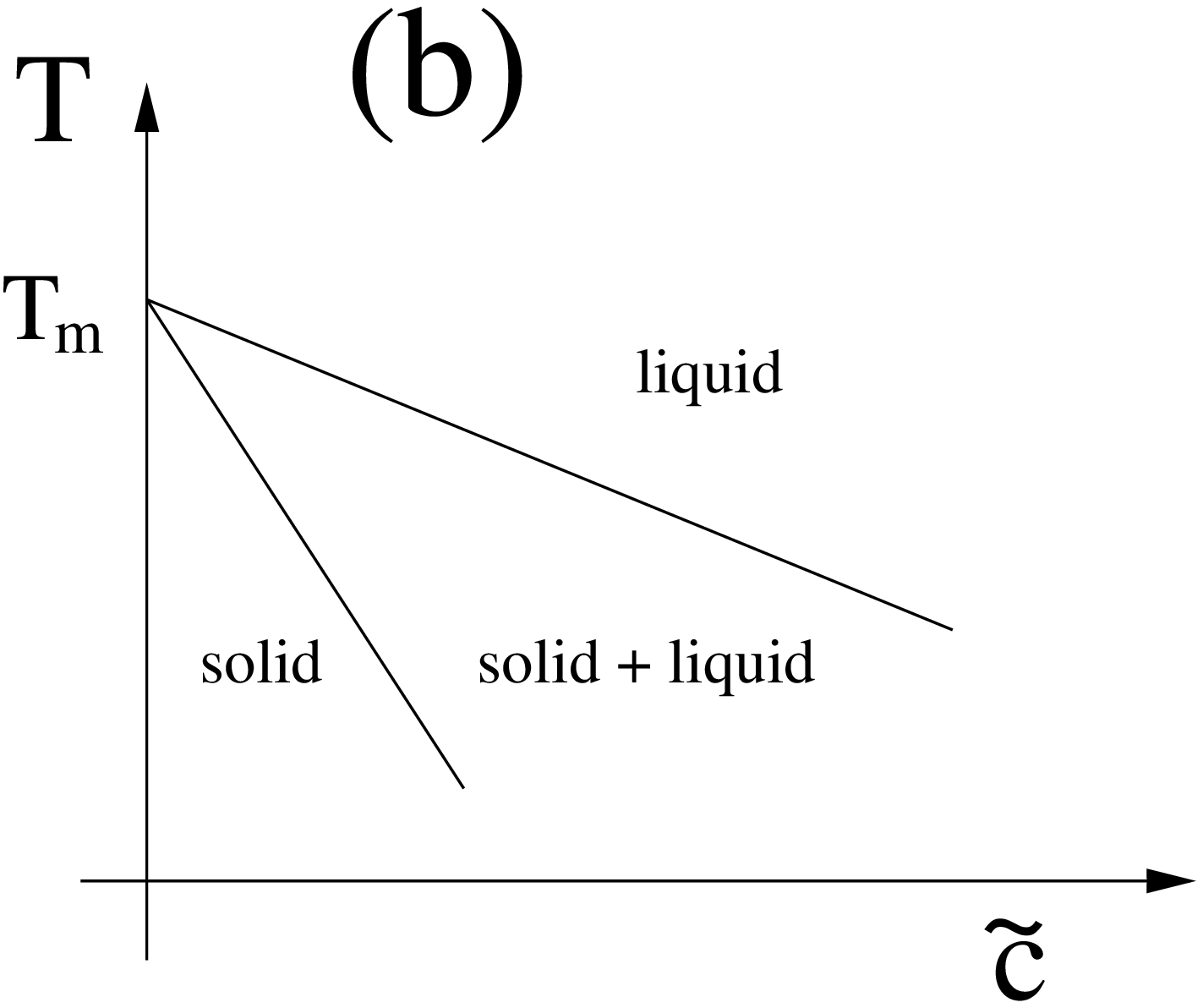,width=.2\textwidth}} 
\centerline{ 
  \psfig{file=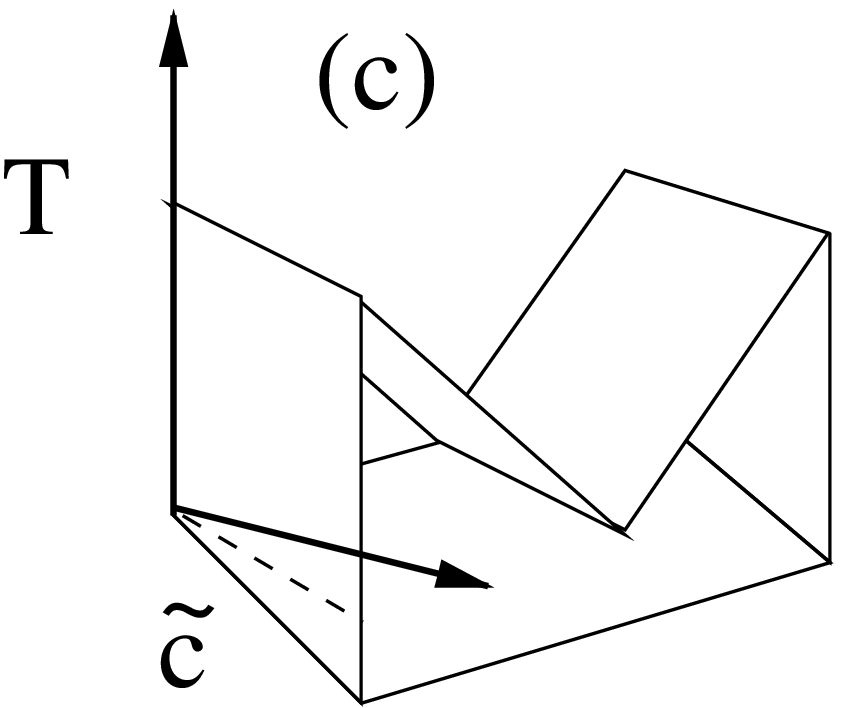,width=.2\textwidth}\hfill 
  \psfig{file=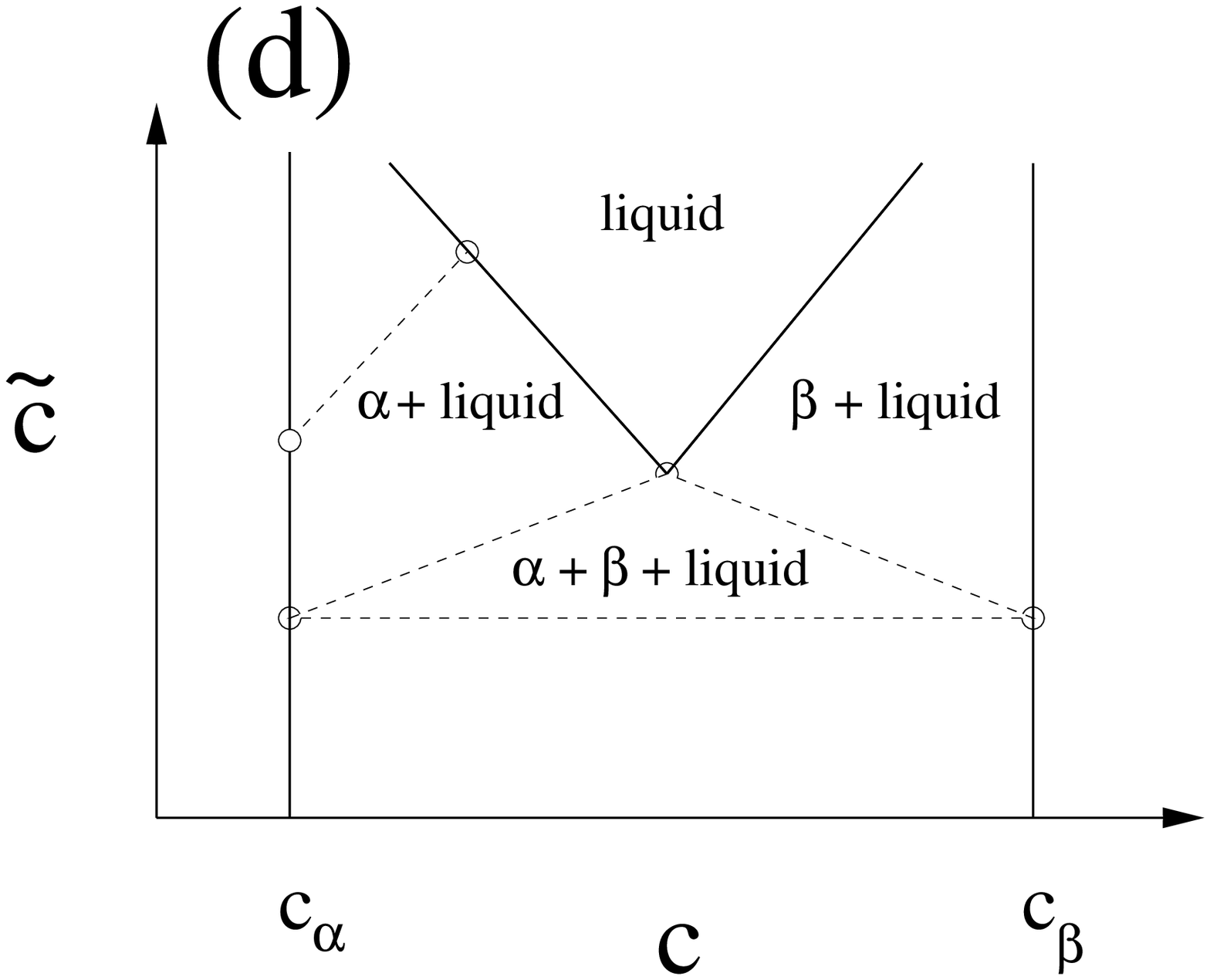,width=.225\textwidth}} 
\medskip 
\caption{Phase diagram of an idealized ternary eutectic alloy. 
(a): binary eutectic phase diagram ($\ctil = 0)$. $T_m$ is the melting  
temperature of the pure phases, $T_E$ the eutectic temperature,  
and $c_E$ the eutectic composition. (b): cut through the 
ternary diagram along the eutectic valley ($c=c_E$).
(c): liquidus and solidus surfaces in the space $(c,\ctil,T)$. The dashed  
line is the projection of the eutectic valley on the solidus surface. The 
liquidus surfaces have metastable extensions beyond the eutectic 
valley (not shown). (d): Coexistence curves for a fixed 
temperature below $T_E$ in the $(c,\ctil)$-plane.} 
\label{figphasediag}
\end{figure}

For coupled eutectic growth, the temperature of the lamellar 
front is close to the eutectic temperature, and the composition 
on the liquid side of the solid-liquid interface is close to 
the eutectic point \cite{Jackson66}. This allows to introduce 
two further simplifications. First, we may approximate the solidus  
and liquidus surfaces in the ternary phase diagram by planes 
around the eutectic point. We denote by $m_s$ and $\mtil_s$ 
($s = \alpha,\beta$) the magnitude of the liquidus slopes 
along the $c$- and $\ctil$-axis, respectively. Second, we will  
assume that $c_\alpha$ and $c_\beta$ are independent of temperature  
and impurity concentration. As already argued by DL, this should 
only slightly affect the final results, because in the temperature 
range explored by the front the relative variations of the 
concentration jumps across the interfaces are negligible. On 
the other hand, our calculations are considerably simplified 
as we can relate the volume fraction $\eta$ of the 
$\alpha$ solid to the composition of the melt far ahead 
of the interface, $c_\infty$, via the relation 
\begeq 
c_\infty = c_\alpha \eta + c_\beta (1-\eta), 
\endeq 
independently of the concentration of the ternary 
impurity and the interface 
temperature. For the impurity, we will work in a dilute 
alloy approximation where the impurity concentrations 
on the solid and liquid sides of the interface 
are related in equilibrium by 
\begeq 
\ctil_s = \ktil_s \ctil_L,\quad s = \alpha,\beta. 
\endeq 
The resulting phase diagram in the space $(c,\ctil,T)$ is  
sketched in Fig. \ref{figphasediag}. The line of intersection of 
the two liquidus surfaces is usually termed the {\em eutectic 
valley} or {\em eutectic trough}. Along this line, 
$c$, $\ctil$, and $T$ are related by the equations 
\begeq 
c-c_E = {\mtil_\beta-\mtil_\alpha\over m_\alpha+m_\beta} \ctil, 
\label{eutecticshift} 
\endeq 
\begeq 
T-T_E = \Mtil \ctil \qquad  
\label{euvalleyt} 
\endeq 
\begeq 
\Mtil = {\mtil_\alpha m_\beta + m_\alpha\mtil_\beta \over 
              m_\alpha + m_\beta}. 
\label{Mtildef} 
\endeq 
Equation (\ref{eutecticshift}) defines a monovariant line in the  
ternary phase diagram where, except for the special case  
$\mtil_\alpha=\mtil_\beta$, the eutectic composition is 
shifted with respect to the binary eutectic point in the 
presence of a ternary impurity. $\Mtil$ is the liquidus slope along the  
eutectic valley. A liquid satisfying Eqs. (\ref{eutecticshift}) 
and (\ref{euvalleyt}) can be in simultaneous 
equilibrium with two solids. 
 
In a typical directional solidification experiment, the sample is 
pulled in a temperature gradient $G$ with a constant pulling 
speed $v_p$. We assume that heat diffusion is much faster than 
chemical diffusion and that the thermal conductivities of 
solid and liquid are of comparable magnitude. Under 
this set of assumptions (commonly referred to as 
the frozen temperature approximation) , the temperature 
is given by  
\begeq 
T(z) = T_E + Gz, 
\endeq 
where we have chosen the origin  
of the $z$ axis at the eutectic 
temperature. 
 
In the absence of convection, the growth of the solid is limited  
by chemical diffusion of the constituents. 
We assume zero diffusivity in the solid (one-sided model). 
In the liquid, the diffusion equations in the laboratory frame 
(moving with velocity $v_p$ with respect to the sample) are
\beglett
\label{diffu0}
\begin{eqnarray}
{1\over D}{\partial c\over\partial t} & = &  
     {2\over l} \partial_z c + \lapl c, 
\label{diffue} \\ 
{1\over \Dtil}{\partial\ctil\over\partial t} & = &  
     {2\over \ltil} \partial_z \ctil + \lapl \ctil,
\label{diffui} 
\end{eqnarray}
\endlett
with the diffusion lengths $l=2D/v_p$ and $\ltil = 2\Dtil/v_p$, 
$D$ and $\Dtil$ being the diffusivities of the eutectic components 
and the ternary impurity, respectively. 
 
For lamellar eutectic growth in thin-film geometry, the problem 
is essentially two-dimensional. Let the position of the solid-liquid  
interface be described by the curve $\zeta(x,t)$.  
During the phase transformation, impurities and the minor component  
of the growing solid are rejected into the liquid.  
The condition of mass conservation implies that at the interface 
\beglett 
\begin{eqnarray} 
      -D \partial_n c & = & v_n \left[c(x,\zeta)-c_\alpha\right] 
          \quad (\alpha{\rm L}-{\rm interface}),\\
      -D \partial_n c & = & v_n \left[c(x,\zeta)-c_\beta\right]
	  \quad (\beta{\rm L}-{\rm interface}), \\
      -\dtil \partial_n \ctil & = & v_n (1-\ktil_\alpha)\ctil(x,\zeta) 
          \quad (\alpha{\rm L}-{\rm interface}), \\   
      -\dtil \partial_n \ctil & = & v_n (1-\ktil_\beta)\ctil(x,\zeta) 
          \quad (\beta{\rm L}-{\rm interface}), 
\end{eqnarray} 
\endlett 
where $v_n$ and $\partial_n$ denote the normal velocity of the 
interface and the derivative normal to the interface, respectively. 
 
We are interested in a regime of relatively low solidification 
velocity where the solid-liquid interface can be considered to be 
in local equilibrium, in which case the temperature and concentration 
fields at the interface are related  
by the Gibbs-Thomson conditions 
\beglett 
\label{githo} 
\begin{eqnarray} 
T & = & T_E - m_\alpha(c_L-c_E) - \mtil_\alpha\ctil_L \nonumber \\ 
  &   & \mbox{} - \Gamma_\alpha K[\zeta]\quad(\alpha\text{L-interface}), \\ 
T & = & T_E + m_\beta(c_L-c_E) - \mtil_\beta\ctil_L \nonumber \\ 
  &   & \mbox{} - \Gamma_\beta K[\zeta]\quad(\beta\text{L-interface}). 
\end{eqnarray} 
\endlett 
Here, $K[\zeta]$ is the local interface curvature, and 
\begeq 
\Gamma_s = T_E \gamma_{sL} / L_s,\quad s=\alpha,\beta, 
\endeq 
are the Gibbs-Thomson constants, with 
$\gamma_{sL}$ and $L_s$ denoting, respectively, the  
liquid-solid surface tensions and the latent heats of the  
two solids at the eutectic point. 
 
Finally, local equilibrium also implies that at the trijunction  
points, where the three phases are in contact, the angles between  
the three interfaces are fixed by the balance of 
surface tension forces, which for isotropic  
interfaces yields the two conditions 
\beglett 
\begeq 
\gamma_{\alpha L} \sin \theta_\alpha + \gamma_{\beta L} \sin \theta_\beta 
     =  \gamma_{\alpha\beta},
\endeq 
\begeq 
\gamma_{\alpha L}\cos\theta_\alpha = \gamma_{\beta L}\cos\theta_\beta, 
\endeq 
\endlett 
where the definition of the angles $\theta_\alpha$ and  
$\theta_\beta$ is illustrated in Fig. \ref{figeutstat}. 
 
\section{Steady-state solution} 
\nobreak 
\noindent 
For a binary eutectic, the steady-state problem has been treated by  
Jackson and Hunt \cite{Jackson66}. Their method has been extended 
to ternary systems by McCartney, Hunt, and Jordan \cite{McCartney80}. 
We need the steady-state solution as the starting point for 
our stability analysis. We will only summarize here 
the essential steps of the calculation; 
more details can be found in Refs. \cite{Jackson66,Datye81,McCartney80}. 
 
\begin{figure} 
\centerline{ 
  \psfig{file=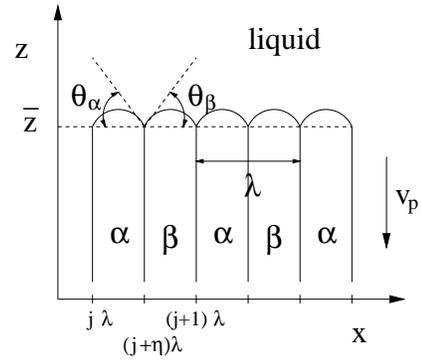,width=.3\textwidth}} 
\medskip 
\caption{Sketch of a steady-state array of lamellae growing 
parallel to the $z$-axis. The lamella pairs are numbered 
by the integer $j$; $\eta$ is the volume fraction of 
$\alpha$-phase, and $\bar z$ is the $z$-coordinate of 
the trijunction points.} 
\label{figeutstat} 
\end{figure} 
A typical configuration for a lamellar eutectic growing at constant  
solidification speed $v_p$ is sketched in Fig. \ref{figeutstat}.  
Alternating lamellae of the $\alpha$ and $\beta$ phase are regularly  
spaced. The width of one lamella pair, $\lambda$, is usually of  
the order of 10 $\mu m$. We must find 
the interface shape for which the diffusion equations and the 
Gibbs-Thomson conditions at the interface are both satisfied. 
The JH approximation starts by solving the diffusion problem  
for a flat interface 
(consisting of lamellae with $\theta_\alpha=\theta_\beta=0$). 
This can be achieved by expanding the diffusion field in Fourier 
modes along the $x$ axis: 
\begeq 
c(x,z)=c_\infty + \sum_{n=-\infty}^\infty B_n \exp[iq_nx-\bar q_n(z-\bar z)], 
\label{diffufield} 
\endeq 
with $q_n = 2\pi n/\lambda$, and $\bar z$ the $z$ coordinate of the 
trijunction points. The constants $\bar q_n$ are obtained by substituting 
the above sum into the diffusion equation. Since  
time derivatives are zero in steady-state,  
this yields at once: 
\begeq 
\bar q_n = 1/l + \sqrt{1/l^2+q_n^2}. 
\label{barqs} 
\endeq 
Inserting the expansion (\ref{diffufield}) into the mass conservation 
condition allows us to determine all Fourier coefficients except for $B_0$. 
The average undercooling for each lamella can then be calculated 
using the Gibbs-Thomson conditions. Finally, the condition that  
the two phases must grow at equal undercooling
determines $B_0$ and the average undercooling $\Delta T$ of the 
solid-liquid interface as a function of $\lambda$ and $\eta$. 
 
For a eutectic with a ternary impurity, we must in addition treat  
the diffusion of the ternary impurity in the liquid phase.  
For this purpose, we use the same Fourier expansion as above 
\begeq 
\ctil(x,z)= \ctilinf +  
     \sum_{n=-\infty}^\infty \Btil_n \exp[iq_nx-\qtil_n(z-\bar z)], 
\endeq 
where $\ctilinf$ is the impurity concentration far from the interface, 
and the constants $\qtil_n$ are equivalent to $\bar q_n$ with $l$ 
replaced by $\ltil$. 
This expansion is inserted into the condition for 
impurity conservation at the interface. To extract an equation 
for the Fourier coefficients $\Btil_n$, both sides of the equation are 
then multiplied by $\exp(-iq_m x)$ and integrated over $x$ from 
$0$ to $\lambda$. The result for the coefficient $\Btil_0$, 
which gives the magnitude of the overall diffusion boundary 
layer, can be written in the form 
\begeq 
\Btil_0=\ctilinf\left({1\over k_E}-1\right), 
\label{btilzero} 
\endeq 
with an effective partition coefficient 
\begeq 
k_E = \eta\ktil_\alpha+(1-\eta)\ktil_\beta. 
\label{kedef} 
\endeq 
For $n \neq 0$, we obtain 
\begin{eqnarray} 
\qtil_n\Btil_n & = &  
   {2\over\ltil}\left(1-{\ktil_\alpha+\ktil_\beta\over 2}\right)\Btil_n + 
   {2\over\ltil}\left(\ktil_\beta-\ktil_\alpha\right)\nonumber\\ 
 && \mbox{}\times  
    \biggl\{\ctilinf{e^{-iq_n\eta\lambda}-1\over -i\lambda q_n} 
        + (\eta-{1\over2})\Btil_n \nonumber \\ 
 && \quad \mbox{} +  
       \sum_{m \neq n}{e^{i(q_n-q_m)\eta\lambda}-1\over i\lambda(q_n-q_m)} 
     \Btil_m \biggr\}. 
\label{btilequation} 
\end{eqnarray} 
We are interested in a growth regime where the diffusion length 
is much larger that the lamellar spacing, and the P{\'e}clet  
number, ${\rm Pe} = \lambda/ l \ll 1$. Consequently, for $n\neq 0$, 
we have $\qtil_n \gg 1/\ltil$, and all terms containing the  
coefficients $\Btil_n$ on the RHS of Eq. (\ref{btilequation})  
can be neglected, which yields 
\begeq 
\Btil_n = {4\ctilinf\left(\ktil_\beta-\ktil_\alpha\right) 
           e^{-iq_n\eta\lambda/2}\sin(q_n\eta\lambda/2) \over 
           \lambda q_n\qtil_n\ltil}. 
\endeq 
Note that, if we want to go beyond this approximation, 
the problem becomes considerably more involved, because then all the 
Fourier coefficients are coupled. 
 
We proceed now as JH by calculating the average impurity 
concentration in front of each phase, that is, by evaluating 
\beglett 
\label{avdef} 
\begin{eqnarray} 
\langle\ctil\rangle_\alpha & = & 
   {1\over\eta\lambda}\int_0^{\eta\lambda}\ctil(x,z) dx, \\ 
\langle\ctil\rangle_\beta & = & 
   {1\over(1-\eta)\lambda}\int_{\eta\lambda}^\lambda\ctil(x,z) dx 
\end{eqnarray} 
\endlett 
at $z=\bar z$. The results are 
\beglett 
\label{ctilav} 
\begin{eqnarray} 
\left<\ctil\right>_\alpha & = & 
    {\ctilinf\over k_E} +  
    {2\lambda\over\ltil\eta}P(\eta)\ctilinf(\ktil_\beta-\ktil_\alpha) 
    \label{ctilavalp} \\ 
\left<\ctil\right>_\beta & = & 
    {\ctilinf\over k_E} - 
    {2\lambda\over\ltil(1-\eta)}P(\eta)\ctilinf(\ktil_\beta-\ktil_\alpha) 
    \label{ctilavbet} 
\end{eqnarray} 
\endlett
with
\begeq 
P(\eta) = \sum_{n=1}^\infty {1\over {(\pi n)}^3} \sin^2(\pi\eta n). 
\endeq 
 
Averaging the Gibbs-Thomson condition 
over individual lamellae, we obtain the mean undercooling of the 
solid-liquid interface: 
\beglett 
\begin{eqnarray} 
\left<\Delta T\right>_\alpha & = &  
    m_\alpha(\left<c\right>_\alpha-c_E) + 
    \mtil_\alpha\left<\ctil\right>_\alpha +  
    \Gamma_\alpha\left<K\right>_\alpha \\ 
\left<\Delta T\right>_\beta & = &  
    - m_\beta(\left<c\right>_\beta-c_E) + 
    \mtil_\beta\left<\ctil\right>_\beta +  
    \Gamma_\beta\left<K\right>_\beta, 
\end{eqnarray} 
\endlett 
where the average values $\left<\Delta T\right>_s$, $\left<c\right>_s$,  
and $\left<K\right>_s$ $(s=\alpha,\beta)$ are defined by expressions 
analogous to Eqs. (\ref{avdef}) for $\left<\ctil\right>_s$. 
The averages for the composition, $\left<c\right>_s$, 
and for the curvatures $\left<K\right>_s$, are \cite{Datye81}: 
\beglett 
\label{caverage} 
\begin{eqnarray} 
\left<c\right>_\alpha-c_E & = & c_\infty+B_0+{2\lambda\Delta c\over l\eta}P(\eta) \\ 
\left<c\right>_\beta-c_E & = & c_\infty+B_0-{2\lambda\Delta c\over l(1-\eta)}P(\eta)  
\end{eqnarray} 
\endlett 
\beglett 
\label{kaverage} 
\begin{eqnarray} 
\left<K\right>_\alpha & = & {2\over\eta\lambda}\sin\theta_\alpha  \\ 
\left<K\right>_\beta & = & {2\over(1-\eta)\lambda}\sin\theta_\beta. 
\end{eqnarray} 
\endlett 
The last step is to apply the condition that neighboring lamellae 
should grow at equal undercooling, $\left<\Delta T\right>_\alpha =  
\left<\Delta T\right>_\beta$. This determines the only degree 
of freedom left in the problem: the eutectic boundary layer $B_0$. 
The solution is 
\begin{eqnarray} 
B_0 & = & -(c_\infty-c_E) + {\ctilinf\over k_E}  
     {\mtil_\beta-\mtil_\alpha\over m_\alpha+m_\beta} \nonumber \\ 
  && \mbox{}+ 
        {2\lambda P(\eta)\Delta c\over m_\alpha+m_\beta} 
       \left[{1\over l} 
        \left({m_\beta\over 1-\eta}-{m_\alpha\over\eta}\right) 
         \right. \nonumber \\ 
  && \quad\mbox{} - \left.{1\over\ltil} 
       \left({\mtil_\alpha\over\eta}+{\mtil_\beta\over 1-\eta}\right) 
       {\ctilinf\over\Delta c}\left(\ktil_\beta-\ktil_\alpha\right)\right]. 
\label{bzero} 
\end{eqnarray} 
There are two terms that are not present in the binary eutectic. 
The second term on the RHS of Eq. (\ref{bzero}) gives, according 
to Eq. (\ref{eutecticshift}), the shift of the eutectic 
composition corresponding to an impurity concentration $\ctilinf/k_E$. 
It is remarkable that, even for a system that started from an 
initial state with three-phase equilibrium, a eutectic boundary 
layer must develop in order that the composition  
condition for three-phase equilibrium 
at the trijunction points can be met. The second new term is 
the last term in brackets on the RHS of Eq. (\ref{bzero}), 
involving the difference of the partition coefficients. This term 
is due to the unequal rejection of impurities into the liquid. 
 
The interfacial undercooling as a function of the lamellar spacing  
can be written in a form very similar to the JH result for a simple binary 
eutectic: 
\begeq 
\Delta T = \Mtil {\ctilinf\over k_E} + {1\over 2} \Delta T_{min} \left( 
    {\lambda\over\lmin}+ {\lmin\over\lambda} \right). 
\label{jhunt} 
\endeq 
The first term, according to Eq. (\ref{euvalleyt}), gives the  
undercooling of the point in the eutectic valley corresponding  
to an impurity concentration $\ctilinf/k_E$. 
The minimum undercooling $\dtmin$ 
and the corresponding spacing $\lmin$ are
\begeq 
\dtmin = {4 \Delta c\over \eta (1-\eta)} 
        \left( m_\alpha m_\beta\over m_\alpha + m_\beta\right) 
	\sqrt{f(\eta)p(\eta,\ctilinf)} 
\label{deltatmin} 
\endeq
\begeq 
\lmin = \sqrt{f(\eta)/p(\eta,\ctilinf)} 
\label{lambdamin} 
\endeq 
with 
\begeq 
f(\eta) = {(1-\eta)\Gamma_\alpha\sin\theta_\alpha\over m_\alpha\Delta c} + 
              {\eta\Gamma_\beta\sin\theta_\beta\over m_\beta\Delta c} 
\endeq 
\begin{eqnarray} 
p(\eta,\ctilinf) & = & 
             {P(\eta)\over l}+{P(\eta)\over\ltil} 
                {\ctilinf\over\Delta c}\left(\ktil_\beta-\ktil_\alpha\right) 
                  \nonumber \\ 
 & & \mbox{} \times \left[(1-\eta){\mtil_\alpha\over m_\alpha} 
                 -\eta{\mtil_\beta\over m_\beta}\right]. 
\label{peqn} 
\end{eqnarray} 
Note that, as in a binary eutectic, we have $\dtmin \sim \sqrt{v_p}$ and 
$\lmin\sim 1/\sqrt{v_p}$. In the special case $\ktil_\alpha=\ktil_\beta$,  
we recover the classic JH result. The two $\lambda$-dependent terms  
on the RHS of Eq. (\ref{jhunt}) represent the effects of diffusion 
and surface tension. For finer lamellae, the diffusion between 
adjacent lamellae is faster, and the undercooling due to the 
concentration term in the Gibbs-Thomson relation is smaller. 
This gives the term proportional to $\lambda$. On the other hand, 
for finer lamellae the average curvature is higher due to the 
constraints at the trijunction points, leading to the term proportional 
to $\lambda^{-1}$. 
 
For our subsequent stability analysis, we will now simplify the 
problem. The main effect of the impurities, rejected by both solid 
phases, is the buildup of the impurity boundary layer of amplitude 
$\Btil_0$. The diffusion of impurities between neighboring 
lamellae leads to corrections in $\lmin$ and $\dtmin$. In the 
dilute limit, however, where $\ctilinf \ll \Delta c$, these corrections 
are small. In addition, the partition coefficients for ternary 
impurities are often close to zero, and we have  
$\ktil_\alpha - \ktil_\beta \ll 1$. Thus, it seems well 
justified to make the approximation  
$p(\eta,\ctilinf) \approx P(\eta)/l$ in Eq. (\ref{peqn}) and 
to drop the term involving the difference of the partition 
coefficients in Eq. (\ref{bzero}). This means that we neglect 
the interlamellar impurity diffusion modes, $\Btil_n = 0$  
for $n \ne 0$, which is equivalent to the assumption of equal  
impurity partition coefficients, $\ktil_\alpha=\ktil_\beta=k_E$.  
For two very different impurity partition coefficients,  
it might be necessary to go beyond this approximation and 
to include the interlamellar impurity diffusion. 
 
We will also assume equal impurity liquidus slopes $\mtil_\alpha$ 
and $\mtil_\beta$ for most of what follows. Then, the 
eutectic composition and the magnitude of the  
eutectic boundary layer, described by the coefficient $B_0$, 
do not depend on the impurity concentration. We will briefly 
comment on the general case at the end of Sec. VI.  
 
\section{Stability analysis} 
\begin{figure} 
\centerline{ 
  \psfig{file=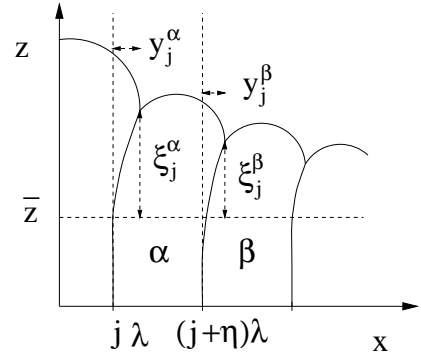,width=.3\textwidth}} 
\medskip 
\caption{Sketch of a perturbed lamellar interface,
 showing the displacements 
$y_j^\alpha$, $y_j^\beta$, $\xi_j^\alpha$, and $\xi_j^\beta$.} 
\label{figeutpert} 
\end{figure}

\nobreak 
\noindent 
The DL method is a perturbation analysis around the JH 
steady-state solution. The fundamental variables in this approach  
are the coordinates of the trijunction points, or 
more precisely the departure of these coordinates from their 
steady-state values.  
As illustrated in Fig. \ref{figeutpert}, the coordinates 
of the trijunction points of the $j$th 
lamella pair are written as
\begeq 
x_j^\alpha = j\lambda + y_j^\alpha(t) \quad x_j^\beta = (j+\eta)\lambda + y_j^\beta(t) 
\endeq 
\begeq 
z_j^\alpha = \bar z + \xi_j^\alpha(t) \quad z_j^\beta = \bar z + \xi_j^\beta(t). 
\endeq 
We assume the system to consist of a total number of $N$ lamella 
pairs, and use periodic boundary conditions. For convenience, we 
define a dimensionless eutectic concentration field by 
\begeq 
u = {c-c_E \over \Delta c}, 
\label{udef} 
\endeq 
and a dimensionless impurity concentration field 
\begeq 
\util = {\ctil-\ctilinf\over \Delta\ctil}, 
\label{utildef} 
\endeq 
with $\Delta\ctil = \ctilinf (1/k_E-1)$. 
 
Let us outline the strategy of the DL calculation. For a slowly 
evolving, slightly perturbed interface, the Gibbs-Thomson condition for 
local equilibrium remains satisfied. The deformation of 
the front modifies the local curvatures and concentrations. 
The perturbed diffusion fields can be calculated using the mass 
conservation conditions. For small displacements, all resulting  
expressions are linearized in $\xi_j^s$ and $y_j^s$. To close the set 
of equations, DL use Cahn's hypothesis, which gives additional 
relations between the $\xi$'s and the $y$'s. Inserting all these 
expressions into the Gibbs-Thomson condition allows one to  
determine the stability spectrum of the interface. 
 
The new element here is the ternary impurity diffusion field. To 
obtain the modifications it generates in the stability spectrum, 
we have to introduce the procedure in more detail. As in the 
JH calculation, a complete solution of the free boundary problem 
is out of reach, and we use quantities that are averaged over 
individual lamellae. The average undercooling of a lamella is  
written as 
\begeq 
\Delta T_j^s(t) = \Delta T + \delta T_j^s(t) \quad (s=\alpha,\beta), 
\endeq 
with $\Delta T$, the steady-state value, given by Eq. (\ref{jhunt}). 
To express the $\delta T_j^s(t)$'s, we have to calculate the deviations 
of curvature and concentrations from their steady-state values, 
that is: 
\beglett 
\begeq 
\delta\Kav_j^\alpha(t) = {1\over x_j^\beta-x_j^\alpha} 
   \int_{x_j^\alpha}^{x_j^\beta} K[\zeta(x,t)]dx - \left<K\right>_\alpha, 
\endeq 
\begeq 
\delta\Kav_j^\beta(t) = {1\over x_{j+1}^\alpha-x_j^\beta} 
   \int_{x_j^\beta}^{x_{j+1}^\alpha} K[\zeta(x,t)]dx - \left<K\right>_\beta, 
\endeq 
\endlett 
\beglett 
\label{deltaus} 
\begeq 
\delta\uav_j^\alpha(t) = {1\over x_j^\beta-x_j^\alpha} 
   \int_{x_j^\alpha}^{x_j^\beta} u(x,\zeta(x,t),t)dx - \left<u\right>_\alpha,  
\endeq 
\begeq 
\delta\uav_j^\beta(t) = {1\over x_{j+1}^\alpha-x_j^\beta} 
   \int_{x_j^\beta}^{x_{j+1}^\alpha} u(x,\zeta(x,t),t)dx - \left<u\right>_\beta, 
\endeq 
\endlett 
with the equivalent quantities for the  
ternary impurity field being obtained by 
replacing $u$ by $\util$ in the last two equations above. The steady-state 
values are obtained 
from Eqs. (\ref{ctilav}) and (\ref{caverage}) using the changes 
of variables (\ref{udef}) and (\ref{utildef}). Furthermore, we 
define the average $z$ position of a lamella by 
\beglett 
\label{xiavdef} 
\begeq 
\xiav_j^\alpha = {1\over 2}\left(\xi_j^\alpha+\xi_j^\beta\right) 
\endeq 
\begeq 
\xiav_j^\beta = {1\over 2}\left(\xi_{j+1}^\alpha+\xi_j^\beta\right). 
\endeq 
\endlett 
In terms of these local averages, the linearized 
Gibbs-Thomson conditions read: 
\beglett 
\label{githopert} 
\begin{eqnarray} 
\delta T_j^\alpha(t) & = &\mbox{} -G\xiav_j^\alpha(t) \nonumber \\ 
   & = & m_\alpha \delta\uav_j^\alpha(t)  
          + \mtil_\alpha \delta\utilav_j^\alpha(t) \nonumber \\ 
    && \mbox{} + \Gamma_\alpha \delta\Kav_j^\alpha(t)  
\end{eqnarray} 
\begin{eqnarray} 
\delta T_j^\beta(t) & = & \mbox{} -G\xiav_j^\beta(t) \nonumber \\ 
   & = &\mbox{} -m_\beta \delta\uav_j^\beta(t)  
           + \mtil_\beta \delta\utilav_j^\beta(t) \nonumber \\ 
    && \mbox{} + \Gamma_\beta \delta\Kav_j^\beta(t). 
\end{eqnarray} 
\endlett 
The next step is to express the  
averages in the two equations above in terms of the displacements 
$\xi_j^s$ and $y_j^s$. For the curvature terms, the
procedure is straightforward, but for the diffusion fields
one needs to introduce a piecewise linear 
representation of the interface shape, as will be 
described in more detail below [see Eq. (\ref{piecewise})].  
Following this step, Eqs. (\ref{githopert}) become  
a system of $2N$ linear equations 
for $4N$ variables and their time derivatives. To complete this  
system, we must specify 
how the trijunction points react to deformations of the growth 
front. Following DL, we will use Cahn's hypothesis and assume  
that the trijunctions always grow perpendicular to the  
eutectic solidification front, which yields the conditions 
\beglett 
\label{grconstraints} 
\begeq 
\dot y_j^\alpha =  
-(\xi_j^\beta - \xi_{j-1}^\beta)v_p/\lambda 
\endeq 
\begeq 
\dot y_j^\beta =  
-(\xi_{j+1}^\alpha - \xi_j^\alpha)v_p/\lambda, 
\endeq 
\endlett 
where the dot denotes the time derivative.  
 
We now transform the problem into an eigenvalue equation by 
analyzing it in terms of time-dependent Fourier modes. We write 
\beglett 
\label{xifourier} 
\begeq 
\xi_j^s=X_k^s \exp(ik\lambda j+\omega t) \quad (s=\alpha,\beta) 
\endeq 
\begeq 
y_j^s=Y_k^s \exp(ik\lambda j+\omega t) \quad (s=\alpha,\beta), 
\endeq 
\endlett 
where the allowed wave vectors $k$ are integer multiples of 
$2\pi/N\lambda$ and lie in the interval $[-\pi/\lambda,\pi/\lambda]$. 
In the limit of an infinite number of lamellae, 
$N\to\infty$, we recover a continuous spectrum. 
The growth constraints (\ref{grconstraints}) give then immediately 
\beglett 
\begeq 
\omega Y_k^\alpha = 
     -{2iv_p\over \lambda} e^{-ik\lambda/2} \sin(k\lambda/2) X_k^\beta 
\endeq 
\begeq 
\omega Y_k^\beta = 
     -{2iv_p\over \lambda} e^{ik\lambda/2} \sin(k\lambda/2) X_k^\alpha. 
\endeq 
\endlett 
This allows us to eliminate the coefficients $Y_k^\alpha$ and 
$Y_k^\beta$, and the only unknowns left in the problem are  
$X_k^\alpha$, $X_k^\beta$, and $\omega$. It is  
useful to write each of the terms  
appearing in the Gibbs-Thomson condition in 
the compact forms 
\begin{eqnarray} 
G\xiav_j^s(t) & = & e^{ik\lambda j+\omega t} 
   \sum_{s'=\alpha,\beta} {\bf G}^{s,s'}(k,\omega)X_k^{s'} \\ 
\delta\Kav_j^s(t) & = & e^{ik\lambda j+\omega t} 
   \sum_{s'=\alpha,\beta} {\bf K}^{s,s'}(k,\omega)X_k^{s'} \\ 
\delta\uav_j^s(t) & = & e^{ik\lambda j+\omega t} 
   \sum_{s'=\alpha,\beta} {\bf U}^{s,s'}(k,\omega)X_k^{s'} \\ 
\delta\utilav_j^s(t) & = & e^{ik\lambda j+\omega t} 
   \sum_{s'=\alpha,\beta} {\bf\tilde U}^{s,s'}(k,\omega)X_k^{s'}. 
\end{eqnarray} 
Then, the conditions (\ref{githopert}) can be written 
as an eigenvalue equation: 
\begeq 
\sum_{s'=\alpha,\beta} {\bf A}^{s,s'}X_k^{s'} = 0, 
\label{githomatrix} 
\endeq 
where the matrix ${\bf A}$ is given by 
\begin{eqnarray} 
{\bf A} & = & {\bf G} + 
  \left(\begin{array}{cc} \Gamma_\alpha & 0 \\ 0 & \Gamma_\beta \end{array} 
       \right) {\bf K} + 
  \left(\begin{array}{cc} m_\alpha\Delta c & 0 \\ 0 & -m_\beta\Delta c \end{array}\right) 
       {\bf U} \nonumber \\ 
 & & \mbox{} +   
  \left(\begin{array}{cc} \mtil_\alpha\Delta\ctil & 0 \\ 0 & \mtil_\beta\Delta\ctil \end{array} 
       \right) {\bf \tilde U}. 
\label{matrix} 
\end{eqnarray} 
The first three terms on the RHS of Eq. (\ref{matrix}) are identical 
to those calculated by DL, whereas the last term is due to the 
presence of the ternary impurity. 
For a given wave number $k$, Eq. (\ref{githomatrix}) is fulfilled 
only for special values of $\omega$. The solvability condition, 
\begeq 
\det{\bf A} = 0, 
\endeq 
gives the dispersion relations $\omega(k)$. 
 
The core of the problem is the calculation of the matrices appearing 
in Eq. (\ref{matrix}). The first two, $\bf G$ and $\bf K$, are 
relatively easy (see Ref. \cite{Datye81} and appendix A). The most difficult 
is the matrix $\bf U$. We start by writing the perturbed diffusion 
field as 
\begeq 
u(x,z,t) = u_0(x,z) + \delta u(x,z,t), 
\endeq 
where $u_0(x,z)$ is the steady-state solution. Next, $\delta u$ is 
expanded in Fourier modes, this time with the periodicity of the 
whole system, i.e. with wave numbers $p = 2\pi n/N\lambda$: 
\begeq 
\delta u(x,z,t) = \sum_p b_p \exp[ipx - \bar q_p(z-\bar z)+\omega t]. 
\label{deltaufourier} 
\endeq 
The $p$'s, unlike the $k$'s, range from $-\infty$ to $\infty$, because 
the diffusion field is continuous.
To simplify our calculations, we use the quasistationary
approximation for the diffusion equation, i.e. we drop
the time derivative in Eq. (\ref{diffue}).
Physically, this means that we assume
that the diffusion field adjusts instantaneously to any
change in the interface configuration. For a perturbation
such as that given by Eq. (\ref{deltaufourier}), this is justified
if $|\omega|\ll Dp^2$ and $|\omega|\ll Dp/l$, conditions 
that will be checked {\em a posteriori} 
at the end of Sec. \ref{compsec}. 
In the range of wavelength of interest here, i.e. for
$\lambda < 2\pi/p < l$, we find that these conditions
are generally well satisfied.
Within this approximation, the constants $\bar q_p$ 
are given by the analog of Eq. (\ref{barqs}). 

To obtain the Fourier coefficients $b_p$, we 
proceed as in the steady-state calculation and insert the expansion 
in the mass conservation condition. To make the problem tractable, 
the actual interface shape $\zeta(x,t)$ is replaced by the piecewise  
constant function 
\begeq 
\zeta(x,t)-\bar z=\left\{\begin{array}{l} 
    \xiav_j^\alpha \quad x_j^\alpha < x < x_j^\beta \\ 
    \xiav_j^\beta \quad x_j^\beta < x < x_{j+1}^\alpha. 
    \end{array}\right. 
\label{piecewise} 
\endeq 
One needs to be careful because the gradients of the steady state 
concentration field diverge at the trijunction points. Details can 
be found in DL's article. Finally, the result is inserted in  
Eqs. (\ref{deltaus}). There are two types of contributions  
to linear order in $\xi_j^s$ and $y_j^s$: the steady-state diffusion  
field is evaluated at the position of the perturbed interface, 
and the perturbed diffusion field is taken at the location of the 
steady-state interface. These different terms, containing sums over the 
steady-state interlamellar diffusion modes, lead to quite complicated 
expressions, summarized in appendix A. 
 
The calculation of $\bf \tilde U$ is somewhat easier because,  
with the assumption of equal solute partition coefficients, the 
steady-state impurity diffusion field is translationally invariant along 
$x$, and hence the horizontal displacements $y_j^s$ drop out of 
the calculation. It would be possible to include the interlamellar 
impurity diffusion in the expression for $\bf \tilde U$ by  
following the lines of the DL calculation for $\bf U$. We expect, 
however, as argued in the preceding section, that this would only 
lead to minor corrections. 
 
The perturbed impurity diffusion field is written as 
\begeq 
\util(x,z,t) = \util_0(z) + \delta\util(x,z,t) 
\endeq 
and expanded in Fourier modes, 
\begeq 
\delta\util(x,z,t)= \sum_{p} \btil_p\exp[ipx-\qtil_p(z-\bar z)+\omega t]. 
\label{utilfourier} 
\endeq 
To calculate the constants $\qtil_p$, we use the 
quasistationary approximation of the impurity diffusion equation.
In terms of the dimensionless impurity field $\util$, the  
continuity equation at the interface takes the form 
\begeq 
-\dtil\left.\partial\util\over\partial z\right|_{z=\zeta} = 
\left(v_p+\dot\zeta\right)\left[(1-k_E)\util + k_E\right]. 
\endeq 
We want to keep only terms that are linear in the displacements 
$\xi_j^s$ or their time derivatives. Such terms come from several 
sources: the time derivative of $\zeta$, the 
corrections $\delta u$ in the diffusion field and its gradient, and 
from evaluating the steady-state diffusion field 
at the new interface position $\zeta(x,t)$. The equation of order 
1 in the displacements becomes 
\begin{eqnarray} 
&&-\dtil\left.\partial^2\util_0\over\partial z^2\right|_{z=\bar z} 
             \left(\zeta-\bar z\right)  
-\dtil\left.\partial\delta\util\over\partial z\right|_{z=\bar z} = \mbox{} 
  \nonumber\\ 
&&\qquad \dot\zeta\left[(1-k_E)\util_0(x,\bar z) + k_E\right] \nonumber \\ 
&&\qquad \mbox{} + v_p (1-k_E)
  \left.\partial\util_0\over\partial z\right|_{z=\bar z} 
  \left(\zeta-\bar z\right)  \nonumber\\ 
&&\qquad \mbox{} + v_p (1-k_E)\,\delta\util(x,\bar z,t). 
\end{eqnarray} 
We now insert the Fourier expansion (\ref{utilfourier}) in 
the above equation, multiply both sides by $\exp(-ip'x)/N\lambda$ 
and integrate over $x$ from $0$ to $N\lambda$. With 
$\util_0(x,z)= \exp(-2(z-\bar z)/\ltil)$ this leads to 
\begin{eqnarray} 
&&\btil_p e^{\omega t}
   \left[\qtil_p - {2\over\ltil}(1-k_E)\right] = \nonumber \\ 
&&\qquad{4k_E\over\ltil^2}{1\over N\lambda} 
        \int_0^{N\lambda} e^{-ipx}\zeta(x,t) dx \nonumber \\ 
&&\qquad \mbox{} + {1\over\dtil} 
   {1\over N\lambda}\int_0^{N\lambda} e^{-ipx}\dot\zeta(x,t) dx. 
\end{eqnarray} 
To perform the integrals over $x$, we use the piecewise constant 
expression Eq. (\ref{piecewise}) for $\zeta(x,t)$. As $\zeta-\bar z$ is 
already of order $\xi$, we can neglect the $y$'s in the integration 
boundaries. 
For $\xi_j^s$ given by a Fourier mode of wave number $k$ according 
to Eqs. (\ref{xifourier}), we obtain 
\begin{eqnarray} 
\btil_p & = & \left[\qtil_p - {2\over\ltil}(1-k_E)\right]^{-1} 
       \left({4k_E\over\ltil^2}+{\omega\over\dtil}\right) 
       {e^{-ip\eta\lambda/2}\over p \lambda}\nonumber \\ 
   & & \mbox{}\times \left[\sin{p\eta\lambda\over 2} 
        \left(\tilde X_p^\alpha + \tilde X_p^\beta\right) \right.\nonumber \\ 
   & & \,\mbox{}+ \left.\sin{p(1-\eta)\lambda\over 2} 
        \left(e^{ip\lambda/2}\tilde X_p^\alpha +  
              e^{-ip\lambda/2}\tilde X_p^\beta\right)\right], 
\label{btilk} 
\end{eqnarray} 
where 
\begin{eqnarray} 
\tilde X_p^s & = & {1\over N}\sum_{j=0}^{N-1} 
          e^{-ip\lambda j -\omega t}\xi_j^s \nonumber \\ 
  & = & \sum_n \delta_{p,k+2\pi n/\lambda} X_k^s \quad (s=\alpha,\beta). 
\end{eqnarray} 
Note that, even if we start with a set of $\xi$'s given by 
a single Fourier mode of wave number $k$,  
the use of the piecewise constant 
interface shape of Eq. (\ref{piecewise}) induces perturbations 
in the diffusion field at all wave numbers shifted with respect 
to $k$ by a multiple of $2\pi/\lambda$. This effect is unavoidable 
if we want to replace the full free boundary problem by equations 
for a discrete set of variables.  
 
The next step is the calculation of the average concentrations 
in front of each lamella: 
\beglett 
\begin{eqnarray} 
&&\delta\utilav_j^\alpha(t) = {1\over\eta\lambda} \nonumber \\ 
&&\qquad \mbox{}\times\int_{j\lambda}^{(j+\eta)\lambda} 
         \left(\left.{\partial\util_0\over\partial z} 
         \right|_{z=\bar z}\xiav_j^\alpha 
         +\delta\util(x,\bar z,t)\right) dx \\ 
&&\delta\utilav_j^\beta(t) = {1\over(1-\eta)\lambda} \nonumber \\ 
&&\qquad \mbox{}\times\int_{(j+\eta)\lambda}^{(j+1)\lambda} 
         \left(\left.{\partial\util_0\over\partial z} 
         \right|_{z=\bar z}\xiav_j^\beta 
         +\delta\util(x,\bar z,t)\right) dx. 
\end{eqnarray} 
\endlett 
The first term inside each integral is due to the displacement of the 
interface in the steady state diffusion field and the second arises  
from the perturbed diffusion field evaluated at the 
steady-state interface position. The final result for the matrix  
$\bf\tilde U$ is 
\beglett 
\begin{eqnarray} 
{\bf \tilde U}^{\alpha,\alpha} & = & 
   -{1\over\ltil} + {1\over \eta\ltil} 
   \left({2k_E\lambda\over\ltil}+{\omega\lambda\ltil\over 2\Dtil}\right) 
   \nonumber \\ 
 && \mbox{} \times 
   \left[\Stil_1(\kappa,\eta)+\Stil_2^\star(\kappa,\eta)\right] \\ 
{\bf \tilde U}^{\alpha,\beta} & = & {\bf \tilde U}^{\alpha,\alpha\star} \\ 
{\bf \tilde U}^{\beta,\beta} & = & 
   -{1\over\ltil} + {1\over (1-\eta)\ltil} 
   \left({2k_E\lambda\over\ltil}+{\omega\lambda\ltil\over 2\Dtil}\right) 
   \nonumber\\ 
 &&  \mbox{} \times 
   \left[\Stil_1(\kappa,1-\eta)+\Stil_2^\star(\kappa,1-\eta)\right] \\ 
{\bf \tilde U}^{\beta,\alpha} & = &  
    e^{ik\lambda}{\bf \tilde U}^{\beta,\beta\star}, 
\end{eqnarray} 
\endlett 
where ${\bf \tilde U}^{\alpha,\alpha\star}$ denotes the expression 
obtained from ${\bf \tilde U}^{\alpha,\alpha}$ by complex conjugation 
of all quantities except for $\omega$ (for $\omega$ real, this 
is the usual complex conjugation), and we have defined 
\begeq 
\kappa  =  k\lambda/2\pi 
\endeq 
\begeq 
r = l/\ltil = D/\Dtil 
\endeq 
\begeq 
\rhotil_n(\kappa) =  
   \sqrt{r^2{\rm Pe}^2 + 4\pi^2(n+\kappa)^2} - r{\rm Pe} +2r{\rm Pe}k_E  
\label{rhotildef} 
\endeq 
\begin{eqnarray} 
\Stil_1(\kappa,\eta) & = &  
   \sum_{n=-\infty}^\infty {\sin^2[\pi\eta(n+\kappa)]\over 
                            \pi^2(n+\kappa)^2\rhotil_n(\kappa)} \\ 
\Stil_2(\kappa,\eta) & = & 
   \sum_{n=-\infty}^\infty e^{-i\pi(n+\kappa)} \nonumber \\ 
  && \mbox{} \times 
     {\sin[\pi\eta(n+\kappa)]\sin[\pi(1-\eta)(n+\kappa)]\over 
      \pi^2(n+\kappa)^2\rhotil_n(\kappa)}. 
\end{eqnarray} 
These notations have been chosen in analogy to some of DL's 
results (see appendix A). The ratio $r$ of the eutectic and 
impurity diffusion lengths is usually close to one, and 
$\kappa$ is the dimensionless wave number. For small 
P{\'e}clet numbers, and perturbation wavelengths much  
larger than the lamellar spacing ($\kappa\ll 1$),  
the sums $\Stil_1$ and $\Stil_2$ are dominated by the 
term with $n=0$. In this limit, we can neglect all the 
other terms in the sums, which corresponds to keeping only 
the Fourier coefficient $\btil_p$ with $p=k$, and hence 
to a single-mode approximation of the perturbed impurity diffusion 
field. When $\kappa$ is larger, however, and in particular near  
the ``Brillouin zone'' boundary, $\kappa=0.5$, we have to 
consider the full sums. 
To obtain the stability spectra, we must now combine this result with 
DL's calculations for the other matrices and solve the characteristic 
equation for $\omega$. 
 
\section{Symmetric phase diagram at eutectic composition} 
\subsection{Stability spectrum}
\nobreak 
\noindent 
In general, the characteristic equation of 
the stability spectrum is a polynomial  
of degree four in $\omega$ with real coefficients. 
The solutions could be obtained algebraically, but this method 
leads to complicated expressions that are difficult 
to interpret. For this reason, we will restrict our 
attention in this section to the special case of an alloy of eutectic 
composition in a model system where the 
phase diagram is symmetrical about the eutectic composition. 
In this case, the characteristic equation can be factored into 
two quadratic equations, thus greatly simplifying the interpretation. 
As we shall see when we treat the general case in Sec. VI,  
all qualitative features of the instability  
are already contained in this special case. 
 
There are general relations between the elements of the 
matrix {\bf A}, due to the existence of two planes of 
mirror symmetry in the steady state, one in the middle of each  
type of lamella ($\alpha$ or $\beta$).  
Hence we can change the sign of $k$ and relabel 
the trijunction points without affecting the final result. This 
leads to the relations 
\beglett 
\begeq 
{\bf A}^{\alpha,\beta}={\bf A}^{\alpha,\alpha *} 
\endeq 
\begeq 
{\bf A}^{\beta,\alpha}=e^{ik\lambda}{\bf A}^{\beta,\beta *}. 
\endeq 
\endlett 
Here, an asterisk again denotes complex conjugation of all quantities 
except $\omega$. If we consider a model eutectic with a completely 
symmetric phase diagram, i.~e. $m_\alpha=m_\beta=m$, $\mtil_\alpha 
= \mtil_\beta=\mtil$, $\Gamma_\alpha=\Gamma_\beta=\Gamma$, 
$\theta_\alpha=\theta_\beta=\theta$, and $u_\beta=-u_\alpha=1/2$,  
at its eutectic composition, $\eta=1/2$, we have in addition 
${\bf A}^{\alpha,\alpha}={\bf A}^{\beta,\beta}$. The solvability 
condition, $\det {\bf A} = 0$, can then be factored into two  
equations, 
\begeq 
\Re\left(e^{-ik\lambda/4}{\bf A}^{\alpha,\alpha}\right) = 0 
\quad {\rm and} \quad 
\Im\left(e^{-ik\lambda/4}{\bf A}^{\alpha,\alpha}\right) = 0, 
\label{quadraticeqs} 
\endeq 
both of them quadratic in $\omega$. 
 
To proceed, we will rewrite the equations in a dimensionless form.  
For the sake of subsequent generalization, we will give expressions 
for the parameters that are valid for any phase diagram. We define 
\begin{eqnarray} 
M & = & {m_\alpha m_\beta\over (m_\alpha+m_\beta)/2} \\ 
\Omega &=& \omega\lambda/v_p \\ 
g &=& \frac{Gl}{M\Delta c} = \frac{2DG}{v_pM\Delta c} \\ 
w & = & \frac{\mtil\Delta\ctil}{M\Delta c} \\ 
\Lambda & = & \lambda/\lmin. 
\end{eqnarray} 
Here, $M$ is a mean liquidus slope (for the symmetric phase 
diagram, $M=m$); $1/\Re\Omega$ is the distance  
along the $z$ direction that the interface needs to travel  
(in units of lamellar spacing) for the amplitude  
of the perturbation to grow by 
a factor of $e$, and $2\pi/\Im\Omega$ is the length  
traveled by the interface during one 
oscillation cycle. The parameter $w$ is the ratio of the 
two freezing ranges and is proportional to the impurity 
concentration since $\Delta \ctil=(1/k_E-1)\ctilinf$. 
 
It may be useful to comment on typical experimental 
values of the dimensionless parameters. 
The best studied system in this context is the organic 
eutectic CBr$_4$-C$_2$Cl$_6$ used originally by Hunt and Jackson 
\cite{Hunt66}. This system contains naturally residual gas that
acts as an impurity, leading to colony formation.  
For this system, Akamatsu and Faivre \cite{Akamatsu96} 
have estimated $w$ to be of order $0.1$, with a 
distribution coefficient $k_E\approx 0.02$. 
The P{\'e}clet number ${\rm Pe}$ is typically between $0.01$ and 
$0.1$ in the low-velocity regime used to investigate 
colony formation. 
 
After multiplication by $l\Omega/(M\Delta c)$, the first of  
Eqs. (\ref{quadraticeqs}) becomes, expressed in the above dimensionless 
quantities (see appendix A for more details),
\begin{eqnarray} 
0 & = &\Omega {g\over 2} \cos{\pikap\over 2}\nonumber\\ 
&&\mbox{}+{2P(\eta)\over\eta\Lambda^2}\sin(\pikap/2)\sin(\pikap) 
   \left(\Omega{\rm cot}\theta - 2/\eta\right)\nonumber\\ 
&&\mbox{}+\Omega^2 \Re\left(e^{-i\pikap/2}U_1^\alpha(\kappa,\eta)\right)  
   \nonumber\\ 
&&\mbox{}+\Omega \Re\left(e^{-i\pikap/2}U_2^\alpha(\kappa,\eta)\right) 
   \nonumber\\ 
&&\mbox{}+2\sin(\pikap) \Re\left(ie^{i\pikap/2} U_3(\kappa,\eta)\right) 
   \nonumber\\ 
&&\mbox{} -wr\Omega\cos{\pikap\over 2} + {wr\Omega\over \eta} 
   \left(2k_E r{\rm Pe}+\Omega\right) \nonumber \\ 
&&\mbox{} \times\left[\Stil_1(\kappa,\eta)\cos{\pikap\over 2} 
   + \Re\left(e^{-i\pikap/2}\Stil_2^\star(\kappa,\eta)\right)\right]. 
\label{theequation} 
\end{eqnarray} 
where we have chosen to display $\eta$, for clarity, 
although the factorization is only possible for $\eta=1/2$. 
The first term of Eq. (\ref{theequation}) arises from the matrix 
$\bf G$; the factor $\cos(\pikap/2)$  
is simply due to the averaging over 
the two trijunction points limiting a lamella. The matrix ${\bf K}$ 
contributes the next two terms, proportional to $\Lambda^{-2}$. 
The first, containing $\Omega$, describes the change of curvature 
due to the bending of the interface over a large scale, and is 
therefore equivalent to the capillary term in the dilute 
binary alloy problem. The second gives the change  
in average curvature upon variation of the local lamellar  
spacing. All terms containing the functions  
$U_n^\alpha(\eta,\kappa)$, defined in appendix A, are due to the  
eutectic diffusion field. Finally, the terms proportional to $w$  
arise from the matrix ${\bf \tilde U}$. 
 
Equation (\ref{theequation}) is exact and can be solved numerically. 
But it is also useful to simplify this equation in order 
to render the physical interpretation of the instability more 
transparent. To this end, let us group the terms 
with equal powers of $\Omega$ and rewrite  
Eq. (\ref{theequation}) as 
\begeq 
a(\kappa)\Omega^2-b(\kappa)\Omega+c(\kappa)=0. 
\label{abcequation} 
\endeq 
To obtain a simplified expression, we expand the coefficients 
$a(\kappa)$, $b(\kappa)$, and $c(\kappa)$ in powers 
of $\kappa$. Details on this procedure can be found in 
appendix B. It turns out that an expansion up to 
order $\kappa^2$ is sufficient to obtain a satisfying 
agreement with the direct numerical solution 
of Eq. (\ref{theequation}). We obtain 
\beglett 
\label{approxcoeff} 
\begin{eqnarray} 
a(\kappa) & = & 2P(1/2)+{wr\over \rhotil_0(\kappa)}  
\label{akappa} \\ 
b(\kappa) & = & wr - g/2 - \left({2\pi^2P(1/2)\cot\theta\over\Lambda^2} 
                +{R_0\over 2}\right)\kappa^2 \nonumber \\ 
          & & \mbox{} - {2w{\rm Pe}r^2k_E\over\rhotil_0(\kappa)} 
\label{bkappa} \\ 
c(\kappa) & = & 8\pi^2P(1/2)\left(1-{1\over\Lambda^2}\right)\kappa^2, 
\end{eqnarray} 
\endlett 
where $\rhotil_0(\kappa)$ is defined by Eq. (\ref{rhotildef}), 
and $R_0$ is a constant given by Eq. (\ref{rzerodef}).  
We note that, to 
obtain these expressions, we have only kept the leading order 
terms of each of the contributions in Eq. (\ref{theequation}) 
and dropped several terms of order $|\kappa|$ and $\kappa^2$ 
that turn out to give  
negligible contributions at the onset of instability 
for the reasons detailed in appendix B.  
The simplified stability spectrum defined by  
the equations above will be used below to derive  
simple analytical expressions for the onset 
velocity and wavelength. Moreover, it will allow  
us to identify the terms that contribute to 
the effective surface tension in the long-wavelength 
free boundary formulation presented in Sec. VI. 
In the rest of this paper, the results based on this 
simplified spectrum will be systematically checked against 
the direct numerical solution of Eq. (\ref{theequation}). 

\begin{figure} 
\centerline{ 
  \psfig{file=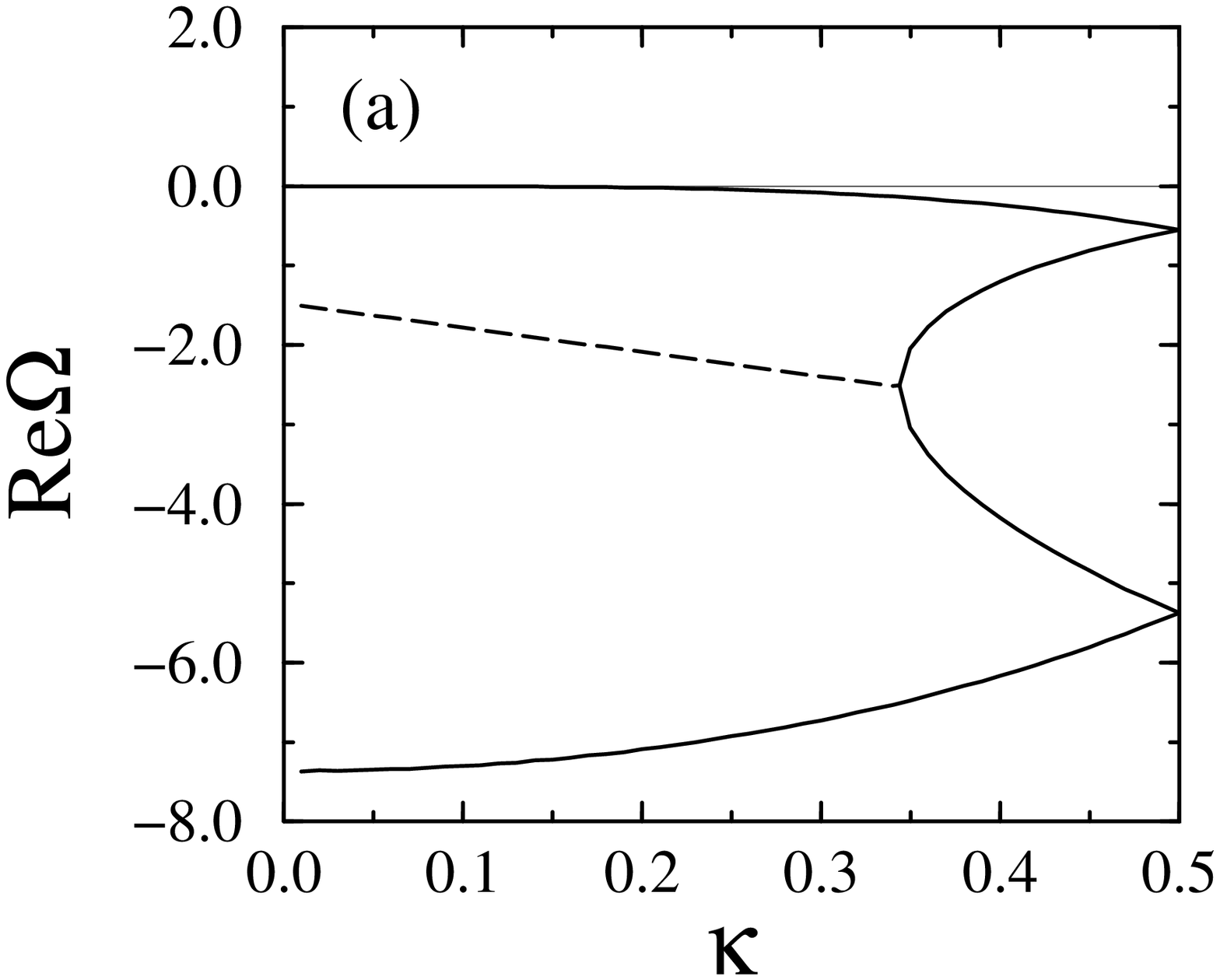,width=.3\textwidth}} 
\centerline{ 
  \psfig{file=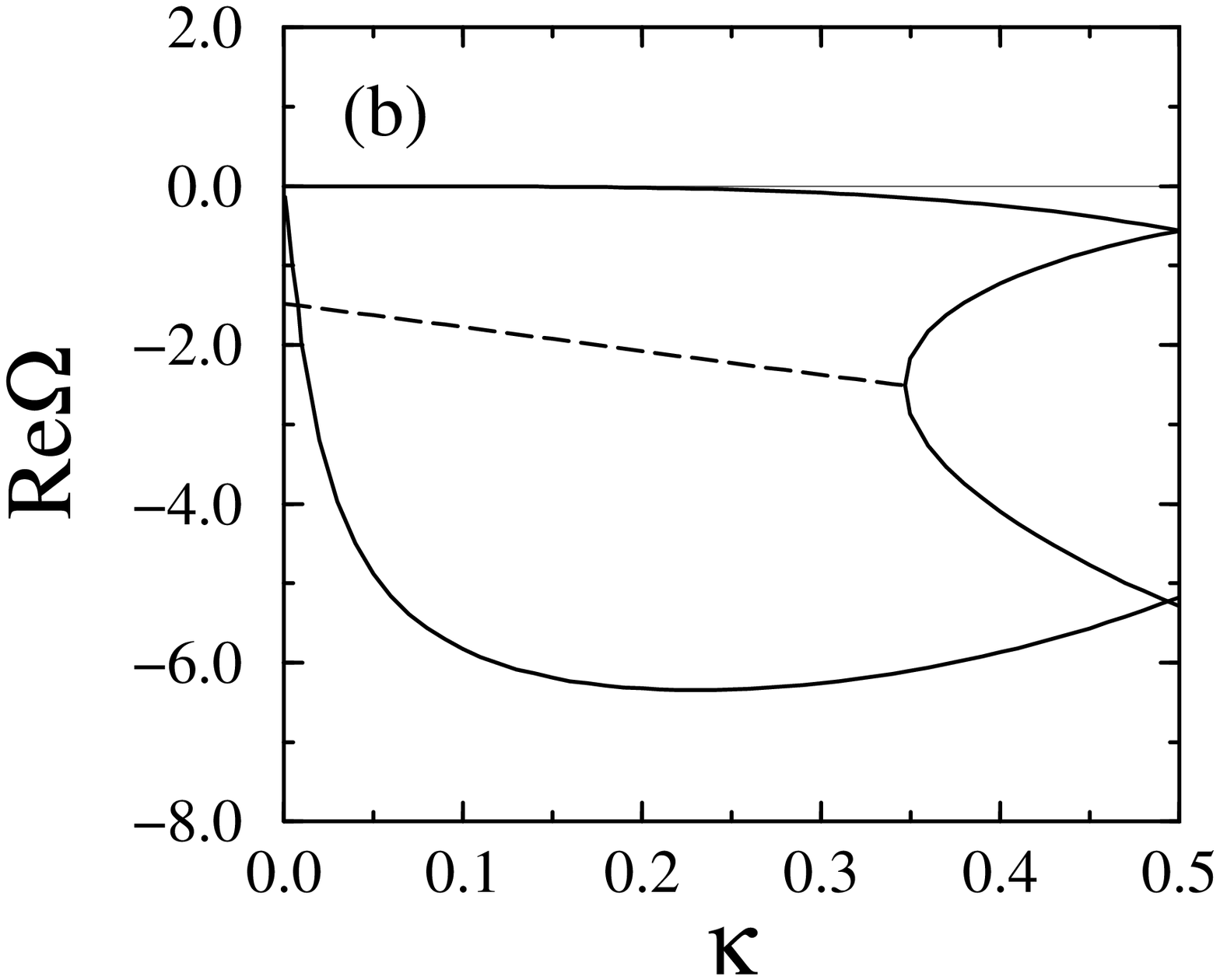,width=.3\textwidth}} 
\caption{Stability spectra of the symmetric eutectic 
at $\Lambda=1$, ${\rm Pe}=0.01$, $g=1$, $r=1$, $k_E=0.05$, 
and $\theta=45^\circ$, (a) without impurity ($w=0$),  
(b) with a small amount of ternary impurity ($w=0.01$).  
Full lines: real modes, dashed lines: complex modes.} 
\label{figeffect}
\end{figure} 
\subsection{Binary versus ternary eutectic: the limit $\kappa\to 0$} 
\nobreak 
\noindent 
Let us first examine the limit $\kappa\rightarrow 0$ of the 
stability spectrum, which governs the relaxation to 
steady-state growth after an infinitesimal translation of the  
solidification front along the $z$ axis. 
Although this is not the limit of interest 
for morphological instability, it is worth 
a brief discussion to understand more fully 
the subtle effects of the ternary impurity on 
the complete stability spectrum. 
 
Since, as mentioned above, the complete 
eigenvalue equation is of degree four in $\Omega$ with real coefficients, 
there are four branches of the dispersion relation, and modes 
can be real or occur in complex conjugate pairs. In the completely 
symmetric case, each of the two equations (\ref{quadraticeqs}) gives a 
pair of branches. To calculate the stability spectra, we must 
first evaluate numerically the sums occurring in the functions  
$U_n(\kappa,\eta)$ to obtain the coefficients of the polynomial 
in $\Omega$, which can then be solved for each $\kappa$. 
Fig. \ref{figeffect} shows a comparison between the stability 
spectrum of a binary eutectic ($w=0$) and a eutectic with a small 
amount of impurities, ``small'' meaning that we stay far below  
the threshold of instability. We have chosen  
the parameters $g$ and $\Lambda$ in order to reproduce, for the 
binary eutectic, Fig. 4(b) of DL's article. 
Without impurities, we can distinguish two types of branches. There is  
always a characteristic diffusive branch, which is real and satisfies 
$\Omega \propto {\cal D}(\Lambda)\kappa^2$ for $\kappa\to 0$:  
the spacing can be locally adjusted by ``diffusion'' in  
$\lambda$ space \cite{Langer80}. This branch is related to the 
long-wavelength lamella elimination instability for $\lambda<\lmin$: 
the effective diffusion coefficient ${\cal D}(\Lambda)$ is 
negative for $\Lambda<1$. The companion of this branch is also real, with  
strongly negative growth rates. The second pair of branches is  
complex for small $\kappa$. This mode gives rise to the  
$2\lambda$-O instability for sufficiently off-eutectic  
compositions: it becomes complex up to the ``Brillouin zone''  
boundary $\kappa = 0.5$ and its real part becomes positive. 
 
When we add a small amount of impurities ($w=0.01$), we find that  
the diffusive and oscillatory branches are nearly unaffected;  
however, the companion of the diffusive branch undergoes 
a drastic change. Physically speaking, this strong reaction to 
a seemingly small perturbation is due to the introduction of a 
new conservation law. In the binary eutectic, this mode describes 
the relaxation of an interface by the motion of the trijunction points 
with respect to the temperature gradient (in the $z$ direction). 
At the eutectic composition, this relaxation involves only short 
range interlamellar diffusion and is therefore fast. On the other 
hand, for a flat interface, the impurities must diffuse over a 
length scale of the order of the diffusion length to achieve 
relaxation. This leads to much slower decay rates for small 
wave numbers. 
 
To see more formally how this change arises, let us consider 
Eqs. (\ref{approxcoeff}) in the limit $\kappa\to 0$. Without the 
impurity terms ($w=0$), $a(\kappa)$ and $b(\kappa)$ stay finite in 
the limit $\kappa\to 0$, whereas $c(\kappa) \propto \kappa^2$. Hence  
$\Omega \propto \kappa^2$ leads to a balance between the last two  
terms of Eq. (\ref{abcequation}). In the impurity terms, we 
have $\rhotil_0(\kappa)\to 2k_Er{\rm Pe}$ for $\kappa\to 0$, and  
for $\kappa=0$ we obtain 
\begeq 
\left(2P(1/2)+{w\over 2k_E{\rm Pe}}\right)\Omega^2 - {g\over 2}\Omega = 0, 
\label{limiteq} 
\endeq 
yielding the solutions, $\Omega=0$, for the diffusive branch and, 
$\Omega = -g/[4P(1/2)+w/(k_E{\rm Pe})]$, for its companion. In the latter 
expression, the impurity contribution is dominant 
for small P{\'e}clet number and small impurity partition coefficient, 
and hence this branch is strongly influenced by the addition of 
impurities. As the terms proportional to $\rhotil_0(\kappa)^{-1}$  
also appear in the classical Mullins-Sekerka
analysis of a monophase solidification front, we will 
hereafter refer to this branch as the MS branch.  
 
The oscillatory branch is little affected by the addition of 
impurities. The reason is that the second of Eqs. (\ref{quadraticeqs})  
does not contain the MS terms at $\kappa=0$. 
We will discuss the relation between long- and short-wavelength  
instabilities in more detail in Sec. VII.  

The derivation of Eqs. (\ref{theequation}) and (\ref{approxcoeff})
is based on the quasistationary approximation of the two diffusion
equations. This approximation relies on the assumption that the wavelength
of the perturbation is smaller than the diffusion length, and
hence breaks down for $\kappa \ll Pe$. 
In the framework of the DL
formalism, however, the calculation becomes
extremely tedious if this assumption is relaxed and the
results of this calculation will not be displayed here. The effective
interface formulation to be presented in Sec. VI, however,
easily allows to relax this assumption and to include
the dynamics of the diffusive boundary layer.
As a result, the MS branch of the spectrum 
becomes complex for $\kappa\ll {\rm Pe}$, corresponding to an
oscillatory relaxation of the interface to steady-state
growth driven by oscillations of the impurity
boundary layer that is already well-known 
for a monophase front. As we shall see below, 
the morphological instability leading to colony formation 
involves only modes with $\kappa \gg {\rm Pe}$ for which the
quasistationary approximation is valid.
 
\subsection{Onset of instability} 
\nobreak 
\noindent 
Let us now examine the onset of instability and compare our 
findings to the well-known results for dilute binary alloys. 
For the one-sided model, the constitutional supercooling (CS)  
criterion is fairly accurate. This criterion states that 
a monophase solidification front is unstable if the diffusion 
length is less than twice the thermal length $l_T=\mtil\Delta\ctil/G$. 
In our dimensionless variables, this is equivalent to 
$g < 2wr$. The Mullins-Sekerka analysis shows \cite{Mullins64} 
that the actual critical velocity differs from CS by corrective 
factors that are usually small. The critical wavelength 
$\lambda_c$ at the onset of instability scales as  
$\lambda_c \sim (d_0l_T\ltil)^{1/3}$, where $d_0$ is the 
capillary length. 
 
Let us briefly comment on some consequences of this scaling. 
From Eq. (\ref{lambdamin}) we can deduce that the eutectic  
spacing $\lmin$ scales as $\lmin \sim (d_0 l)^{1/2}$.  
Therefore, for low P{\'e}clet numbers 
we always expect $\lmin \ll \lambda_c$, and hence we can 
consider the limit of small $\kappa$ for the determination  
of the onset of instability. On the other hand, $\lambda_c$ 
will always be smaller than the diffusion length, and we 
have $k\ltil/2\pi\gg 1$, or equivalently $\kappa\gg {\rm Pe}$. 
Therefore, we may use the simplification 
$\rhotil_0(\kappa) \approx 2\pi|\kappa|$ 
in Eqs. (\ref{approxcoeff}). 

The occurrence of unstable modes is determined by the behavior of 
$b(\kappa)$: if $wr-g/2$ is positive and large enough, $b(\kappa)$ 
becomes positive for a certain range in $\kappa$. We want to determine 
the critical value $g_c$ of the parameter $g$ where the first 
unstable mode occurs, and the wave number $\kappa_c$ of this mode.  
The two solutions of the quadratic equation (\ref{abcequation}) are 
\begeq 
\Omega_{\pm} = {b(\kappa) \pm \sqrt{b^2(\kappa)-4a(\kappa)c(\kappa)} 
                 \over 2a(\kappa)}. 
\label{omeqn} 
\endeq 
As the product $a(\kappa)c(\kappa)$ is always positive and finite, 
the solutions become complex when $b(\kappa)$ tends to $0$, and 
we have $\Re\Omega = b(\kappa)/2a(\kappa)$. Hence the two conditions 
to obtain $g_c$ and $\kappa_c$ are simply $b(\kappa)=0$ and  
$db(\kappa)/d\kappa = 0$. 
Let us rewrite the first of these conditions in the dimensional 
quantities and divide through by $wr$. The result is 
\begeq 
1 - {G\Dtil\over\mtil\Delta\ctil v_p} -  
    {\Gamma\ltil\cos\theta\over\mtil\Delta\ctil}{k^2\over 2} -  
    {R_0\lambda^2\over 4\pi^2wr} {k^2\over 2} - {2k_E\over k\ltil} = 0. 
\label{onsetcon} 
\endeq 
There are two terms proportional to $k^2$. The first is  
the surface tension term which stabilizes the interface. 
It is analogous to the surface tension term 
in the monophase MS spectrum, except that  
it is multiplied by a geometrical factor $\cos\theta$.  
This factor is present because the 
eutectic interface is made of an array of arcs, 
each one linking two trijunctions, which renders the front 
less stiff than a flat monophase interface. The second term 
arises from the eutectic diffusion field.  
It can be directly interpreted by noting that the above equation 
becomes identical to the result for a monophase 
planar interface if we define the effective capillary length  
\begeq 
\bar d_0 = {\Gamma\cos\theta\over\mtil\Delta\ctil} +  
           {R_0{\rm Pe}\lambda\over 4\pi^2w}. 
\label{capilleff} 
\endeq 
The above expression implies that the interlamellar eutectic 
diffusion field has a stabilizing effect. This is rather  
remarkable since it implies that  the two 
diffusion fields (associated with the eutectic components 
and the ternary impurities, respectively) play 
antagonistic roles in the instability. 
 
From the condition $db(\kappa)/d\kappa = 0$ we obtain 
the expression 
\begeq 
k_c^3 = {2k_E\over \bar d_0\ltil^2} 
\endeq 
for the critical wave number.  
Furthermore, substituting this expression 
in Eq. (\ref{onsetcon}), we find 
\begeq 
v_c = {G\Dtil\over\mtil\Delta\ctil} 
        {1\over 1-3\left(\bar d_0 k_E^2/2\ltil\right)^{1/3}} 
\label{critvel} 
\endeq 
for the critical pulling speed. This expression is equivalent to 
the constitutional supercooling criterion 
($v_{CS} = G\Dtil/\mtil\Delta\ctil$) up to 
the capillary correction in the denominator of 
the second term on the RHS of Eq. (\ref{critvel}). 
This correction is typically negligibly small, except for very low 
impurity concentrations where the effective capillary length 
becomes large. Consequently, as for a monophase front, 
there is a critical impurity concentration 
below which the eutectic front  
is completely stable. 
 
The above results show that the instability 
of the eutectic interface is qualitatively similar 
to the standard MS instability of a monophase front, 
as far as the expressions for $k_c$ and $v_c$ 
are concerned, up to a renormalization 
of the surface tension as described by Eq. (\ref{capilleff}).  
One main difference, however, 
is that the branch of the spectrum that becomes 
unstable is complex, which suggests the existence of 
oscillatory patterns with a wavelength much  
larger than the lamellar spacing. To decide whether such 
patterns are observable  
experimentally, we need to examine next how this complex 
branch evolves above the onset of instability. 

\begin{figure} 
\centerline{ 
  \psfig{file=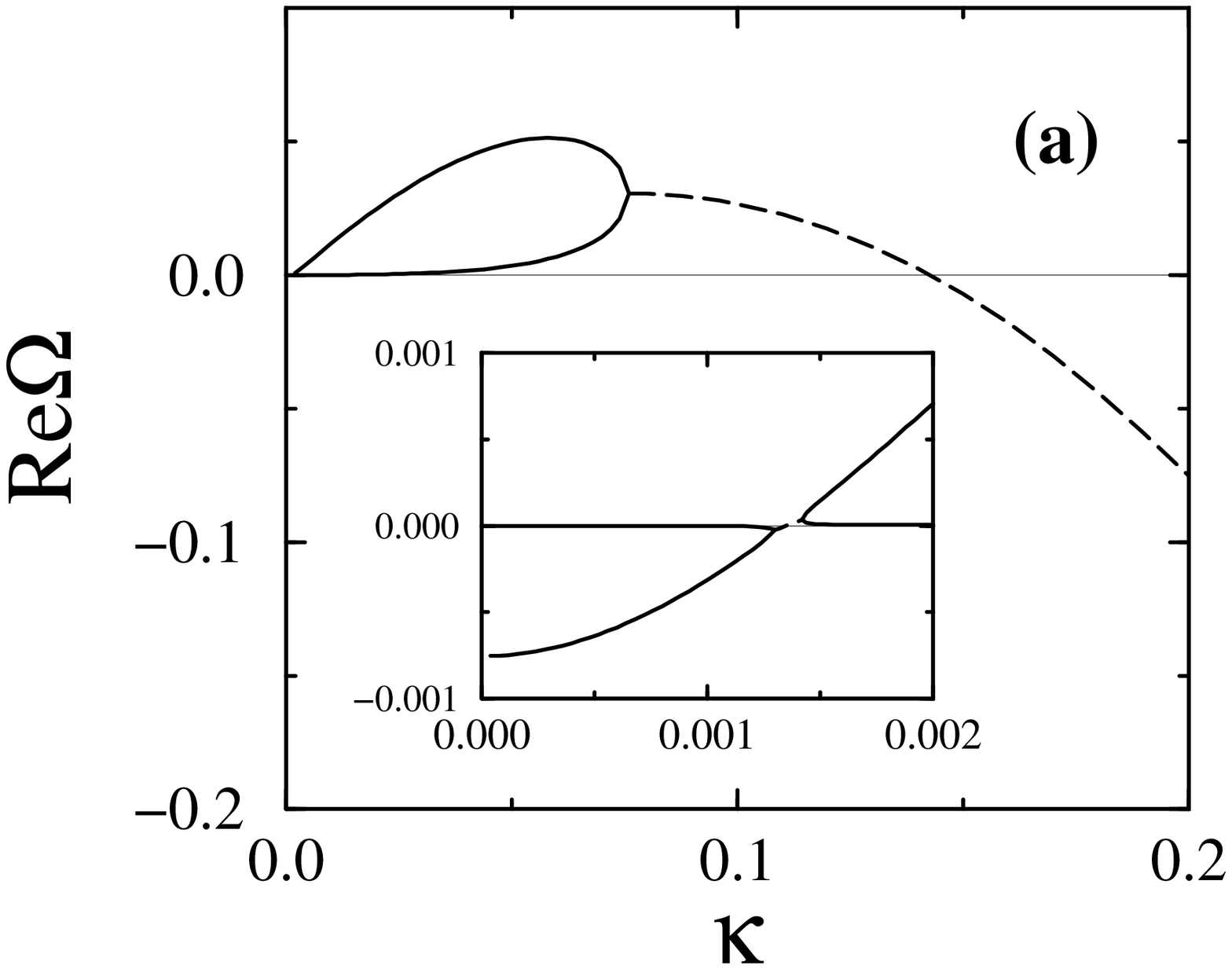,width=.25\textwidth} 
  \psfig{file=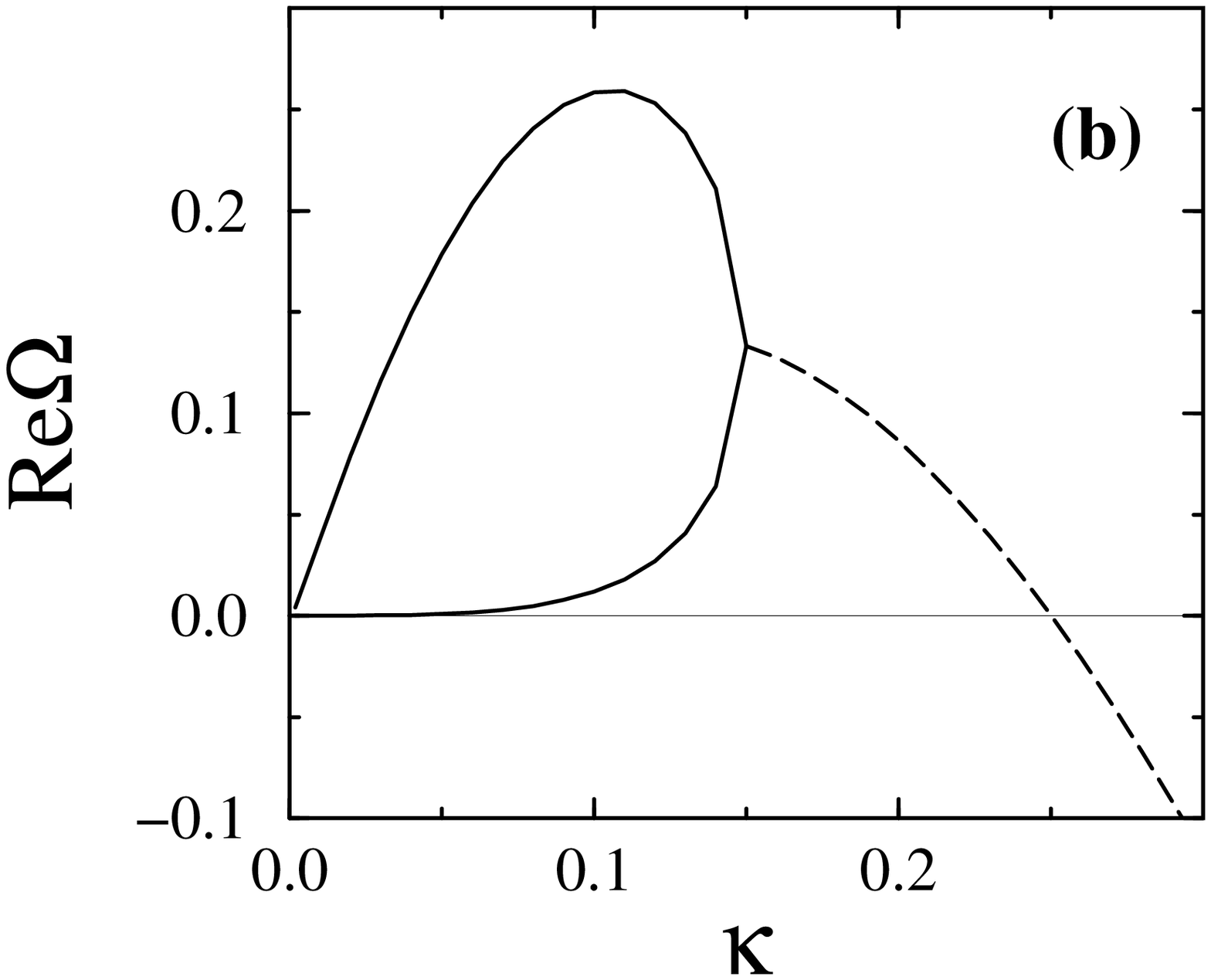,width=.25\textwidth}} 
\centerline{ 
  \psfig{file=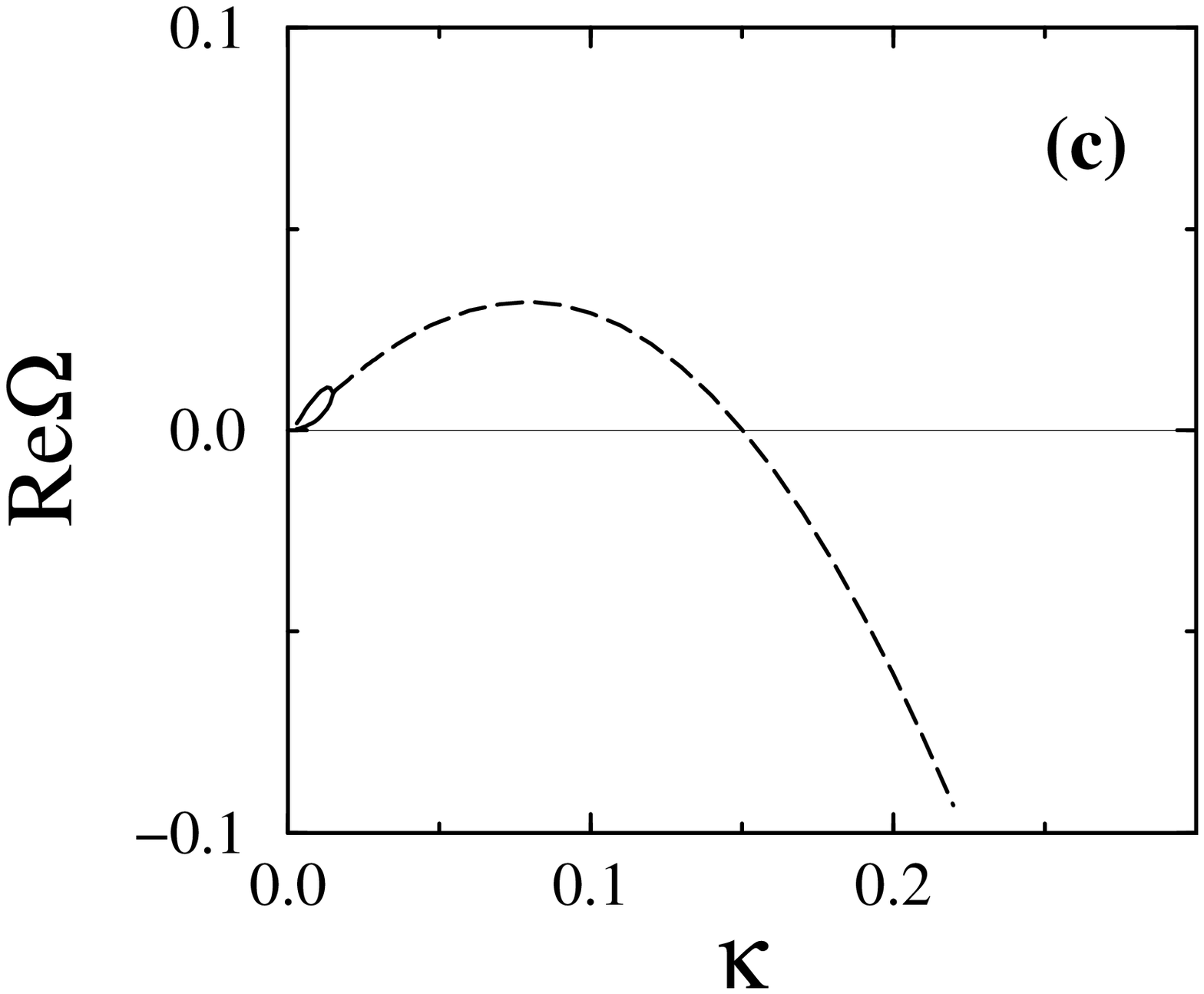,width=.25\textwidth} 
  \psfig{file=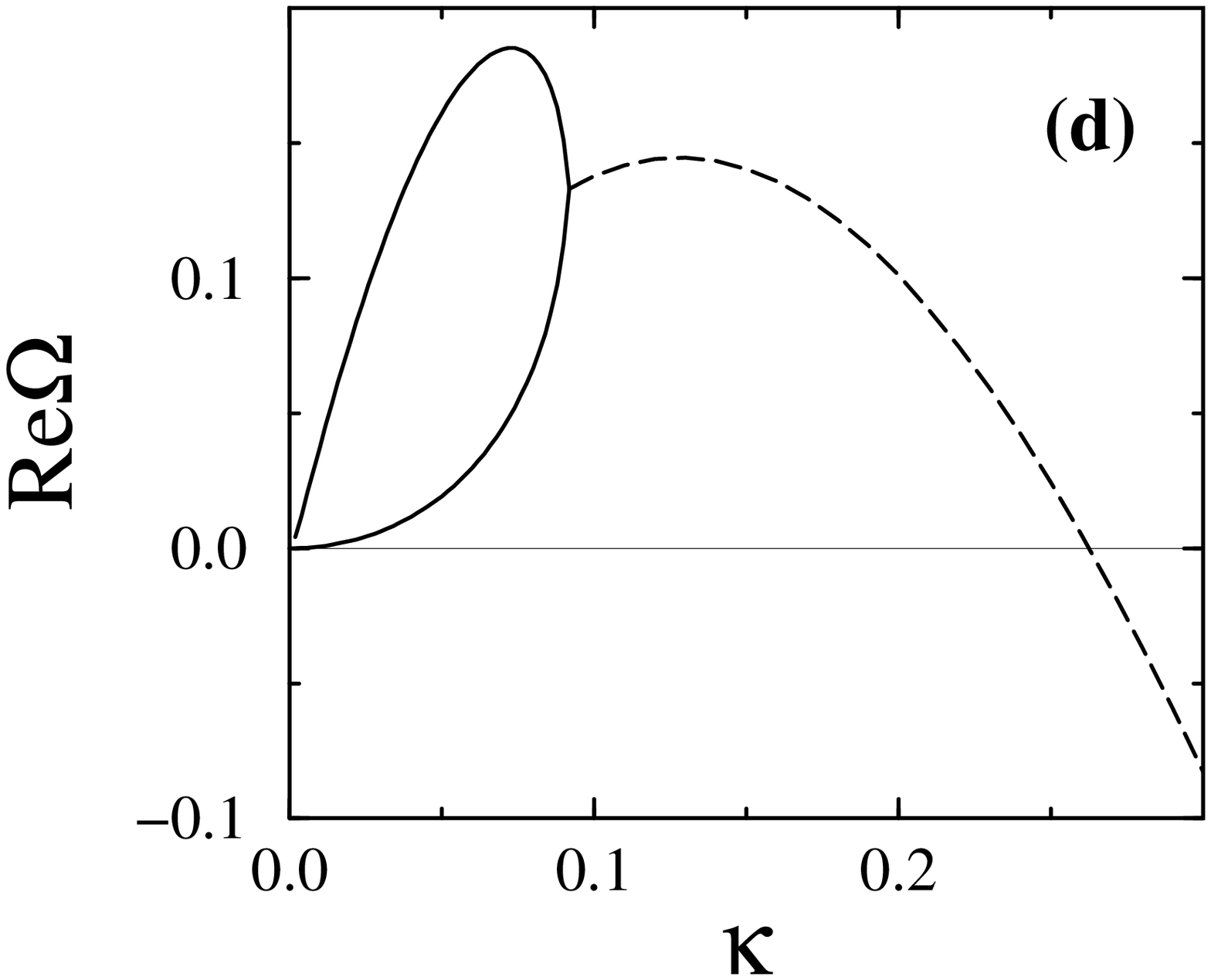,width=.25\textwidth}} 
\centerline{ 
  \psfig{file=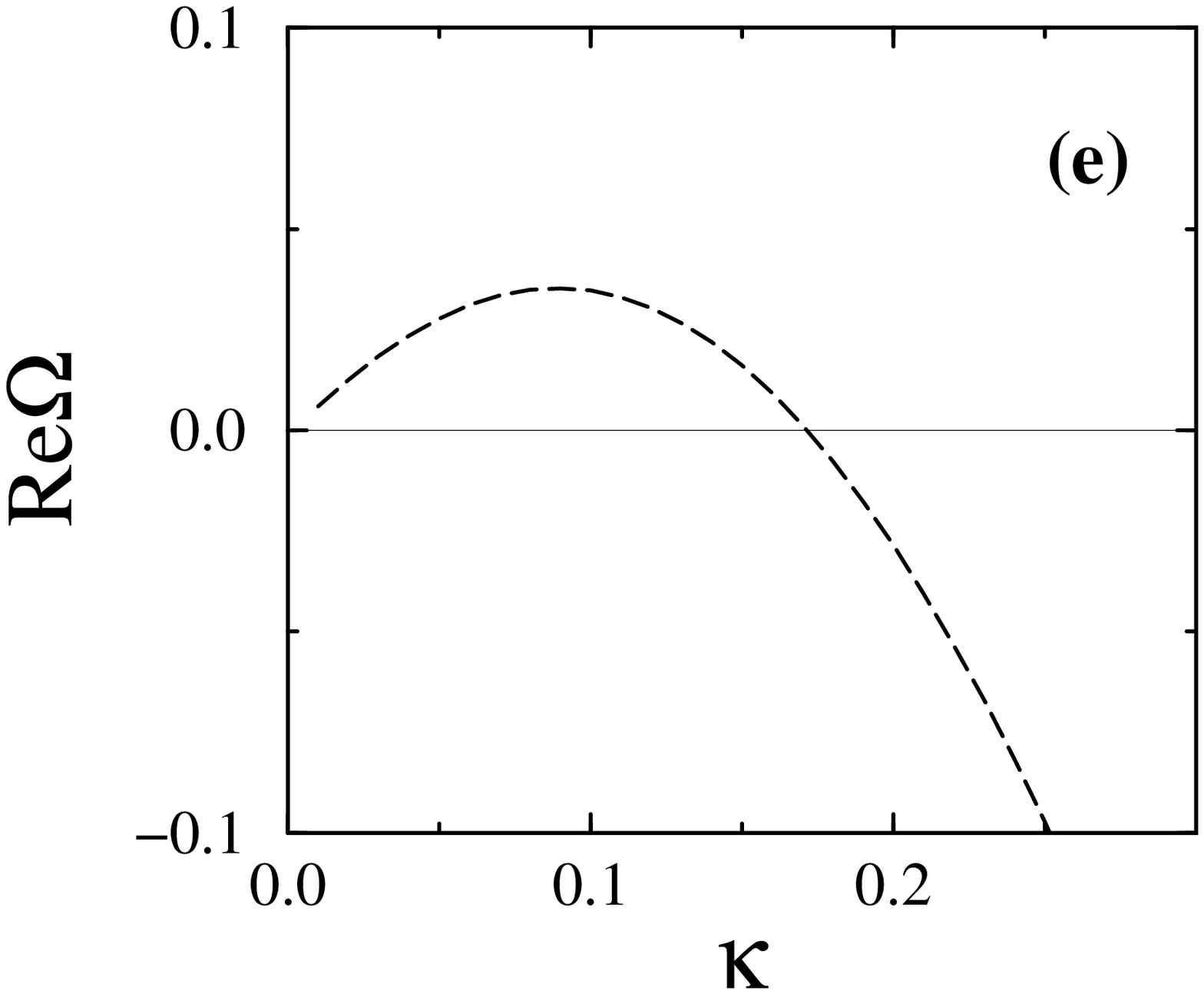,width=.25\textwidth} 
  \psfig{file=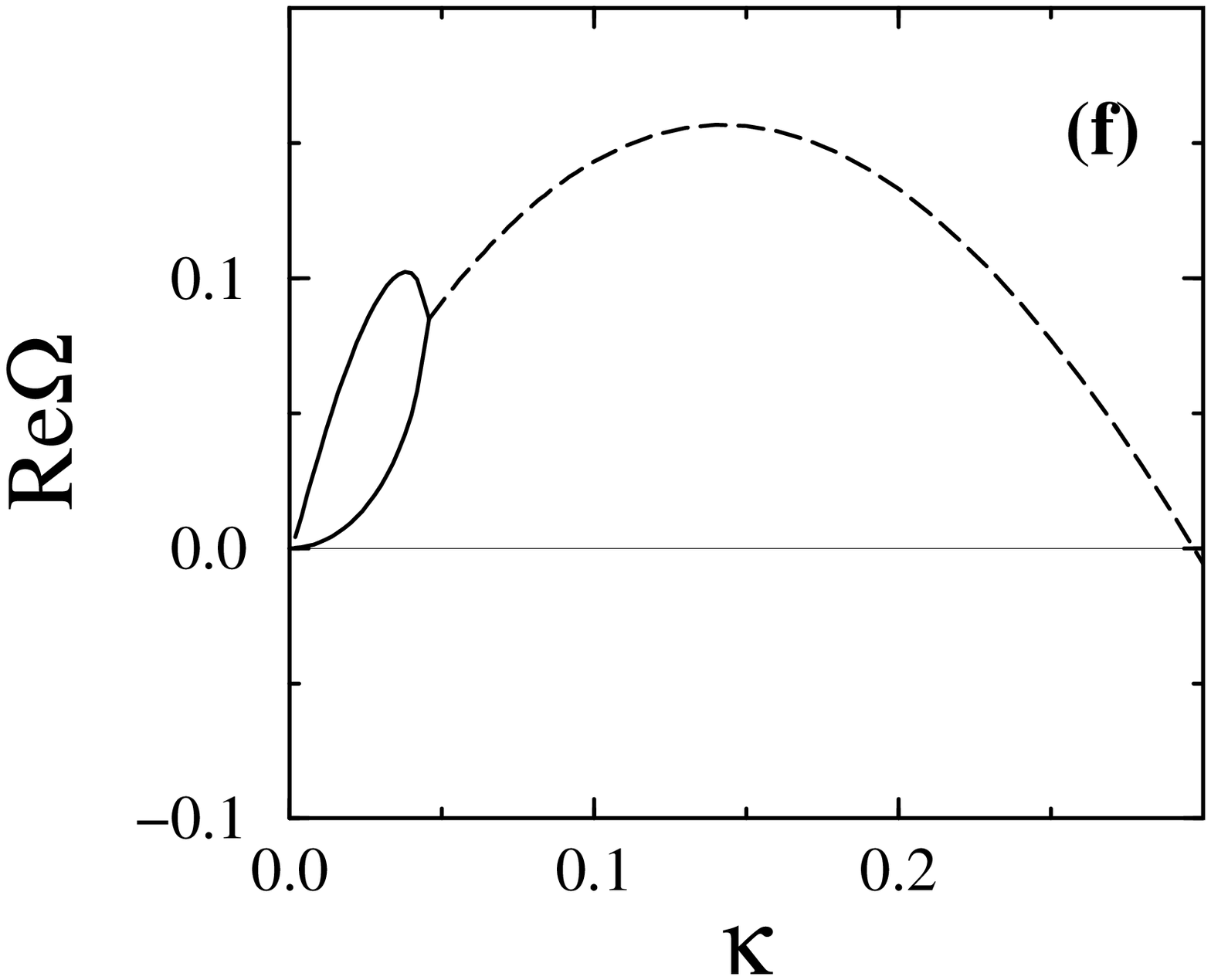,width=.25\textwidth}} 
\caption{Stability spectra for varying $g$ and $\Lambda$
and $w=0.1$, ${\rm Pe}=0.01$, $r=1$, $k_E=0.05$ and $\theta=45^\circ$.
(a) $g=0.15$, $\Lambda=1$; (b) $g=0.05$, $\Lambda=1$;
(c) $g=0.15$, $\Lambda=1.1$; (d) $g=0.05$, $\Lambda=1.1$;
(e) $g=0.15$, $\Lambda=1.5$; (f) $g=0.05$, $\Lambda=1.5$.
Solid lines: real modes; dashed lines: complex modes. Note 
that we show only the diffusive and MS-branches in the 
region of small $\kappa$ where they are unstable. The inset 
in (a) shows in more detail the part of the spectrum 
at small $\kappa$.} 
\label{figinst} 
\end{figure} 
\subsection{Above onset}
\label{compsec}
\nobreak 
\noindent 
For $v>v_c$, the solutions  
for $\Omega$ still satisfy the quadratic equation 
(\ref{abcequation}). The nature of the mode, whether complex 
or real, is determined by the sign of the discriminant, 
$b^2(\kappa)-4a(\kappa)c(\kappa)$. The most important parameter, 
besides $g$, that controls the discriminant is $\Lambda$, because 
$c(\kappa)$ strongly depends on $\Lambda$ near $\Lambda=1$.  
We are interested only in the case $\Lambda>1$,  
since spacings below $\lmin$ are intrinsically 
unstable. For $\Lambda$ near $1$, $c(\kappa)$ is small, and real 
modes should appear. In contrast, 
for larger $\Lambda$, complex modes  
should be favored. 

To check this prediction, we plot in Fig. \ref{figinst}
a series of stability spectra, calculated using
Eq. (\ref{theequation}), for varying $g$ and $\Lambda$.
We chose $w=0.1$, which, using the constitutional
supercooling criterion, gives a critical temperature
gradient of $g_{CS}=0.2$. We display spectra for two
values of $g$, one close to onset ($g=0.15$), and one
far above the onset ($g=0.05$).

Let us first comment on the structure of 
the spectrum at small values of $\kappa$. The inset of 
Fig. \ref{figinst}(a) shows that the growth rate 
still satisfies Eq. (\ref{limiteq}) for $\kappa=0$, but now 
as $\kappa$ grows the MS branch and the diffusive branch meet  
to form a complex conjugate pair, because $b(\kappa)$ 
approaches zero and the discriminant becomes negative.  
When $\kappa$ grows further, the real part of $\Omega$ 
becomes positive, and the branch may stay complex or 
split again in two real branches, both with positive growth  
rates. For $\Lambda = 1$ [Fig. \ref{figinst} (a) and (b)], 
the spectra always exhibit real modes 
above the onset, and the mode with the maximum growth rate is 
real for both values of $g$. For 
$\Lambda=1.1$ [Fig. \ref{figinst} (c) and (d)], near the  
onset only a narrow band of real modes is present,  
and the fastest growing mode is complex. This changes at  
lower $g$: the $\kappa$ range of real modes has increased,  
and the fastest growing mode is real. Note that in 
both spectra there are two maxima of the growth rate, one real 
and the other complex. At a certain value of $g$, the two maxima 
are of equal height; at this point, we have a finite jump in the 
wave number of the fastest growing mode when $g$ is varied, 
and two perturbations of different wave numbers grow with the 
same rate. Finally, for 
$\Lambda=1.5$ [Fig. \ref{figinst} (e) and (f)], there is a range of $g$ 
where the spectrum is entirely complex, and only far above the onset 
real modes appear. The fastest growing mode is always complex. 
 
\begin{figure} 
\centerline{ 
  \psfig{file=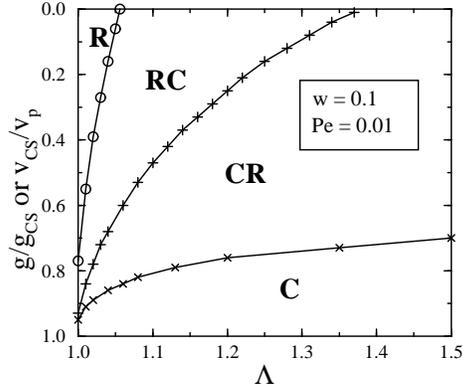,width=.4\textwidth}} 
\caption{Structure of the dispersion relation 
as function of the parameters $\Lambda=\lambda/\lmin$ and 
$g/g_{\rm CS} = v_{\rm CS}/v_p$. The onset of instability  
(constitutional supercooling) is at the bottom of the diagram. 
The meanings of R, RC, CR, and C are defined in the text.} 
\label{figdiag} 
\end{figure} 
We can summarize these results in a diagram that shows the nature 
of the spectrum in the plane $(\Lambda,g)$ (Fig. \ref{figdiag}).
For this diagram, we normalize the temperature gradient by
its critical value according to the constitutional supercooling
criterion, $g_{\rm CS} = 2rw$.
Note that we have $g/g_{\rm CS} = v_{\rm CS}/v_p$, where
$v_{\rm CS}=G\Dtil/\mtil\Delta\ctil$ is the critical velocity.
We classify the spectra into four categories, according to the 
occurrence of maxima. We denote by R spectra with only a real 
maximum (Fig. \ref{figinst}a); spectra with a real and 
a complex maximum are denoted by RC when the fastest growing  
mode is real (Fig. \ref{figinst}d), and CR when it is complex 
(Fig. \ref{figinst}c). Finally, the entirely complex spectra 
are denoted by C. Figure \ref{figdiag} shows for which parameters 
we can expect predominantly real or complex modes; around the 
line between RC and CR, we can have the competition of two different  
modes. This diagram was determined using Eq. (\ref{theequation}) 
with the parameters $w=0.1$, ${\rm Pe}=0.01$, and $k_E=0.05$.  
Increasing the concentration of impurities (increasing $w$)  
favors real modes: all curves are shifted to the  
right and to the bottom in the diagram. The P{\'e}clet 
number and the partition coefficient influence the diagram only 
in the part near the onset of instability. 
 
\begin{figure} 
\centerline{ 
  \psfig{file=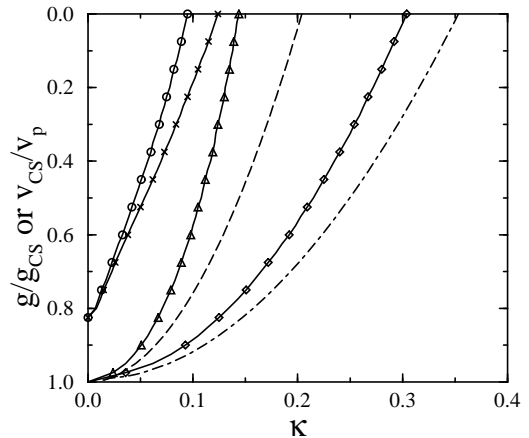,width=.45\textwidth}} 
\caption{Wave numbers of some characteristic points in the 
stability spectrum as a function of $g/g_{\rm CS}$
for $w=0.1$, ${\rm Pe}=0.01$, $k_E=0.05$, and $\Lambda=1.1$. 
Solid lines with circles: real maximum, with crosses: limit 
between real and complex modes, with triangles: complex 
maximum, and with diamonds: marginally stable mode. The 
dashed and dash-dotted lines are the approximations for the 
complex maximum and the marginal mode given by Eqs. 
(\ref{scaling}) and (\ref{approxk}), respectively.} 
\label{figkmax} 
\end{figure} 
To describe a spectrum in more detail, we may use several characteristic 
wave numbers: the wave numbers of the fastest growing modes, 
the limit between real and complex modes, and the wave number 
of the marginally stable mode $\kappa_m$. Figure \ref{figkmax}  
shows these quantities as a function of $g$ for $\Lambda=1.1$. 
As can be seen from Figs. \ref{figinst}c and \ref{figinst}d, 
for this value of $\Lambda$ real modes become dominant only 
well beyond the CS threshold. 
 
\begin{figure} 
\centerline{ 
  \psfig{file=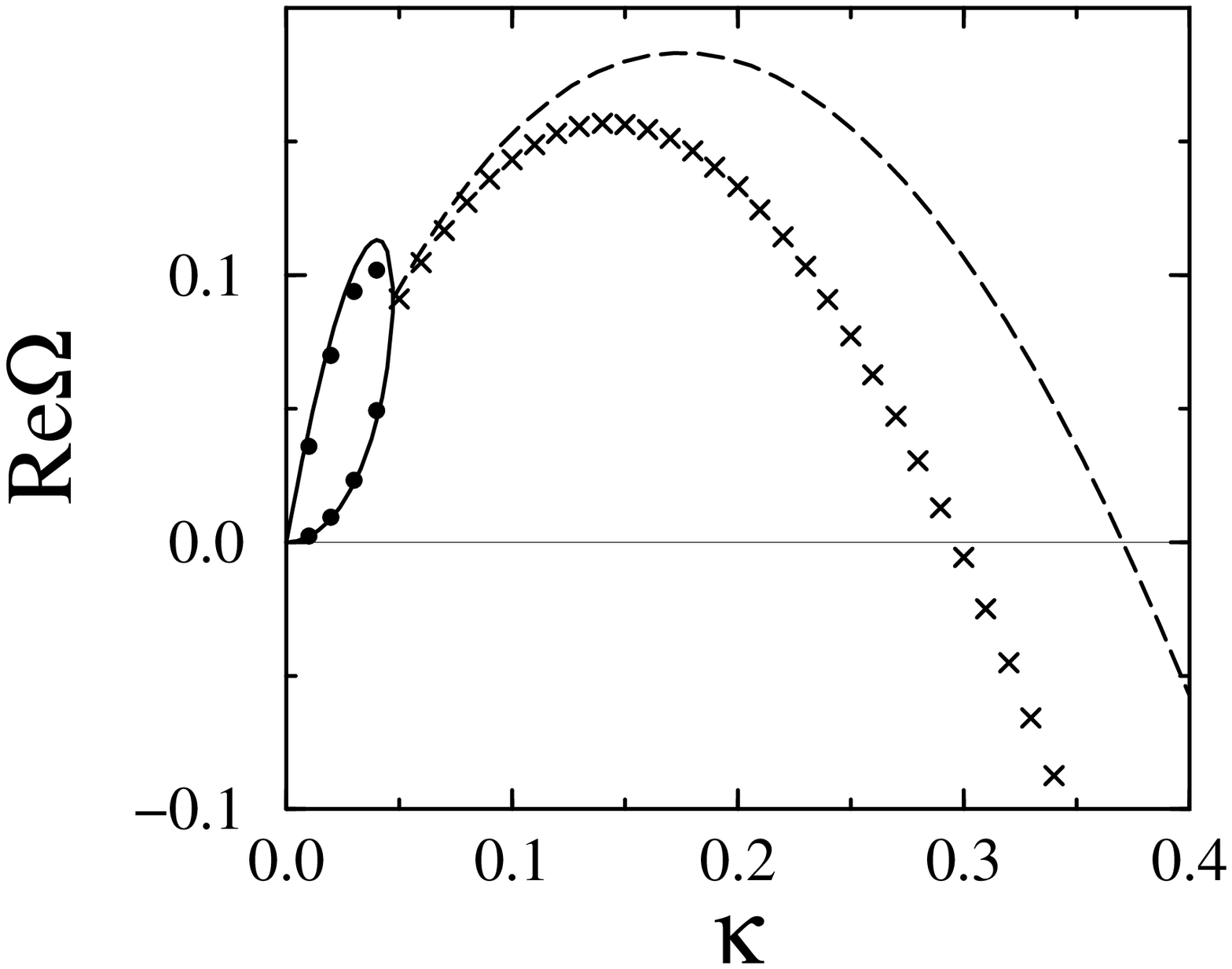,width=.32\textwidth}} 
\centerline{ 
  \psfig{file=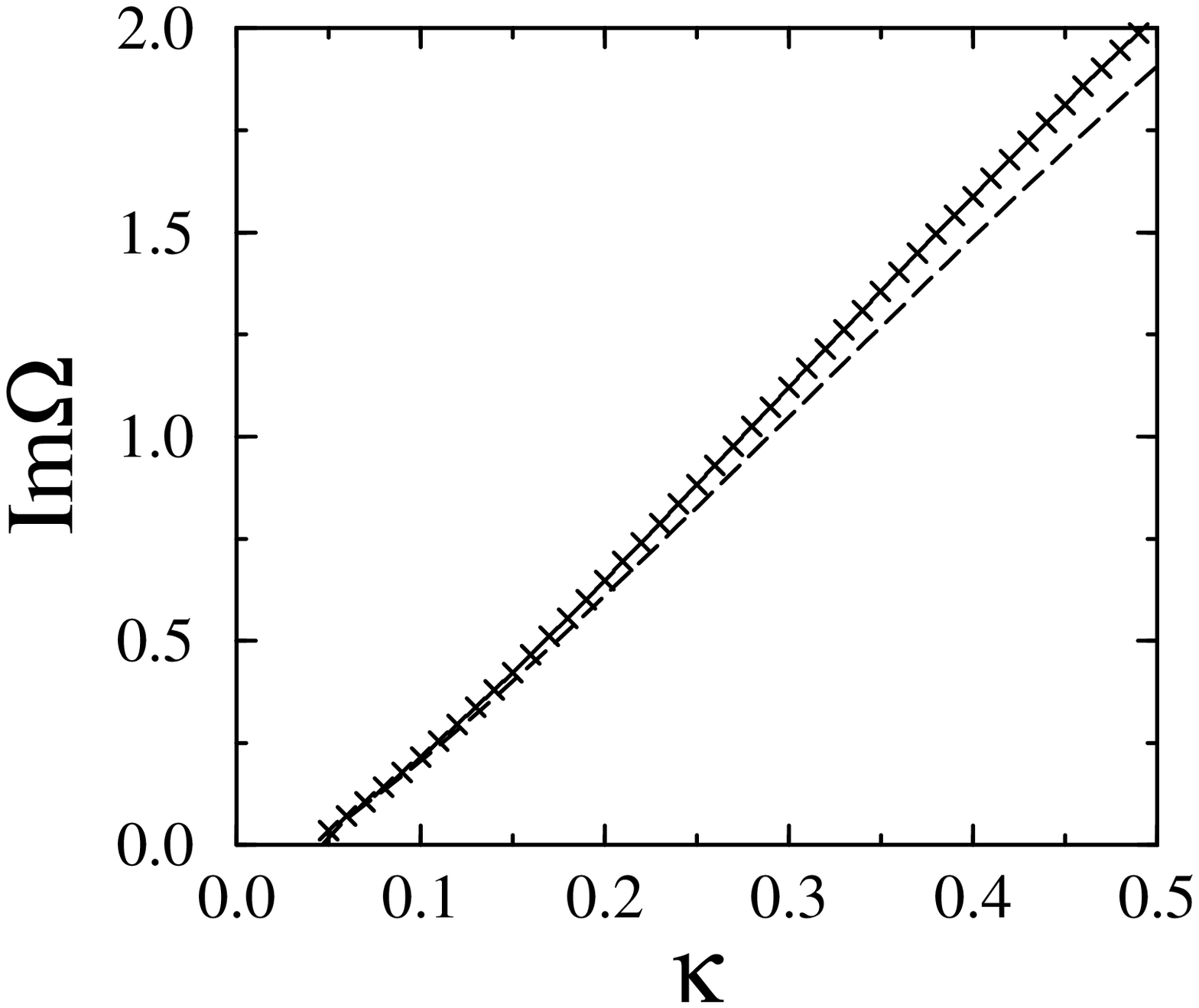,width=.32\textwidth}} 
\caption{Comparison of stability spectra obtained by 
the full calculation, Eq. (\ref{theequation}), and 
by Eq. (\ref{abcequation}) using Eqs. (\ref{approxcoeff}) for 
the coefficients $a$, $b$, and $c$. Symbols: full calculation 
(circles: real modes, crosses: complex modes). Lines: approximation 
(solid lines: real modes, dashed lines: complex modes). 
The parameters are the same as for Fig. \ref{figinst}(f).} 
\label{figcomp} 
\end{figure}

To investigate systematically these quantities as  
functions of the control parameters, it is cumbersome to 
use Eq. (\ref{theequation}) because of the sums involved 
in the functions $U_n$. Therefore, it is convenient 
to obtain approximate expressions 
for the wave numbers of the marginal mode and the complex 
maximum using the approximate spectrum  
defined by Eq. (\ref{abcequation})  
and Eqs. (\ref{approxcoeff}). 
We will show in Sec. VI that this simplified spectrum 
can be recovered from an 
effective interface approach, which  
applies to an arbitrary phase diagram 
and composition. It is therefore worthwhile to 
investigate the quality of this approximation in the present 
symmetric case by quantitative comparison with 
the exact spectrum of Eq. (\ref{theequation}).  
Figure \ref{figcomp} shows that, for the example of  
the last spectrum in Fig. \ref{figinst}, the qualitative  
aspect of the spectrum is well reproduced. We checked several  
cases and always found that the approximation 
shifts the marginally stable mode and  
the maxima to larger $\kappa$. The error, however,  
never exceeded about 30\%.  
The imaginary part of $\Omega$ is very well approximated 
over the whole range of $\kappa$. We also recalculated the 
diagrams of Figs. \ref{figdiag} and \ref{figkmax}, and found 
that the lines are slightly shifted, but the qualitative 
structure of the diagrams stays the same. It seems therefore 
valid to use Eqs. (\ref{approxcoeff}) for a general analysis. 
 
The marginally stable mode is always complex, and $\kappa_m$ 
can be determined by the condition $b(\kappa_m) = 0$. 
In addition, far enough above the onset the term in $1/\kappa$  
in Eq. (\ref{bkappa}) can be neglected. We find 
\begeq 
\kappa_m^2 = {1\over\alpha}\left(wr-{g\over 2}\right) 
\label{scaling} 
\endeq 
with 
\begeq 
\alpha = {2\pi^2P(1/2)\cot\theta\over\Lambda^2}+R_0 =  
{4\pi^2wl\over \lambda^2} \bar d_0, 
\label{alphadef} 
\endeq 
where $R_0$, as before, is given by Eq. (\ref{rzerodef}). 
We can also give an approximate expression for the wave number of the 
complex maximum. For complex modes, $\Re\Omega=b(\kappa)/2a(\kappa)$. 
If we neglect the first term in Eq (\ref{akappa}) for $a(\kappa)$,  
we obtain 
\begeq 
\kappa_{max}\approx \kappa_m/\sqrt 3 =  
   \sqrt{{1\over 3\alpha}\left(wr-{g\over 2}\right)}. 
\label{approxk} 
\endeq 
The resulting curves for $\kappa_m$ and $\kappa_{max}$ are 
shown in Fig. \ref{figkmax}. For the real maximum, all terms 
in the quadratic equation for $\Omega$ have to be retained, 
and no simple expression for the wave number of the maximum 
can be obtained. 
 
Some remarks seem to be in order here to clarify the meaning of  
Eq. (\ref{scaling}) in the dimensional variables. Since 
$g/g_{CS}=v_{CS}/v_p$, the limit of high pulling speeds 
corresponds to $g\to 0$. In this limit, the structure of the 
spectrum varies very little with the pulling speed, because 
both the characteristic length of the MS problem, 
$\sqrt{d_0\ltil}$, and the spacing $\lmin$ vary as $1/\sqrt{v_p}$ 
far above the onset. Hence, if we increase the pulling speed and 
{\em at the same time} change the lamellar spacing such that 
$\Lambda$ stays constant, the only parameter that changes 
(except $g$) is the P{\'e}clet number. But ${\rm Pe}$ appears in the 
problem only in the last term of Eq. (\ref{bkappa}), which we 
have neglected in order to obtain Eq. (\ref{scaling}). This term becomes 
important only at very high pulling speeds and leads to the  
absolute stability of the interface, as in the case of 
a dilute binary alloy. 
Substituting $\kappa=k\lambda/2\pi$ in Eq. (\ref{alphadef}) 
shows that, for constant $\Lambda$, far above the 
CS threshold the wavelength of the fastest growing complex mode 
scales as $\sqrt{\bar d_0 l}$, as for monophase solidification.  
On the other hand, in experiments on eutectics unstable states  
are usually reached from stable states by a sudden increase  
of the pulling velocity \cite{Faivre92,Ginibre97}. The lamellar  
spacing immediately after the jump is the same as before, 
but $\lmin$ and consequently $\Lambda$ have changed. Hence, starting 
from the same initial state and varying the final pulling speed 
corresponds to a variation of both $g$ and $\Lambda$. 
Equation (\ref{scaling}) still applies, but in view of 
Eq. (\ref{alphadef}) no simple scaling with 
$v_p$ is expected. A last remark concerns 
the dependence of $\kappa_m$ on the impurity 
content. Taking the limit $g\to 0$ in Eq. (\ref{scaling}) shows that 
$\kappa_m\sim\sqrt{w}\sim\sqrt{\ctilinf}$. The reason for this 
behavior is that the effective capillary length,  
Eq. (\ref{capilleff}), scales as $\bar d_0\sim w^{-1}$, whereas  
$\lmin$ is independent of $w$. This means that the MS instability 
length increases with decreasing $w$, whereas the lamellar 
spacing stays constant, and hence we expect the wavelength of
the primary instability of a eutectic front to decrease with
increasing impurity content.

Finally, let us verify that the quasistationary approximation
we have used to obtain the above results is justified. To
this end, we estimate the order of magnitude of the
three terms in the diffusion equation Eq. (\ref{diffue}).
For a perturbation $\delta u$ of wave number $k>1/\ltil$ 
($\kappa > {\rm Pe}$) growing at rate $\omega$, we have 
$\partial_t \delta u \sim \omega\delta u$, 
$\partial_z\delta u \sim k\delta u$, and
$\lapl \delta u \sim k^2\delta u$. In terms of the
dimensionless variables, the magnitude of these terms
is $|\Omega|{\rm Pe}$, $\kappa {\rm Pe}$, and $\kappa^2$.
From Fig. \ref{figcomp} we see that, for $\kappa>{\rm Pe}$,
$|\Omega| < B\kappa$ with some number $B$ of order unity,
and hence the omission of the time derivative
from the diffusion equation is well justified for the
range $\kappa> {\rm Pe}$ of interest here. It is also 
possible to relax the quasistationary approximation.
Then, the growth rate $\omega$ appears in the denominators
of all the sums $S_n(\kappa)$ and $\Stil_n(\kappa)$. 
An analytic treatment becomes impossible, 
but the equation for $\Omega$ can be iterated numerically.
We have checked that for the present range of parameters,
the use of this complete calculation leads only to 
insignificant changes in the spectra. Note, however,
that for higher P{\'e}clet numbers it may be necessary
to include this effect. 

\subsection{Oscillatory modes}
\nobreak 
\noindent 
The most interesting result of this analysis is evidently 
the existence of complex modes. To illustrate the 
type of microstructures these modes would generate,  
we have calculated the trajectories of the trijunction points for  
a particular example, using the definitions Eqs. (\ref{xifourier})  
and the complex amplitudes $X_k^s$ and $Y_k^s$ obtained from 
the eigenvalue equation. For the symmetric alloy at the eutectic  
composition, the symmetry between the $\alpha$ and $\beta$ phases 
gives immediately $X_k^\beta = \exp(ik\lambda/2)X_k^\alpha$. The 
amplitude and phase of $X_k^\alpha$ may be chosen arbitrarily, 
as this amounts to fixing the origins of the space and time axes. 
The $Y_k^s$ were then calculated using the growth constraints  
Eqs. (\ref{grconstraints}). 
 
\begin{figure} 
\centerline{ 
  \psfig{file=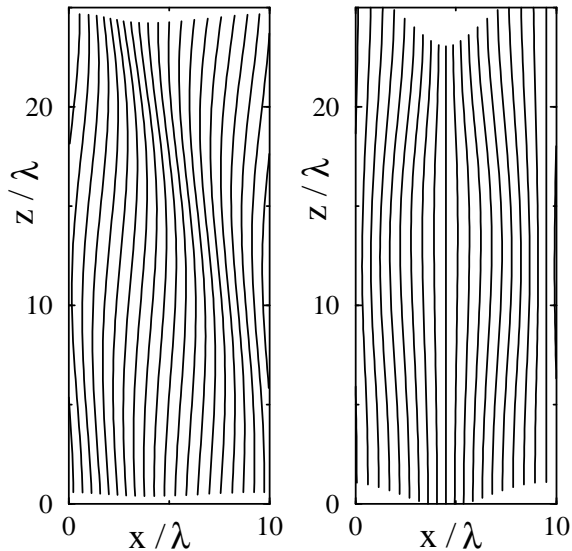,width=.5\textwidth}} 
\caption{Microstructures created by complex linear growth modes. 
Left: a single mode traveling to the left. Right: a standing 
wave mode obtained by the superposition of two complex conjugate 
modes. Growth direction from bottom to top.} 
\label{figstruct} 
\end{figure}

We chose the complex modes with a maximum growth rate of 
Fig. \ref{figinst}(e) ($w=0.1$, $\Lambda=1.5$, $g=0.15$). As 
for every point in the complex part of the spectrum, there are 
two ``degenerate'' modes with complex conjugate growth rates. The 
modes of our example have a wavelength of $10\,\lambda$ ($\kappa=0.1$)  
and growth rates $\Omega = 0.0347 \pm 0.257i$. One of these two 
modes is depicted in the left part of Fig. \ref{figstruct}. We see  
that it has a ``traveling wave'' structure. The wavelength  
in the $z$ direction, 
expressed in lamellar spacings, is $2\pi/\Im\Omega$, and the 
propagation velocity is $v/v_p=\Im\Omega/2\pi\kappa$. 
There are two complex conjugate modes: one propagates to the 
left, and the other to the right. As their growth rates are equal, we can 
create any superposition of the two, in particular a ``standing 
wave'' shown in the right part of Fig. \ref{figstruct}.  
 
The reason for the existence of these oscillatory modes is  
the interplay between the destabilizing impurity diffusion 
field and the dynamical response of the internal lamellar structure. 
A protrusion of the interface rejects impurities more efficiently  
than a flat interface, and hence grows faster. But as the 
curvature of the front increases, the trajectories of the  
trijunctions are more and 
more curved, and the local spacing increases. This leads to a 
decreased efficiency of the interlamellar eutectic diffusion 
and, hence, the interface slows down. As a result, the protrusion  
grows back. The lamellar spacing, however, still increases  
due to the geometric constraints, 
and the process overshoots, leading to a concave deformation
of the front. This gives a geometric interpretation of 
the difference between traveling and standing waves: for the 
traveling wave, the lamellar spacings to the right and to 
the left of the protrusion are different, providing 
a driving force for the propagation of the perturbation. For the 
standing wave, the interface shape and the spacing are ``in phase'',  
and the perturbation oscillates without propagation. Hence, the 
resulting microstructure depends on the initial relation between  
interface shape and lamellar spacing. This implies that in an  
experiment, where the initial perturbations of interface position  
and lamellar spacing have no reason to be in a particular phase  
relation, one should observe all possible superpositions. 
 
\begin{figure} 
\centerline{ 
  \psfig{file=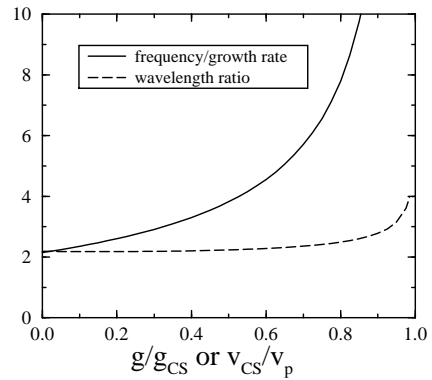,width=.4\textwidth}} 
\caption{Ratio of oscillation frequency to amplification rate, 
$\Im\Omega/\Re\Omega$, and ratio of the wavelengths in 
the $z$ and $x$ direction, $2\pi\kappa/\Im\Omega$, for the oscillatory 
modes with maximum growth rates as a function of $g/g_{\rm CS}$. Other 
parameters are $\Lambda=1.5$, $w=0.1$, ${\rm Pe}=0.01$, $k_E=0.05$.} 
\label{figratios} 
\end{figure} 
There are two characteristic quantities (besides the wave number) 
related to an oscillatory mode: its propagation velocity, or 
equivalently the ratio of the wavelengths in the $z$ and $x$ directions, 
$2\pi\kappa/\Im\Omega$, 
and the ratio of its frequency and its amplification time,  
$\Im\Omega/\Re\Omega$, which determines 
how many oscillations should be observable before the amplification 
leads to lamella elimination. In Fig. \ref{figratios} we show 
both quantities as a function of $g$ for $\Lambda=1.5$, where complex 
modes always dominate. We used Eqs. (\ref{abcequation}) and 
(\ref{approxcoeff}), determined for each value of $g$  
the wave number of the fastest growing mode, 
and calculated $\Re\Omega$ and $\Im\Omega$ at this point. 
Approximate values could be obtained by using 
Eq. (\ref{approxk}) to obtain $\kappa_{max}$.  
Figure \ref{figratios} shows that the ratio of frequency and amplification 
rate diverges when we approach the onset. This is to be expected, as 
the growth rate vanishes whereas the imaginary part of $\Omega$  
remains finite. With decreasing $g$,  
fewer and fewer oscillations are observable before 
lamella terminations occur. However, even at $g=0$, more than 
two oscillations are completed before the amplitude of the mode has  
grown by a factor $e$, which means that such modes should be  
transiently observable. The amplification ratio 
generally increases when $w$ decreases or 
$\Lambda$ increases. The ratio of the wavelengths is fairly constant 
and slightly increases when $\Lambda$ decreases or $w$ increases. 
 
\section{Effective interface approach} 
\nobreak
\noindent
The discrete analysis of Sec. IV
can in principle be used to calculate 
the stability properties of
a eutectic front for an arbitrary phase diagram
and composition. As we have seen, however, the resulting
eigenvalue equation is quite complicated.
It is therefore advantageous
to develop an alternate formulation of the stability problem
by exploiting the fact that the 
instability wavelength is typically
much larger than the lamellar spacing.
The idea, therefore, is to consider 
the shape of the large-scale front 
instead of the actual lamellar interface, and to solve a 
modified free boundary problem for this ``effective 
interface'', with boundary conditions 
that account for the effect of the underlying 
lamellar structure. 

It is useful to present this
approach in two steps. In a first step, we write down
the free boundary problem for the effective interface
in the absence of surface tension effects. This yields
a rigorous long-wavelength limit where the
expression for $\Omega$ agrees up to
order $\kappa^2$ with the one obtained from taking
the small $\kappa$ limit of the full discrete spectrum.
This expression also reduces, in the absence of a ternary impurity,
to the one derived by Langer \cite{Langer80} 
for a binary eutectic, and contains the 
long-wavelength instability leading to lamella 
termination for $\Lambda<1$.
In a second step, we introduce phenomenologically
the effect of surface tension guided by the insights
of Sec. V. In the following, we will allow the volume fraction
$\eta$ and the eutectic liquidus slopes $m_\alpha$ and $m_\beta$ 
to be arbitrary, and we will only require
for brevity of notation that $\mtil_\alpha=\mtil_\beta$. We will 
briefly comment on the general case, where 
$\mtil_\alpha\ne \mtil_\beta$, at the 
end of this section. 

\nobreak 
\noindent 
\subsection{Long wavelength limit} 
\nobreak 
\noindent 
We start by defining the effective interface
as the continuous curve $\xi(x,t)$ that interpolates
between the displacements of the trijunction points $\xi_j^\alpha$ 
and $\xi_j^\beta$ and a continuous field $y(x,t)$ for the  
displacements along the $x$ direction. 
To obtain the free boundary problem that governs 
the large-scale motion of this interface, we start by 
writing the diffusion equation for the ternary impurity 
in the liquid phase and the associated mass conservation 
condition at the phase boundary, which yields the equations
\begin{eqnarray}
\partial_t \ctil&=& D\nabla^2 \ctil\label{diffimp},\\
-\dtil \partial_n \ctil &=& (v_p + \dot\xi)(1-k_E)\ctil\label{massimp},
\end{eqnarray}
where $k_E$ is given by Eq. (\ref{kedef})
and we have used in Eq. (\ref{massimp}) the expression for the
normal interface velocity in the moving frame, 
$v_n = v_p + \dot\xi$, which is valid for small amplitude 
deformations of the interface. Next, we need a
boundary condition for $\ctil$ on this interface.
For this purpose, we note that lamellae can be assumed to 
grow locally in steady state as long as the 
interface deformation is on a scale much larger than the
lamellar spacing. Therefore, we can assume that for 
such deformations, $\ctil$ obeys locally the
Gibbs-Thomson condition
\begeq 
T=T_E - \mtil\ctil - \Delta T_{JH}(\lambda,v_n),
\label{githoeff} 
\endeq 
where the contribution of the eutectic structure
to the interfacial undercooling is given
by the Jackson-Hunt formula
\begeq 
\Delta T_{\rm JH}(\lambda,v_n) = {1\over 2} \dtmin \left( 
    {\lambda\over\lmin}+ {\lmin\over\lambda} \right), 
\endeq 
with $\dtmin$ and $\lmin$ given by Eqs. (\ref{deltatmin}) 
and (\ref{lambdamin}). Finally, to complete the problem, we
need to relate the local lamellar
spacing $\lambda(x,t)$ and the shape of the front $\xi(x,t)$.
This is done, as in Ref. \cite{Langer80},
by noting that the local lamellar spacing is given by 
\begeq 
\lambda(x,t) \approx \lambda_0\left(1+{\partial y\over\partial x}\right), 
\label{lambdaxt} 
\endeq 
for $y \ll \lambda$  
where $\lambda_0$ is the unperturbed spacing. 
The field $y$ can then be 
eliminated by using the geometrical constraint that lamellae
grow locally perpendicular to the solidification front,
which, expressed in terms of the continuous fields, takes
the form
\begeq 
{\partial y(x,t) \over \partial t} = -v_p {\partial\xi\over\partial x}. 
\label{growthconst}
\endeq 
Equations (\ref{diffimp}-\ref{growthconst}),
together with the boundary condition $\ctil=\ctil_\infty$
far from the interface, define the
free boundary problem for small amplitude and long-wavelength
deformations of the effective interface.
 
The stability spectrum can now be obtained by carrying out
a standard linear stability analysis of the above equations,
which is analogous to the analysis for
a monophase front with the added ingredient that the 
Gibbs-Thomson condition is coupled to a slow evolution
equation for $\lambda(x,t)$ obtained by combining Eqs.
(\ref{lambdaxt}) and (\ref{growthconst}).

We start the stability analysis by writing the perturbations 
$\xi$ and $y$ in terms of Fourier modes, 
\beglett 
\begin{eqnarray} 
\xi(x,t) & = & \xi_k \exp(ikx+\omega t) \\ 
y(x,t) & = & y_k \exp(ikx+\omega t). 
\end{eqnarray} 
\endlett 
The impurity diffusion field is expanded 
in Fourier modes according to Eq. (\ref{utilfourier}), and carrying out 
the same steps as from Eqs. (\ref{utilfourier}) to (\ref{btilk}) 
allows us to determine the Fourier coefficients. 
As a result, Eq. (\ref{btilk}) is replaced by 
\begeq 
\btil_k = \left[\qtil_k - {2\over\ltil}(1-k_E)\right]^{-1} 
    \left({\omega\over\Dtil}+{4k_E\over\ltil^2}\right)\xi_k 
\label{utileqn} 
\endeq 
with
\begeq
\qtil_k = {1\over \ltil} + \sqrt{{1\over \ltil^2} + k^2 + {\omega\over\Dtil}}.
\label{nonquasi}
\endeq
We will in the following again use the quasistationary 
approximation of the impurity diffusion equation, which
corresponds to dropping the term $\omega/\Dtil$
on the RHS of Eq. (\ref{nonquasi}).
As discussed before, we are mainly interested in perturbation
wavelengths much smaller than the diffusion length. Note
that in this limit, and within the quasistationary
approximation, we have $\qtil_k \approx |k|$. We will not,
however, make use of this simplification for the
sake of generality. Next, we linearize the JH formula 
around the initial spacing $\lambda_0$ and the 
pulling speed $v_p$: 
\begin{eqnarray} 
\Delta T_{JH}(\lambda,v) & = & \Delta T_{\rm JH}(\lambda_0,v_p)\nonumber\\ 
 & & \mbox{} +  
    \left.\partial\Delta T_{\rm JH}\over\partial\lambda\right|_{\lambda_0,v_p} 
    \lambda_0{\partial y\over \partial x} \nonumber \\ 
 & & \mbox{} +  
    \left.\partial\Delta T_{\rm JH}\over\partial v\right|_{\lambda_0,v_p} 
    \dot \xi, 
\end{eqnarray} 
where we have used $v_n=v_p+\dot\xi$ and  
$\lambda-\lambda_0 = \lambda_0\partial y/\partial x$ 
from Eq. (\ref{lambdaxt}). The Gibbs-Thomson condition,
Eq. (\ref{githoeff}), linearized in the perturbations, becomes 
\begin{eqnarray} 
G\xi_k & = & \mtil\Delta\ctil\left({2\over\ltil}\xi_k - \btil_k\right) 
     - \left.\partial\Delta T_{\rm JH}\over\partial\lambda\right|_{\lambda_0,v_p} 
     {v_p k^2\over \omega}\xi_k\nonumber \\ 
 & & \mbox{} - \left.\partial\Delta T_{\rm JH}\over 
     \partial v\right|_{\lambda_0,v_p} 
    \omega \xi_k. 
\label{protodisp} 
\end{eqnarray}
Inserting Eq. (\ref{utileqn}) for $\btil_k$ with the
quasistationary approximation for $\qtil_k$, and using  
Eqs. (\ref{deltatmin}) and (\ref{lambdamin}) to calculate 
the derivatives of $\Delta T_{\rm JH}$, we obtain
a quadratic equation for $\omega$. In the dimensionless 
quantities defined as before, this equation reads 
\begin{eqnarray} 
& & \left({wr\over \rhotil_0(\kappa)}+ 
     {P(\eta)\over 2\eta(1-\eta)}\right)\Omega^2 \nonumber \\ 
& & \mbox{} - \left(wr -{g\over 2} 
        -{2w{\rm Pe}r^2k_E\over \rhotil_0(\kappa)}\right)\Omega  
          \nonumber \\ 
& & \mbox{} + {2\pi^2P(\eta)\over \eta(1-\eta)} 
              \left(1-{1\over\Lambda^2}\right)\kappa^2 = 0, 
\label{final} 
\end{eqnarray} 
with $\rhotil_0(\kappa)$ defined by Eq. (\ref{rhotildef}). 
\begin{figure} 
\centerline{ 
  \psfig{file=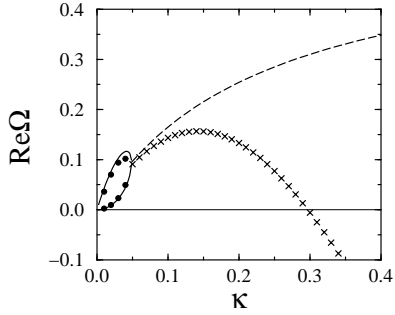,width=.33\textwidth}} 
\caption{Comparison between the full calculation and the 
long-wavelength limit, Eq. (\ref{final}), for the 
stability spectrum. Symbols: full calculation, 
Eq. (\ref{theequation}), circles: real modes, crosses: 
complex modes. Lines: long-wavelength limit, Eq. (\ref{final}), 
solid lines: real modes, dashed lines: complex modes. 
The parameters are the same as for Figs. \ref{figinst}(f)
and \ref{figcomp}.} 
\label{figcomplw}
\end{figure} 
Figure \ref{figcomplw} shows a comparison between this 
formula and the full calculation, Eq. (\ref{theequation}), 
for the same parameters as in Fig. \ref{figcomp}. 
We see that, indeed, all features of the spectrum at 
small $\kappa$ are correctly predicted, including the 
transition from real to complex growth rates with growing 
$\kappa$. This means that the simple calculation outlined 
above is able to capture the qualitatively new feature 
of the instability. Formally, the occurrence of the 
complex growth rates is due to the fact that the 
equation is quadratic in $\omega$, whereas the analogous 
equation of the Mullins-Sekerka calculation is linear. 
This difference arises from the growth constraint
resulting from Cahn's hypothesis. Physically, the
change in the local lamellar spacing resulting from this 
constraint counteracts the destabilization of the front 
by the impurities. This is due to the fact that 
in a convex (i.e. protruding) part of the front, the growing
lamellar spacing leads to an increase in the JH 
undercooling, whereas the inverse is true for concave 
parts. The magnitude of this effect is
proportional to the slope of the JH undercooling
versus spacing curve, which increases with 
lamellar spacing. The oscillations occur because 
only the {\em time derivative} of the spacing 
depends on the instantaneous front shape, 
but not the spacing itself. Therefore, the 
perturbation of the lamellar spacing is phase shifted 
by $\pi/2$ with respect to the perturbation of the front shape.
Consequently, during an oscillation cycle the former has its 
maximum amplitude when the front is planar.
In summary, the long-wavelength oscillations 
are created by the interplay of the 
destabilizing effect of the impurity diffusion field, 
which is the same as for a monophase solidification front, and 
the dynamical response of the underlying lamellar structure. 

\subsection{Inclusion of surface tension} 
\nobreak 
\noindent 
As shown in Fig. \ref{figcomplw}, the spectrum 
derived from the stability analysis of the
effective interface free boundary problem
is in good quantitative agreement with the full
spectrum at small $\kappa$. This approach, however, fails 
to predict the restabilization of the interface at
larger $\kappa$ because it lacks 
capillarity. To add this effect,
we can use the insights of Sec. V for the
symmetric case, where it was noted that the
eutectic stability spectrum could be interpreted
as a planar interface spectrum with an
effective surface tension. This suggests that
we can simply add to the Gibbs-Thomson condition
(Eq. \ref{githoeff}) a capillary
term proportional to the curvature of the
effective interface, which yields the new condition
\begeq 
T=T_E - \mtil\ctil - \Delta T_{\rm JH}(\lambda,v_n) - \Gamma_{\rm eff} K[\xi], 
\endeq 
where $\Gamma_{\rm eff}$ is 
an effective Gibbs-Thomson coefficient.
For the completely 
symmetric case, we can identify $\Gamma_{\rm eff}$ by 
comparing Eq. (\ref{final}) to Eq. (\ref{abcequation}) with 
the approximate expressions Eqs. (\ref{approxcoeff}) for the 
coefficients $a(\kappa)$, $b(\kappa)$, and $c(\kappa)$,
which yields at once the expression
\begeq 
\Gamma_{\rm eff} = \Gamma_E + {M\Delta c {\rm Pe} \lambda_0\over 4\pi^2} R_0, 
\label{gammaeff}
\endeq 
where $\Gamma_E=\Gamma\cos\theta$ 
and $R_0$ is given by Eq. (\ref{rzerodef}).

Let us now consider the extension of this result
to a general alloy phase diagram and an arbitrary
composition. The RHS of Eq. (\ref{gammaeff}) 
contains two contributions that
arise from the large-scale bending of the effective interface.
The first is, in terms of the discrete formalism,
the part of the curvature matrix $\bf K$ that is not
contained in the JH formula. For arbitrary Gibbs-Thomson  
constants, contact angles, and volume fractions,  
simple arguments detailed in appendix C lead to the 
conclusion that the reaction of a composite interface 
to a small curvature can be described by a Gibbs-Thomson 
constant $\Gamma_E$ defined by Eq. (\ref{gedef}) 
that depends on $\Gamma_\alpha$, 
$\Gamma_\beta$, $\theta_\alpha$, $\theta_\beta$ and the 
volume fraction $\eta$. 

The second term on the RHS of Eq. (\ref{gammaeff}) 
originates from the eutectic diffusion field, which 
was shown in Sec. V to have a stabilizing effect 
analogous to a supplementary capillary term. 
For general alloy composition and phase diagram, this 
contribution can in principle be extracted by
expanding the complete spectrum to order $\kappa^2$.
Since, as noted earlier, calculating this spectrum 
involves finding the roots of a fourth order polynomial
in $\Omega$, this expansion is extremely tedious and
was not carried out here. We have found numerically, 
however, that reasonably accurate predictions can be 
obtained for off-eutectic compositions if we simply use 
$\Gamma_{\rm eff}$ defined by Eq. (\ref{gammaeff}) with 
$\Gamma_E$ given by Eq. (\ref{gedef}). In view of the 
large uncertainty in the knowledge of several of the materials 
parameters, notably the Gibbs-Thomson constants and the diffusion 
coefficients, this level of accuracy seems presently sufficient 
to interpret experimental results. Only very precise experiments 
could probe the differences between the full calculation and
this approximation.

Let us state the final result for the stability spectrum 
in two different forms to display the analogies with  
the MS and DL calculations, respectively. First, to  
compare to the MS instability, the dispersion relation 
can be written 
\begin{eqnarray} 
{\omega\ltil^2\over 2\Dtil}& = & 
A(k\ltil)\left(1-{\ltil\over 2 l_T}-
     {\bar d_0\over 2\ltil}(k\ltil)^2\right) - 2k_E\nonumber\\ 
 & & \mbox{}-{\ltil A(k\ltil)\over 4\mtil\Delta\ctil}\dtmin 
       \left[\left(\Lambda-{1\over\Lambda}\right) 
          {v_p k^2\over\omega}+{\omega\Lambda\over v_p}\right]. 
\label{mulsek} 
\end{eqnarray} 
where $A(k\ltil)$, the thermal length $l_T$ and the  
effective capillary length $\bar d_0$ 
are defined by 
\begin{eqnarray} 
A(k\ltil) & = & \sqrt{1+(k\ltil)^2}-1+2k_E, \\ 
l_T & = & \mtil\Delta\ctil / G, \\ 
\bar d_0 & = & {\Gamma_{\rm eff}\over\mtil\Delta\ctil}. 
\end{eqnarray} 
Without the eutectic part on the RHS, Eq. (\ref{mulsek})  
is the classical MS result for the one-sided model.  
 
\begin{figure} 
\centerline{ 
  \psfig{file=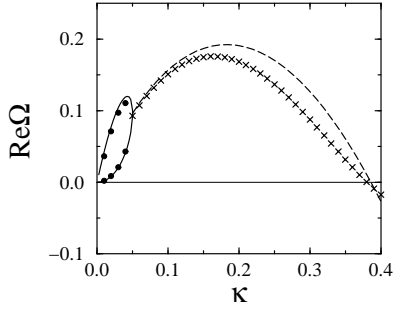,width=.33\textwidth}} 
\caption{Comparison between the full calculation and 
Eq. (\ref{finalfinal}) for an off-eutectic composition
($\eta=0.65$). The other parameters are as for 
Figs. \ref{figcomp} and \ref{figcomplw}. Symbols: full calculation, 
Eq. (\ref{theequation}), (circles: real modes, crosses: 
complex modes). Lines: Eq. (\ref{final}), 
solid lines: real modes, dashed lines: complex modes.}
\label{figcompoff} 
\end{figure} 
In the dimensionless quantities used by DL, the result is 
\begin{eqnarray} 
& & \left({wr\over \rhotil_0(\kappa)}+ 
     {P(\eta)\over 2\eta(1-\eta)}\right)\Omega^2 \nonumber \\ 
& & \mbox{} - \left[wr -{g\over 2}- 
        \left({4\pi^2\gamma(\eta)P(\eta)\over \Lambda^2}+R_0\right) 
         {\kappa^2\over 2}\right. \nonumber \\ 
& & \quad \left. \mbox{}-{2w{\rm Pe}r^2k_E\over \rhotil_0(\kappa)}\right]\Omega  
          \nonumber \\ 
& & \mbox{} + {2\pi^2P(\eta)\over \eta(1-\eta)} 
              \left(1-{1\over\Lambda^2}\right)\kappa^2 = 0, 
\label{finalfinal} 
\end{eqnarray} 
with the function $\gamma(\eta)$ given by 
\begeq 
\gamma(\eta)={\Gamma_E(\eta)(m_\alpha+m_\beta)/2\over 
           (1-\eta)m_\beta\Gamma_\alpha\sin\theta_\alpha + 
           \eta m_\alpha\Gamma_\beta\sin\theta_\beta}. 
\endeq 
Figure \ref{figcompoff} shows both the full  
calculation and the result of Eq. (\ref{finalfinal}) 
for the stability spectrum of the symmetric phase diagram
at an off-eutectic composition,
$\eta=0.65$. The two are in reasonably 
good quantitative agreement, even though the value of
$R_0$ corresponding to $\eta=1/2$ 
was used in Eq. (\ref{gammaeff}).
 
Let us now briefly indicate which modifications will occur 
if the two impurity liquidus slopes differ, 
$\mtil_\alpha \neq \mtil_\beta$. In this case, 
the eutectic composition depends on the  
impurity concentration, and there is a eutectic boundary 
layer with a magnitude depending on the impurity concentration 
at the interface. If the two diffusion coefficients 
$D$ and $\Dtil$ are equal ($r=1$), all results carry over 
if we use the liquidus slope $\tilde M$ defined by  
Eq. (\ref{Mtildef}) instead of $\mtil$ in all equations, 
and redefine the parameter $w=\tilde M\Delta\ctil/M\Delta c$. 
This should usually be a reasonable approximation. If the 
two diffusion lengths are very different, however, one 
would have to consider the eutectic boundary layer separately. 
Then, a separate Fourier expansion has to be used for 
the eutectic boundary layer, and Eq. (\ref{protodisp}) 
is replaced by a more complicated form containing 
both diffusion lengths. 
 
\begin{figure} 
\centerline{ 
  \psfig{file=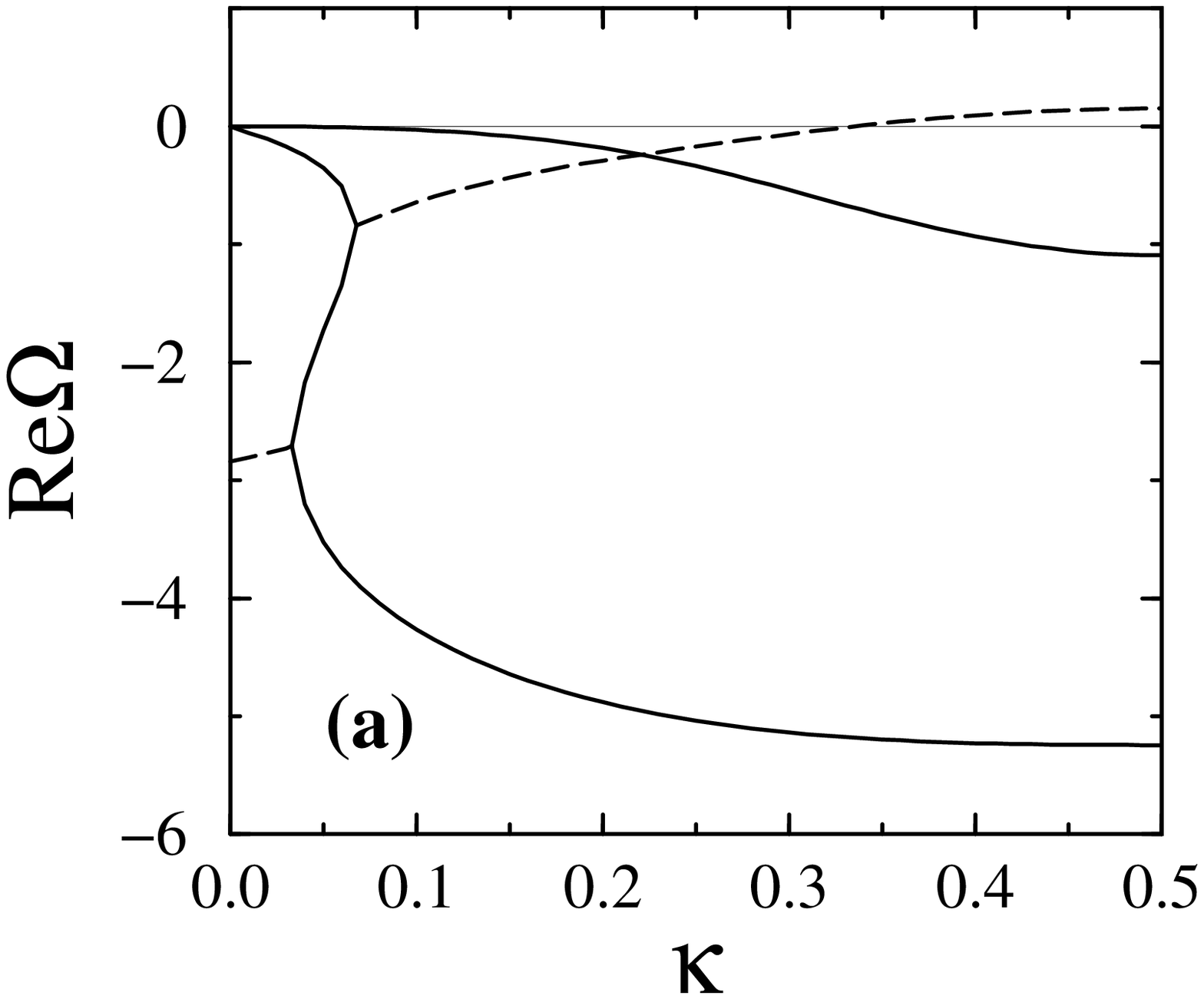,width=.33\textwidth}} 
\centerline{ 
  \psfig{file=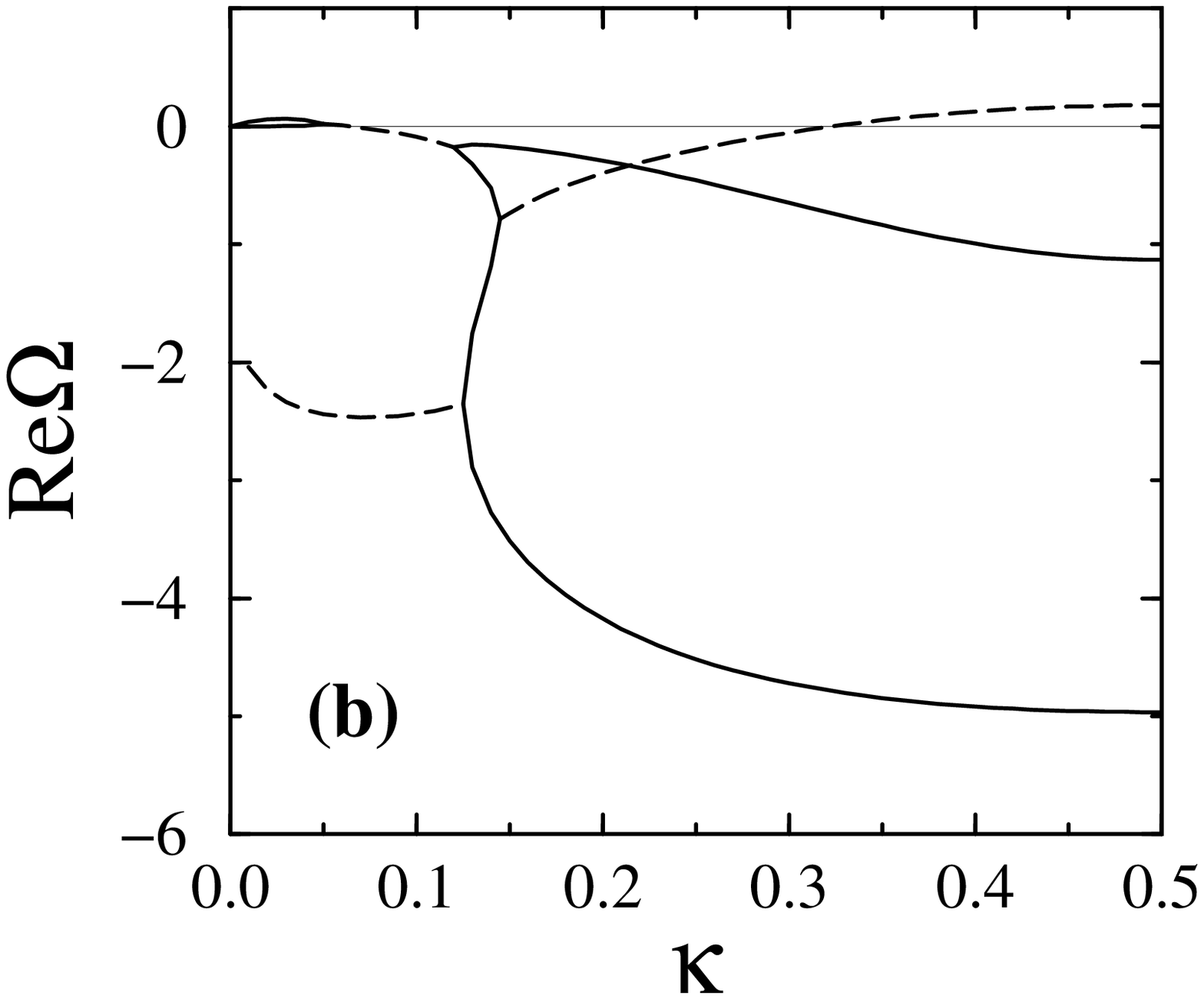,width=.33\textwidth}} 
\centerline{ 
  \psfig{file=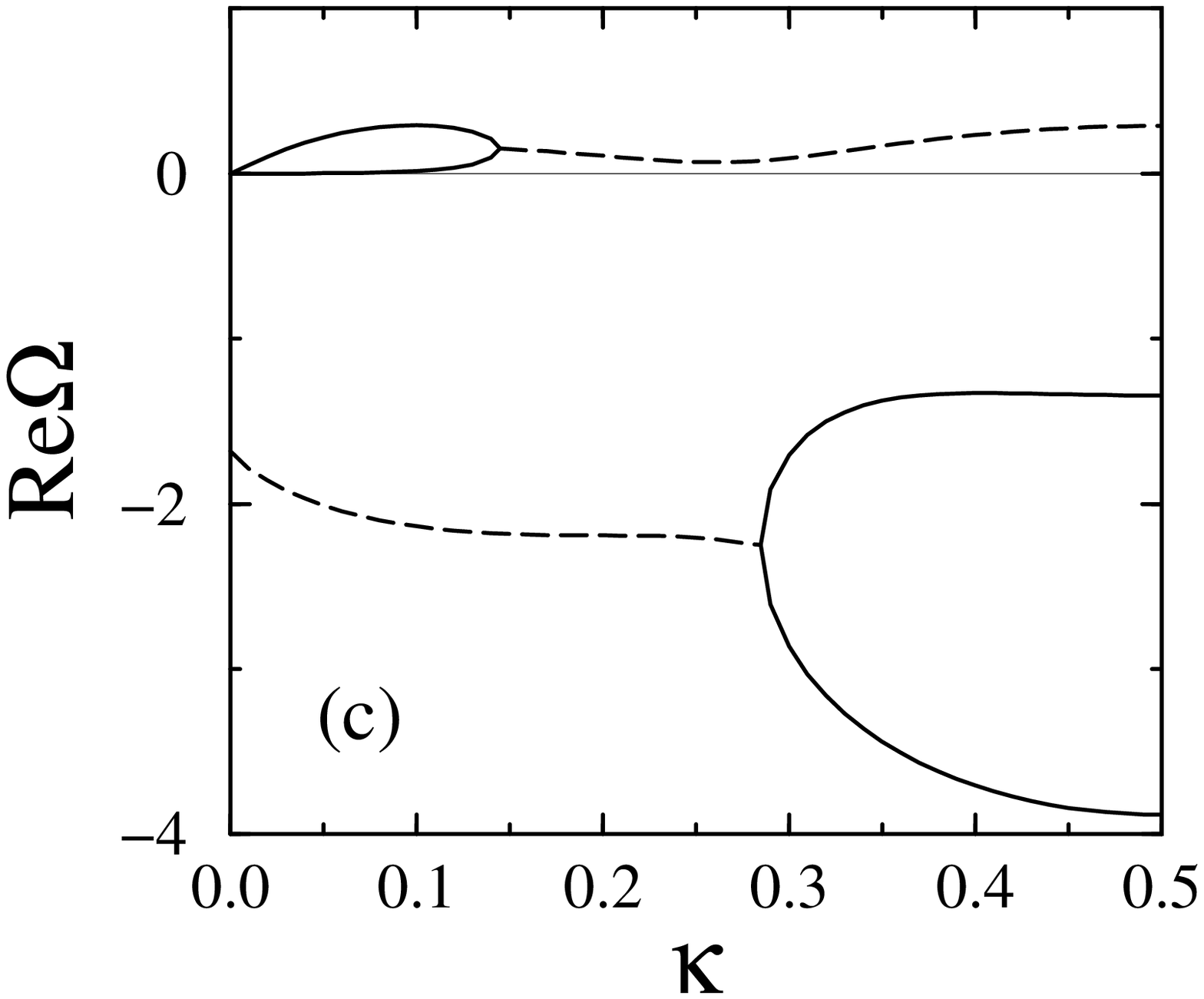,width=.33\textwidth}} 
\centerline{ 
  \psfig{file=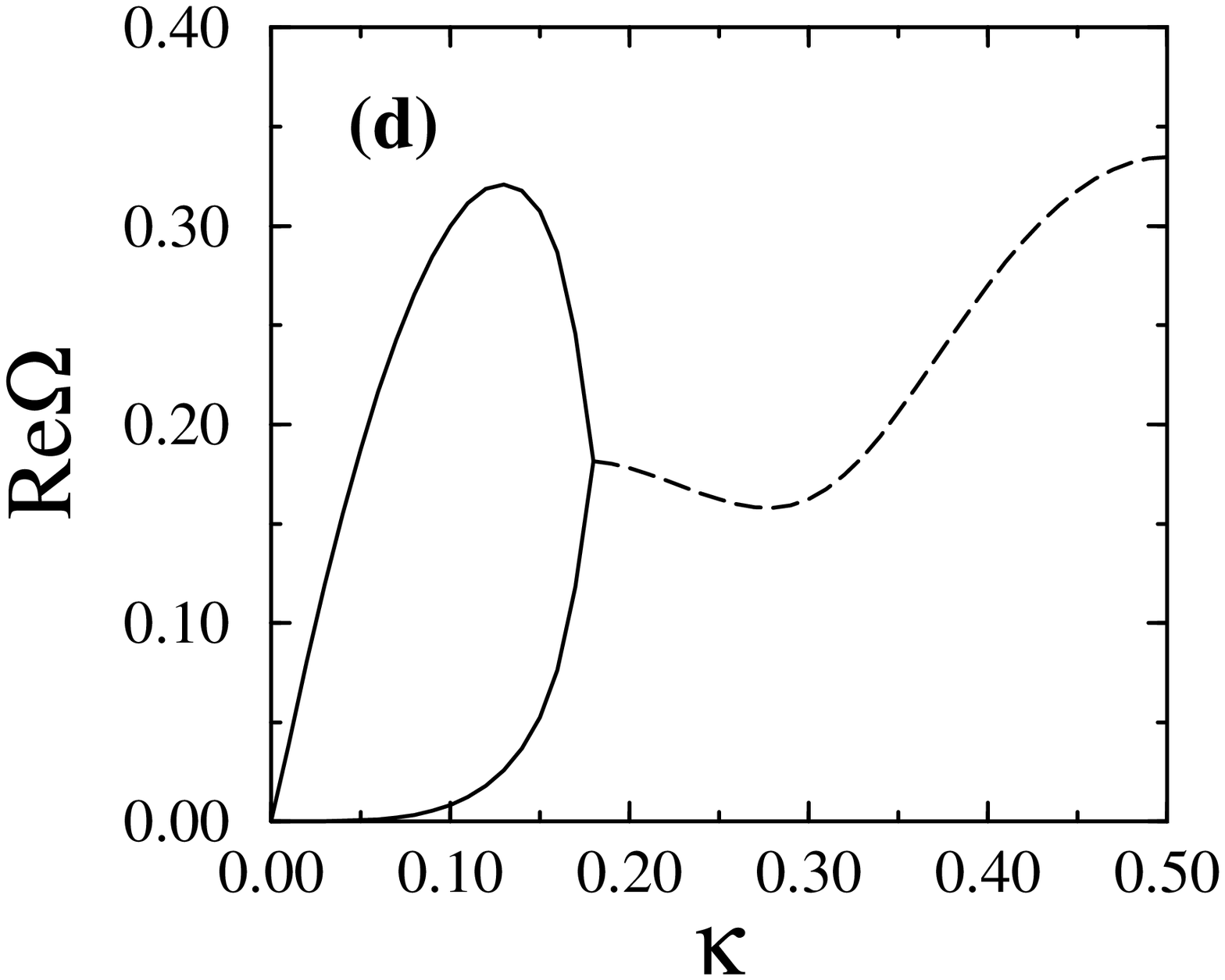,width=.33\textwidth}} 
\caption{Stability spectra for $\eta=0.75$, $\Lambda=1$, 
$\theta=45^\circ$, $k_E=0.05$, and (a) $g=0$, $w=0$, 
(b) $g=0$, $w=0.01$, (c) $g=0$, $w=0.05$, and (d) $g=0.05$, 
$w=0.1$. In (d), only the unstable branch is shown.} 
\label{figtwolam} 
\end{figure} 
\section{Short-wavelength modes} 
\nobreak 
\noindent 
Up to now, we have only considered the instabilities arising 
from the diffusive and MS modes. Let us now turn to the 
effect of the ternary impurity on the $2\lambda$-oscillatory  
($2\lambda$-O) instability at off-eutectic compositions. 
In Fig. \ref{figtwolam}, we show a series of spectra at 
an off-eutectic composition ($\eta=0.8$) with increasing 
impurity concentration. The first spectrum, without 
impurity, again reproduces one of DL's figures. The 
diffusive branch is completely stable, but there is an
unstable complex branch, with the most unstable mode
at $\kappa= 0.5$. For a small impurity 
concentration ($w=0.01$), the long-wavelength 
morphological instability is simultaneously present, but the structure 
of the spectrum stays qualitatively unchanged. For 
still higher impurity concentration ($w=0.05$), however, 
the spectrum becomes quite different: a single 
branch of the spectrum now contains both the long- and 
short-wavelength instabilities, making their distinction somewhat 
arbitrary, whereas the other branch is completely stable. 
We have always found a similar structure for impurity 
concentrations larger than $w\approx 0.02$. Figure \ref{figtwolam}(d)  
shows only the unstable branch for $g=0.05$, $w=0.1$, and 
$\Lambda=1$, for a comparison with the spectrum at the eutectic 
composition shown in Fig. \ref{figinst}(b). The long-wavelength  
part of the two spectra is very similar, but at the 
off-eutectic composition the most unstable mode is 
the $2\lambda$-O mode. The growth rates of the long- and 
short-wavelength instabilities, however, are not very different, 
and we can expect a competition between the two. 
 
\begin{figure} 
\centerline{ 
  \psfig{file=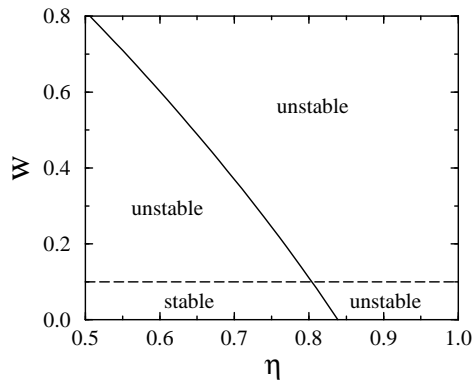,width=.4\textwidth}} 
\caption{Stability diagram for the symmetric eutectic 
alloy in the space $(\eta,w)$ for 
$\Lambda=1$, $g=0.2$, ${\rm Pe}=0.01$, and $k_E=0.05$. 
Solid line: neutral stability limit for the $2\lambda$-O mode. 
Dashed line: constitutional supercooling criterion.} 
\label{figsdiag} 
\end{figure}

At $\kappa=0.5$, the matrix elements of $\bf A$ become 
real. The characteristic equation can again be factored 
in two quadratic equations, which are simply 
\begeq 
{\bf A}^{\alpha,\alpha}=0 \quad \text{and} \quad 
{\bf A}^{\beta,\beta}=0. 
\endeq 
This allows us, in particular, to obtain an equation for the 
neutral stability boundaries where the $2\lambda$-O mode 
first becomes unstable. For the model alloy with the 
symmetric phase diagram, we have on the $\alpha$-rich  
side of the phase diagram ($\eta>0.5$) 
\begin{eqnarray} 
g & = & 2wr-{wr\over \eta} 2k_Er{\rm Pe}\Stil_1(0.5,\eta) + 2\eta -1\nonumber \\ 
  & & \mbox{} + {2\over\eta} \left(R_2(\eta)-S_3(0.5,\eta)- 
        {2\cot\theta\over\Lambda^2}P(\eta)\right), 
\label{twolameq} 
\end{eqnarray} 
where $R_2(\eta)$ and $S_3(\kappa,\eta)$ are defined in 
appendix A. This is a direct generalization of DL's result 
for the binary eutectic. Figure \ref{figsdiag} shows the 
resulting stability diagram for $g=0.2$ and $\Lambda=1$. 
The dashed line is the constitutional supercooling 
criterion, and the long-range instability is present 
everywhere above this line. The solid line was calculated 
using Eq. (\ref{twolameq}), and the $2\lambda$-O mode is 
unstable to the right of this line. We see that 
when the impurity concentration increases, the range in 
volume fraction for which the eutectic front is stable  
decreases. This means that, not surprisingly, 
the ternary impurity boundary layer enhances the oscillatory  
instability. Furthermore, there is a large region in parameter  
space where the two instabilities compete and the fastest growing
linear mode needs to be identified from a plot of the spectrum.
 
\section{Conclusion} 
\nobreak 
\noindent 
We have performed a linear 
stability analysis of a thin 
lamellar eutectic interface in the presence
of a ternary impurity to investigate 
the initial stages of colony formation. The extension of Datye  
and Langer's method has allowed us to calculate the complete 
stability spectrum of the steady-state interface. 
From previous numerical studies of the binary
eutectic case \cite{Karma96}, we expect this discrete stability
analysis to be quantitatively accurate for 
spacings near the JH minimum 
undercooling spacing $\lmin$.

The most dramatic 
conclusion resulting from our analysis is that
the morphological instability of the eutectic interface
induced by the ternary impurity is oscillatory, in
contrast to the standard MS instability of a planar
interface for a dilute binary alloy, which is non-oscillatory.
We have seen that oscillatory modes 
originate from the interplay between the diffusive 
instability driven by the ternary impurity 
and the `dynamical feedback' of the local change in
lamellar spacing on the front motion. In a transient 
regime, these modes should create oscillatory 
microstructures with a wavelength of several 
lamellar spacings such as the ones displayed in Fig. \ref{figstruct}.
There indeed seems to be recent experimental evidence for 
large-scale oscillatory structures of this type in a 
transparent organic with a dilute ternary impurity \cite{Faivre98},
but a more detailed comparison between theory and experiment
is now needed.

Aside from its oscillatory character, the morphological 
instability of the eutectic interface is qualitatively 
similar, near onset, to the standard
MS instability of a monophase front. In particular,
we find that the expressions for the critical onset 
velocity and morphological instability wavelength 
are analogous to those for the classic Mullins-Sekerka 
instability of a planar interface.
The main difference is that the
restabilization of the interface at short wavelength is
controlled by an effective surface tension that depends on the 
geometry of the lamellar interface and on 
interlamellar diffusion, which has a restabilizing
effect. One consequence of this result is that the
constitutional supercooling criterion that has been
commonly used in the metallurgical literature to
predict the onset of instability is indeed applicable for 
typical alloy compositions. Note, however, that 
this criterion becomes inaccurate for very small
concentrations of the ternary impurity.

Above the onset of instability, the stability spectrum
can exhibit both real and complex modes. The 
scaling of the wavelength of the fastest growing mode 
with pulling velocity depends on the nature of the mode. 
For complex modes, far above the onset this wavelength 
scales as the geometric mean of the capillary length 
and the diffusion length. For real modes, 
the situation is more complicated. In both cases, for fixed 
velocity and lamellar spacing, the wavelength far above the 
onset and at sufficiently high impurity concentration  
scales as the inverse square root of the impurity 
concentration. Note that all these statements concern the 
{\em primary instability} of the eutectic front and not the 
finally selected colony spacing. After the first colonies 
have formed, the nonplanar front may undergo a complicated 
sequence of cell elimination or tip splitting events, as during
the development of monophase cellular structures.
 
Furthermore, we have shown that the eutectic front dynamics 
on scales much larger than the lamellar spacing 
can be formulated as a free boundary problem 
with a modified Gibbs-Thomson condition 
that is coupled to a slow evolution 
equation for the lamellar spacing. 
This formulation provides a deeper physical
understanding of the eutectic front dynamics on
this scale. In addition, we have shown that it can
be used to calculate an approximate stability spectrum
that is well suited to interpret experimental data. 
The effective capillary length appearing in this spectrum
contains contributions both from an averaging over the 
material properties of the two phases, weighted by geometric  
factors, and from the eutectic interlamellar diffusion 
field, which acts as a stabilizing force. 
 
Finally, we have found that the 
short-wavelength oscillatory instability, already
present in a binary eutectic, is enhanced 
by the ternary impurity boundary layer. This reduces the composition 
range for stable lamellar eutectic growth, even below constitutional 
supercooling. Above constitutional supercooling and for 
sufficiently off-eutectic compositions, the long-range 
and $2\lambda$-oscillatory instabilities are both present 
and may compete with each other. 
 
In conclusion, we have shown that the instability of a 
lamellar eutectic interface in the presence of a ternary impurity 
is in some respects similar to the Mullins-Sekerka instability 
of a dilute binary alloy, but also presents striking differences. 
There are two interesting future prospects.
First, dynamical simulations of the complete equations of motion 
are necessary to go beyond this linear stability analysis 
and to investigate the subsequent stages of the instability, 
as well as to determine what structures are ultimately formed. 
Work on these issues using the phase-field method  
is currently in progress. Second, it seems
worthwhile to extend the effective interface
approach to a nonlinear regime to model the shape
and dynamics of fully developed colonies, as depicted 
in Fig. \ref{fighunt}.
 
\acknowledgements 
 
We thank S. Akamatsu and G. Faivre
for many fruitful discussions and for sharing with us
their unpublished experimental results.
This research was supported by the U.S. DOE under
grant No. DE-FG02-92ER45471.

\appendix 
 
\section{Summary of Datye and Langer's results} 
\nobreak 
\noindent 
We will state here DL's results for the matrices $\bf G$, $\bf K$,
and $\bf U$ and transform them into our notations. The matrix 
$\bf G$ can be simply read off the definitions of the average 
interface positions $\xiav_j^s$, Eqs. (\ref{xiavdef}), and of the  
Fourier expansion Eq. (\ref{xifourier}): 
\begeq 
{\bf G} = {G\over 2} 
  \left(\begin{array}{cc} 1 & 1 \\ e^{ik\lambda} & 1 \end{array}\right). 
\endeq 
The curvature matrix contains two contributions. The first arises 
from the change in the local lamellar spacing due to the horizontal 
displacements $y_j^s$. The second appears when the interface is bent 
on a scale of several lamellae. Then, the trijunction points are 
turned by small angles with respect to their steady-state orientation. 
These angles can be related to the time derivatives of the 
horizontal displacements. Finally, we use the growth 
constraints Eq. (\ref{grconstraints}) and obtain 
\begeq 
{\bf K}^{\alpha,\alpha} = 
   {2ie^{ik\lambda/2}\sin(k\lambda/2)\over\lambda^2} 
   \left({2v_p\sin\theta_\alpha\over\omega\lambda\eta^2} -  
         {\cos\theta_\alpha\over\eta}\right) 
\endeq 
\begeq 
{\bf K}^{\beta,\beta} = 
   {2ie^{ik\lambda/2}\sin(k\lambda/2)\over\lambda^2} 
   \left({2v_p\sin\theta_\beta\over\omega\lambda(1-\eta)^2} -  
         {\cos\theta_\beta\over 1-\eta}\right) 
\endeq 
\begeq 
{\bf K}^{\alpha,\beta} = {\bf K}^{\alpha,\alpha *} 
\endeq 
\begeq 
{\bf K}^{\beta,\alpha} = e^{ik\lambda}{\bf K}^{\beta,\beta *}, 
\endeq  
where the asterisks in the last two equations denote complex 
conjugation of all the coefficients, but not $\omega$. 
 
For the symmetric phase diagram, we may conveniently rewrite  
this matrix in the dimensionless parameters defined in Sec. IV. 
We remark that in this case we have 
\begeq 
\lmin^2 = {l\Gamma\sin\theta\over M\Delta c P(\eta)}. 
\endeq 
This relation can be used to eliminate $\sin\theta$ in the matrix; 
the reduced lamellar spacing $\Lambda$ appears, and, for 
example, we obtain for ${\bf K}^{\alpha,\alpha}$ 
\begeq 
{\bf K}^{\alpha,\alpha} = 
   2ie^{i\pikap}\sin(\pikap){M\Delta c P(\eta)\over l\Lambda^2\Gamma} 
   \left({2\over\Omega\eta^2} - {\cot\theta\over\eta}\right). 
\endeq 
Note that in the general case the scaling with respect to the 
physical parameters would remain the same; however, additional 
coefficients depending on the angles $\theta_\alpha$ and $\theta_\beta$ 
and the liquidus slopes $m_\alpha$ and $m_\beta$ would appear. 
 
For the matrix $\bf U$, we will just state DL's results; for more 
details, see \cite{Datye81}. The calculation is straightforward 
but tedious because we have to treat the interlamellar 
diffusion. This brings in various sums over the Fourier modes 
of the steady-state expansion, Eq. (\ref{diffufield}). We define 
\begeq 
\Delta_\alpha = B_0 + u_\infty - u_\alpha, 
\endeq 
\begeq 
\Delta_\beta = u_\beta - B_0 - u_\infty, 
\endeq 
\begeq 
R_1(\eta) = \sum_{n=1}^\infty {\sin(2\pi\eta n)\over(\pi n)^2}, 
\endeq 
\begeq 
R_2(\eta) = \sum_{n=1}^\infty {2\sin^2(\pi\eta n)\over (\pi n)^2}, 
\endeq 
\begeq 
\rho_n(\kappa) = \sqrt{4\pi^2(n+\kappa)^2 + {\rm Pe}^2} - {\rm Pe}, 
\endeq 
\begeq 
S_1(\kappa,\eta) = 
   \sum_{n=-\infty}^\infty {\sin^2[\pi\eta(n+\kappa)]\over 
                            \pi^2(n+\kappa)^2\rho_n(\kappa)}, 
\endeq 
\begin{eqnarray} 
S_2(\kappa,\eta) & = & 
   \sum_{n=-\infty}^\infty e^{-i\pi(n+\kappa)} \nonumber \\ 
  && \mbox{} \times 
     {\sin[\pi\eta(n+\kappa)]\sin[\pi(1-\eta)(n+\kappa)]\over 
      \pi^2(n+\kappa)^2\rho_n(\kappa)}, 
\end{eqnarray} 
\begin{eqnarray} 
S_3(\kappa,\eta) & = & 
   4 \sum_{m=-\infty}^\infty 
     {\sin[\pi\eta(m+\kappa)]\over\pi(m+\kappa)\rho_m(\kappa)}  
      \nonumber \\ 
 & & \times\mbox{}\sum_{n=-\infty;\neq 0}^\infty 
        {|n|\over n}\sin(\pi\eta n)e^{i\pi(n-m-\kappa)/2} 
      \nonumber \\ 
 & & \times\mbox{} 
        {\sin[\pi(n-m-\kappa)/2]\over \pi(n-m-\kappa)}, 
\end{eqnarray} 
\begeq 
S_4(\kappa,\eta) = 
   \sum_{n=-\infty}^\infty e^{i\pi\eta(n+\kappa)} 
      {\sin[\pi\eta(n+\kappa)]\over\pi(n+\kappa)\rho_n(\kappa)}. 
\endeq 
Using these quantities, the matrix elements of $\bf U$ are 
\begin{eqnarray} 
{\bf U}^{\alpha,\alpha} & = & {1\over l}\biggl(\Omega U_1^\alpha(\kappa,\eta) 
        +U_2^\alpha(\kappa,\eta) \nonumber \\ 
  & & \mbox{} + 
        {2ie^{i\pikap}\over\Omega}\sin(\pikap) U_3(\kappa,\eta)\biggr), 
\end{eqnarray} 
\begin{eqnarray} 
{\bf U}^{\beta,\beta} & = & {-1\over l}\biggl(\Omega U_1^\beta(\kappa,\eta) 
        +U_2^\beta(\kappa,\eta) \nonumber \\ 
  & & \mbox{} + 
        {2ie^{i\pikap}\over\Omega}\sin(\pikap) U_3(\kappa,1-\eta)\biggr), 
\end{eqnarray} 
\begeq 
{\bf U}^{\alpha,\beta} = {\bf U}^{\alpha,\alpha *} 
\endeq 
\begeq 
{\bf U}^{\beta,\alpha} = e^{ik\lambda}{\bf U}^{\beta,\beta *}, 
\endeq 
with 
\begeq 
U_1^\alpha(\kappa,\eta) = {1\over\eta}\left[ 
     \Delta_\alpha S_1(\kappa,\eta)-\Delta_\beta S_2^*(\kappa,\eta)\right] 
\endeq 
\begeq 
U_1^\beta(\kappa,\eta) = {1\over 1-\eta}\left[ 
     \Delta_\beta S_1(\kappa,1-\eta)-\Delta_\alpha S_2^*(\kappa,1-\eta)\right] 
\endeq 
\begeq 
U_2^\alpha(\kappa,\eta) = - B_0 + {1\over\eta}\left[ 
     S_3^*(\kappa,\eta)-R_2(\eta)\right] 
\endeq 
\begeq 
U_2^\beta(\kappa,\eta) = B_0 + {1\over(1-\eta)}\left[ 
     S_3^*(\kappa,1-\eta)-R_2(1-\eta)\right] 
\endeq 
\begeq 
U_3(\kappa,\eta) = {1\over\eta}\left[ 
     2P(\eta)/\eta-R_1(\eta)-2S_4^*(\kappa,\eta)\right]. 
\endeq 
 
\section{Limit of the DL sums for $\kappa\to 0$} 
\nobreak 
\noindent 
For the detailed study of the symmetric phase diagram at eutectic 
composition, we need to know the leading order behavior of the 
functions $U_1^\alpha$, $U_2^\alpha$, and $U_3$, for $\kappa\to 0$. 
To this end, we have to expand the sums $S_n$ in $\kappa$ and 
resum the resulting terms. We will systematically neglect terms 
of relative magnitude ${\rm Pe}$, as was already done in DL's calculations 
leading to the results of the preceding appendix. 
 
We have to single out terms containing the function $\rho_0(\kappa)$ 
in the denominator, because these terms will be singular in the 
limit $\kappa\to 0$. To see this, note that we have 
\begeq 
\rho_0(\kappa)= \left\{\begin{array}{ll} 
      2\pi^2\kappa^2/{\rm Pe} + \order(\kappa^4) & {\rm for}\quad 2\pi\kappa\ll {\rm Pe} \\ 
      2\pi|\kappa| & {\rm for}\quad 2\pi\kappa\gg {\rm Pe} 
                \end{array}\right. 
\endeq 
We will be interested in a regime where the wavelength of the 
perturbation is larger than the lamellar spacing, but much smaller 
than the diffusion length; hence the latter limit applies. 
 
Similarly, the function $\rhotil_0(\kappa)$ will become small 
when $\kappa$ tends to 0: 
\begeq 
\rhotil_0(\kappa) = \left\{\begin{array}{ll} 
      r{\rm Pe}k_E + 2\pi^2\kappa^2/{\rm Pe} + \order(\kappa^4) &  
           \quad 2\pi\kappa\ll {\rm Pe} \\ 
      2\pi|\kappa| & \quad 2\pi\kappa\gg {\rm Pe} 
                \end{array}\right. 
\endeq 
Hence the terms proportional to $\rhotil_0^{-1}$ in the
impurity contributions have to be considered separately. 
 
Expanding $S_1(\kappa,\eta)$ and $S_2(\kappa,\eta)$, we obtain 
\begin{eqnarray} 
S_1(\kappa,\eta) & = & {\eta^2-{1\over 3}\eta^4\pi^2\kappa^2 
          \over\rho_0(\kappa)} + P(\eta) \nonumber \\ 
 & & \mbox{} + \left[\eta^2 R_3(\eta) - 3\eta R_4(\eta) +  
              6 R_5(\eta) \right] \pi^2\kappa^2 \nonumber \\ 
 & & \mbox{} + \order(\kappa^4/\rho_0) + \order(\kappa^4) \\ 
S_2(\kappa,\eta) & = & \left\{1-i\pi\kappa -{1\over2}\left[ 
          1+{1\over 3}\left(\eta^2+{(1-\eta)}^2\right)\right] 
          \pi^2\kappa^2\right\} \nonumber \\ 
 & & \mbox{} \times {\eta(1-\eta)\over\rho_0(\kappa)}  
      - (1-i\pi\kappa-\pi^2\kappa^2)P(\eta) \nonumber \\ 
 & & \mbox{} + \left[\eta(1-\eta)R_3(\eta) +  
               3(\eta-1/2)R_4(\eta) \right. \nonumber \\ 
 & & \quad \left. \mbox{} -6 R_5(\eta) \right] \pi^2\kappa^2 \nonumber \\ 
 & & \mbox{} + \order(\kappa^4/\rho_0) + \order(\kappa^4) 
\end{eqnarray} 
with 
\begin{eqnarray} 
R_3(\eta) & = & \sum_{n=1}^\infty {\cos 2\pi\eta n\over (\pi n)^3} \\ 
R_4(\eta) & = & \sum_{n=1}^\infty {\sin 2\pi\eta n\over (\pi n)^4} \\ 
R_5(\eta) & = & \sum_{n=1}^\infty {\sin^2 \pi\eta n\over (\pi n)^5}. 
\end{eqnarray} 
The expansions for the impurity sums $\Stil_1(\kappa,\eta)$ and 
$\Stil_2(\kappa,\eta)$ are obtained by replacing $\rho_0(\kappa)$ 
by $\rhotil_0(\kappa)$ in the above expressions. 
 
With the help of these expressions and the definition of $U_1^\alpha$, 
we find for $\eta=1/2$ and $\Delta_\alpha=\Delta_\beta=1/2$: 
\begin{eqnarray} 
&& \Re\left(e^{-i\pikap/2}U_1^\alpha(\kappa,1/2)\right) = \nonumber \\ 
&& \qquad  2P(1/2) + \left[12 R_5(1/2)-{3\over 4}P(1/2)\right] 
    \pi^2\kappa^2 \nonumber \\ 
&& \qquad \mbox{} + \order(\kappa^4). 
\end{eqnarray} 
Rather remarkably, all the singular terms cancel out. A comparison with 
direct numerical summation shows that the limit behavior is correct and 
that the neglected terms sum up to a correction that does not exceed 
$P(1/2)$, even for large values of $\kappa$. 
 
For the impurities, we need the expression 
\begin{eqnarray} 
&& \Stil_1(\kappa,1/2) \cos{\pi\kappa\over 2} +  
    \Re\left(e^{-i\pi\kappa/2}\Stil^*_2(\kappa,1/2)\right) = \nonumber\\ 
&& \qquad {1-{5\over 24}\pi^2\kappa^2\over 2\rhotil_0(\kappa)} 
    + {1\over 2}\left[P(1/2)+R_3(1/2)\right] \pi^2\kappa^2 \nonumber \\ 
&& \qquad \mbox{} + \order(\kappa^4/\rhotil_0) + \order(\kappa^4). 
\end{eqnarray} 
 
We will simplify our task for the function $U_2^\alpha$,  
which contains the most difficult sum $S_3(\kappa,\eta)$, by directly 
expanding the product $e^{i\pikap/2}S_3(\kappa,1/2)$. Using $B_0=0$ 
for the symmetric phase diagram at eutectic composition, we obtain 
\begeq 
\Re\left(e^{i\pikap/2}U_2^\alpha\right) = R_0 \kappa^2 /2 + \order(\kappa^4) 
\label{kappasquare} 
\endeq 
with 
\begin{eqnarray} 
R_0 & = & 2\pi^2\left[3R_6(1/2)-{1\over 6} R_2(1/2) - R_7(1/2)\right] 
         \nonumber \\ 
    & \approx & 0.4965 
\label{rzerodef} 
\end{eqnarray} 
\begeq 
R_6(\eta) = \sum_{n=1}^\infty {2\sin(\pi\eta n)\over (\pi n)^4} 
\endeq 
\begin{eqnarray}  
R_7(\eta) & = & \sum_{m=1}^\infty {2\sin(\pi\eta m)\over (\pi m)^2} 
        \sum_{n=-\infty;\neq 0;\neq m}^\infty {|n|\over n} 
             \sin(\pi\eta n) \nonumber \\ 
        & & \mbox{} \times {\cos^2(\pi(n-m)/2)\over \pi^2(n-m)^2} 
\end{eqnarray} 
Comparison to direct summation shows that the expression (\ref{kappasquare}) 
is accurate to within 5\% over the whole range of $\kappa$. 
 
Finally, to express $U_3$, we need the expansion 
\begin{eqnarray} 
S_4(\kappa,\eta) & = & {\eta +i\pi\eta^2\kappa\over\rho_0(\kappa)} 
        - {2\pi^2\eta^3\kappa^2+i\pi^3\eta^4\kappa^3\over 3\rho_0(\kappa)} \nonumber \\ 
  & & \mbox{} + R_1(\eta)/2 \nonumber + 2\pi i \left[P(\eta)-\eta R_1(\eta)\right]\kappa \nonumber \\ 
  & & \mbox{} + \left[6R_4(\eta)-8\eta R_3(\eta)-\eta^2R_1(\eta)\right]  
      \pi^2\kappa^2/4 \nonumber \\ 
  & & \mbox{} + \order(\kappa^3). 
\end{eqnarray} 
For Eq. (\ref{theequation}), we need 
\begeq 
2\sin(\pikap)\Re\left(ie^{i\pikap/2}U_3(\kappa,1/2)\right) = 8\pi^2P(\eta)\kappa^2 + \order(\kappa)^4. 
\endeq 
 
Collecting all these results, we can finally write down the 
complete expressions for the coefficients $a(\kappa)$, $b(\kappa)$, 
and $c(\kappa)$ of Eq. (\ref{abcequation}) up to order $\kappa^2$: 
\begin{eqnarray} 
a(\kappa) & = & 2P(1/2) + \left[12 R_5(1/2)-{3\over 4}P(1/2)\right] 
    \pi^2\kappa^2 \nonumber \\ 
 & & \mbox{} + wr \biggl\{{1-{5\over 24}\pi^2\kappa^2\over  
      \rhotil_0(\kappa)} \nonumber \\ 
 & & \quad \mbox{} + \left[P(1/2)+R_3(1/2)\right]  
      \pi^2\kappa^2\biggr\} \\ 
b(\kappa) & = & \left(wr - g/2\right)(1-\pi^2\kappa^2/8) \nonumber \\ 
 & & \mbox{} - \left({2\pi^2P(1/2)\cot\theta\over\Lambda^2} 
                +{R_0\over 2}\right)\kappa^2 \nonumber \\ 
 & & \mbox{} - 2w{\rm Pe}r^2k_E 
      \biggl\{{1-{5\over 24}\pi^2\kappa^2\over  
        \rhotil_0(\kappa)} \nonumber \\ 
 & & \quad \mbox{} + 
         \left[P(1/2)+R_3(1/2)\right] \pi^2\kappa^2\biggr\}  \\ 
c(\kappa) & = & 8\pi^2P(1/2)\left(1-{1\over\Lambda^2}\right)\kappa^2. 
\end{eqnarray} 
To obtain the expressions given by Eqs. (\ref{approxcoeff}), 
we remark that for small $\kappa$ we can neglect most of the 
terms listed above. We have to be careful, however, to keep 
track of all physical effects. For example, formally the 
leading order term in $a(\kappa)$ is of order $\kappa^{-1}$ 
if we use $\rhotil_0(\kappa) = 2\pi|\kappa|$. But this term 
comes with a prefactor $w$, proportional to the impurity 
concentration, whereas the leading order term in the expansion 
of $U_1^\alpha$, arising from the eutectic diffusion field, 
is independent of $w$. Hence for the approximation to be valid 
for arbitrary $w$, we need to keep both terms, leading to 
Eq. (\ref{akappa}) for $a(\kappa)$. Similarly, in $b(\kappa)$, 
we need to keep the dominant terms for {\em each} physical 
effect, even if their order is higher than other terms we  
may neglect. In particular, we must keep the capillary term 
that is of order $\kappa^2$ and has a prefactor of order unity.  
Let us show that, at the onset of instability,  
we can indeed neglect other terms of order 
$\kappa$ and $\kappa^2$. Keep in mind that we are interested 
in a regime where $\kappa$ is small, but not {\em too} small; 
a plausible estimate is $\kappa\approx 0.01$. First, there is  
the correction 
$-\pi^2\kappa^2(wr-g/2)/8$ to the constitutional supercooling 
criterion, of order $\kappa^2$. But as $wr-g/2$ is very 
small near the onset, this term is actually much smaller 
than the capillary term. Next, there are terms of orders 
$\kappa$ and $\kappa^2$ arising from the impurity 
contribution. But all of these come with a prefactor of 
$w {\rm Pe} k_E$; as both $w$ and ${\rm Pe}$ are small quantities and $k_E<1$, 
it seems justified to neglect them. For example, for 
$w=0.1$, ${\rm Pe}=0.01$, and $\kappa=0.01$, the largest 
neglected term is of order $w{\rm Pe}\kappa \sim 10^{-5}$, 
whereas the capillary term is of order $\kappa^2 \sim 10^{-4}$. 
Hence, Eq. (\ref{bkappa}) for $b(\kappa)$ seems well justified 
at the onset of instability. 
 
\section{Effective surface tension} 
\nobreak 
\noindent 
We will give here an expression for the Gibbs-Thomson constant  
of a lamellar eutectic interface $\Gamma_E$ that 
describes the shift of the average interface temperature 
when the composite interface is curved on a scale much larger 
than the lamellar spacing. This analysis is necessary because, 
in a composite material, the interface with the weaker surface 
tension will absorb more of the curvature, leading 
to an effective surface tension that depends on the volume 
fraction. Note that the expression derived here is valid in
thermodynamic equilibrium and contains only the ``geometric part'' 
of the effective Gibbs-Thomson constant $\Gamma_{\rm eff}$ for
a {\em moving} eutectic front, in which the stabilizing effect
of the interlamellar diffusion has to be included.
 
Consider a lamellar interface that is curved such that 
the $\alpha\beta$ solid-solid interfaces on the two sides 
of a lamella pair make a small angle $\phi$. Suppose that the 
$\beta\alpha$ interface between them is turned by an angle 
$\phi_1$. The corrections of the curvature with respect to the 
planar front values are then given by (using the fact that 
$\phi\approx \lambda/R$ for large radii of curvature $R$) 
\begeq 
\delta K_\alpha = {\cos\theta_\alpha\over\eta\lambda} \phi_1 
   \quad {\rm and} \quad 
\delta K_\beta = {\cos\theta_\beta\over(1-\eta)\lambda}  
   \left(\phi-\phi_1\right). 
\endeq 
As the average temperature of neighboring lamellae should be the 
same, we must have 
\begeq 
\delta K_\alpha \Gamma_\alpha = \delta K_\beta \Gamma_\beta. 
\endeq 
From this condition, we can determine the unknown angle $\phi_1$. 
Finally, we obtain the undercooling of the interface as 
\begeq 
\Delta T = \Gamma_E K 
\endeq 
with 
\begeq 
\Gamma_E = {\Gamma_\alpha \Gamma_\beta 
                 \cos\theta_\alpha\cos\theta_\beta \over 
                (1-\eta) \Gamma_\alpha\cos\theta_\alpha + 
                \eta \Gamma_\beta\cos\theta_\beta}. 
\label{gedef} 
\endeq

\end{document}